\documentclass[11pt]{article}

\usepackage[mathscr]{euscript}

\usepackage{amsmath,amsfonts,amsbsy,amssymb,array,accents,dsfont}
\usepackage{enumerate,array,latexsym,graphicx,mathrsfs,verbatim,psfrag}
\usepackage{bm} 
\usepackage[normalem]{ulem}
\usepackage{multirow}
\usepackage{cellspace}
\usepackage{pdflscape}
\usepackage{booktabs} 
\usepackage[usenames]{color}
\usepackage[utf8]{inputenc}
\usepackage[english]{babel}
\usepackage{fancybox}

\usepackage{datetime}

\usepackage[nosort]{cite}
\usepackage{chngpage} 
\usepackage{setspace}
\usepackage{tensor}

\usepackage{enumitem}
\setlist{itemsep=0pt}

\usepackage{geometry}
\usepackage{longtable} 
\allowdisplaybreaks
\usepackage{pdflscape}

\usepackage[vcentermath,stdtext]{youngtab} 
\usepackage{lieart} 

\definecolor{cardinal}{rgb}{0.6,0,0}
\definecolor{darkgreen}{rgb}{0.1,0.4,0}
\definecolor{golden}{rgb}{0.92, 0.7, 0}
\definecolor{midnight}{rgb}{0, 0, 0.5}
\definecolor{darkblue}{rgb}{0.3,0.3,0.7}
\definecolor{darkred}{rgb}{0.7,0,0}

\usepackage{hyperref}
\hypersetup{
  colorlinks=true,
  linkcolor=darkblue,
  citecolor=darkred,
  urlcolor=darkgreen,
  linktocpage=true,
  linktoc=page
}

\usepackage{arydshln}
\usepackage{mathtools}

\newcommand{\captionfonts}{\small}
\makeatletter  
\long\def\@makecaption#1#2{%
  \vskip\abovecaptionskip
  \sbox\@tempboxa{{\captionfonts #1: #2}}%
 \ifdim \wd\@tempboxa >\hsize
    {\captionfonts #1: #2\par}
  \else
    \hbox to\hsize{\hfil\box\@tempboxa\hfil}%
  \fi
  \vskip\belowcaptionskip}
\makeatother   

\DeclareMathSymbol{\medhatsym}{\mathord}{largesymbols}{"62} 

\DeclareMathSymbol{\medtildesym}{\mathord}{largesymbols}{"65}

\newcommand{\comm}[1]{} 

\renewcommand{\arraystretch}{1.2}
\def\IC{\mathbb{C}}

\def\IP{\mathbb{P}}

\def\({\left(}
\def\){\right)}
\def\[{\left[}
\def\]{\right]}

\def\coeff#1#2{{\textstyle \frac{#1}{#2}}}

\def\One{{\hbox{ 1\kern-.8mm l}}}

\def\barray{\begin{array}}
\def\earray{\end{array}}
\def\be{\begin{equation}}
\def\ee{\end{equation}}
\def\bea{\begin{eqnarray}}
\def\eea{\end{eqnarray}}
\def\bal{\begin{align}}
\def\eal{\end{align}}

\def\eo{\overset{_{\phantom{.}\circ}}{e}{}}
\def\go{\overset{_{\phantom{.}\circ}}{g}{}}
\def\sc{\overset{_{\phantom{.}\circ}}{s}{}}

\def\HJ{{Hamilton-Jacobi }}
\def\bfs#1{\mathbf{#1}}

\numberwithin{equation}{section} 
\numberwithin{table}{section} 

\makeatletter
\g@addto@macro\bfseries{\boldmath}
\makeatother

\definecolor{cardinal}{rgb}{0.6,0,0}
\definecolor{darkgreen}{rgb}{0,0.4,0}
\definecolor{purple}{rgb}{0.5, 0, 0.5}
\definecolor{golden}{rgb}{0.92, 0.7, 0}
\definecolor{midnight}{rgb}{0, 0, 0.5}
\definecolor{darkblue}{rgb}{0, 0, 0.8}

\def\DG{\DarkGreen}

\def\IC{\mathbb{C}}
\def\Neql#1{{\cal N}\!=\!{#1}}
\def\coeff#1#2{\relax{\textstyle {#1 \over #2}}\displaystyle}

\def\cB{{\cal B}}
\def\cC{{\cal C}}

\def\cF{{\cal F}}

\def\cI{{\cal I}}
\def\cJ{{\cal J}}

\def\cL{{\cal L}}
\def\cM{{\cal M}}
\def\cN{{\cal N}}

\def\cU{{\cal U}}
\def\cV{{\cal V}}

\def\eql{~=~}

\def\sech{{\mathop{\rm sech}}}

\def\SL{{\rm SL}}

\def\diag{{\rm diag}}

\def\SO{{\rm SO}}

\def\SU{{\rm SU}}

\def\so{\frak{so}}
\def\su{\frak{su}}
\def\sp{\frak{sp}}
\def\gl{\frak{gl}}

\def\fg{\frak{g}}
\def\fh{\frak{h}}

\def\bbR{\mathbb{R}}

\def\cals#1{\mathcal{#1}}

\newcommand{\abs}[1]{\ensuremath{\left|#1\right|}}

\begin{document}


\begin{flushright}
\end{flushright}

\vspace{6mm}

\begin{center}
\begin{adjustwidth}{-7mm}{-7mm} 
\begin{center}
{\fontsize{20}{20} \bf{Separability in Consistent Truncations}} \medskip \\
\end{center}
\end{adjustwidth}
\vspace{10mm}

\centerline{{\bf Krzysztof Pilch$^1$, Robert Walker$^{2}$ and Nicholas P. Warner$^{1,3,4}$}}
\bigskip
\bigskip
\vspace{1mm}

\centerline{$^1$\,Department of Physics and Astronomy,}
\centerline{University of Southern California,} \centerline{Los
Angeles, CA 90089-0484, USA}
\vspace{2mm}
\centerline{$^2$ Instituut voor Theoretische Fysica, KU Leuven,} 
\centerline{Celestijnenlaan 200D, B-3001 Leuven, Belgium}
\vspace{2mm}
\centerline{$^3$ Institut de Physique Th\'eorique,}
\centerline{Universit\'e Paris Saclay, CEA, CNRS, }
\centerline{Orme des Merisiers,  F-91191 Gif sur Yvette, France}
\vspace{2mm}
\centerline{$^4$\,Department of Mathematics,}
\centerline{University of Southern California,} \centerline{Los
Angeles, CA 90089, USA}

\vspace{9mm} 
{\footnotesize\upshape\ttfamily pilch @ usc.edu ,~ robert.walker @ kuleuven.be,~warner @ usc.edu} \\

\vspace{9mm}
 
\textsc{Abstract}

\end{center}
\begin{adjustwidth}{6mm}{6mm} 
 
\vspace{2mm}
\noindent
\noindent
The separability of the \HJ equation has a well-known connection to the existence of Killing vectors and rank-two Killing tensors. This paper combines this connection with the detailed knowledge of the compactification metrics of consistent truncations on spheres. The fact that both the inverse metric of such compactifications, as well as the rank-two Killing tensors can be written in terms of bilinears of Killing vectors on the underlying ``round metric,'' enables us to perform a detailed analyses of the separability of the \HJ equation for consistent truncations. We introduce the idea of a {\it separating isometry} and show that when a consistent truncation, without reduction gauge vectors, has such an isometry, then the \HJ equation is {\it always} separable. When gauge vectors are present, the gauge group is required to be an abelian subgroup of the separating isometry to not impede separability. We classify the separating isometries for consistent truncations on spheres, $S^n$, for $n=2, \dots, 7$, and exhibit all the corresponding Killing tensors. These results may be of practical use in both identifying when supergravity solutions belong to consistent truncations and generating separable solutions amenable to scalar probe calculations. Finally, while our primary focus is the \HJ equation, we also make some remarks about separability of the wave equation.

\bigskip

\end{adjustwidth}

\vspace{8mm}
 

\thispagestyle{empty}

\newpage


\baselineskip=11.5pt
\parskip=1pt
\setcounter{tocdepth}{2}

\tableofcontents

\newpage 

\baselineskip=15pt
\parskip=3pt

\section{Introduction}
\label{sect:introduction}

Geodesics and scalar waves represent the two simplest, and perhaps most useful geometric probes.  The geometric optics approximation relates these probes in that, at high frequencies, or short wavelengths, the normals to wave fronts become tangents to geodesics. The wave equation 
\begin{equation}\label{WaveEqn}
\nabla^M \nabla_M  \Psi  ~=~  \frac{1}{\sqrt{\abs{G}}} \,\partial_{M} \,\Big( \sqrt{\abs{G}} \, G^{MN}\partial_{N}\Psi \Big) ~=~ -m^{2}\Psi\,,
\end{equation}
reduces then to the Hamilton-Jacobi    equation
\begin{equation}\label{HJ-eqs}
G^{MN}\,{\partial S\over\partial x^M}\,{\partial S\over\partial x^N}\eql -m^2\,,
\end{equation}
for the principal function $S$, where $\Psi(x)\sim \exp S$. The analysis of these equations is, of course, greatly simplified if there are isometries. For every Killing vector, $K$, generated by an isometry, there is a conserved momentum, $p_K\equiv K^M \partial_MS$, for the geodesics,  and the equations \eqref{WaveEqn} and  \eqref{HJ-eqs} may be, at least partially, separated by introducing coordinates along the symmetry directions,

More generally, the  wave equation and the corresponding \HJ equation    might have additional separation constants that go beyond those that are a consequence of the isometries.  For the geodesic problem, such separation constants correspond to non-trivial rank two Killing tensors, $\cals K_{MN}$, satisfying
\begin{equation}\label{KillTens}
\nabla_{(M}\cals K_{NP)}\eql 0\,,
\end{equation}
that define non-trivial conserved quantities, which are quadratic in velocities: 
\begin{equation}
\cals K_{MN} \, \frac{d x^M}{d \lambda}\, \frac{d x^N}{d \lambda}   \,.
\label{Ktensor1}
\end{equation}
Perhaps the most celebrated example of this was the discovery of the non-trivial Killing tensor in the Kerr metric \cite{Carter:1968rr,Carter:1968ks}.   

There is a further generalization of this story to the conformal structure of the manifold.  There are circumstances in which only the massless \HJ  equation is separable, then there are only conformal Killing vectors, $\zeta_M$, or conformal Killing tensors, $\xi_{MN}$, defined by
\begin{equation}
\nabla_{(M }\, \zeta_{N)} ~=~ \gamma \,  G_{MN} \,, \qquad  \nabla_{(P }\, \xi_{MN )} ~=~  \eta_{(P }\,  G_{MN )}  \,,
\label{confKtens1}
\end{equation}
where $G_{MN}$ is the metric and
\begin{equation}
 \gamma  ~=~ \frac{1}{n} \, \nabla_{M }\, \zeta^{M}    \,, \qquad   \eta_{M } ~=~ \frac{1}{n+2} \,\big(  \nabla_{M }\, ( \xi^N{}_N) + 2\,  \nabla^{P  }\, \xi_{P M} \big)    \,.
\label{confKtensrelns}
\end{equation}
Since the right-hand side of  \eqref{confKtens1} vanishes when all the indices are contracted with a null tangent vector, such conformal Killing vectors and tensors provide conserved quantities along null geodesics. This still provides invaluable insight into the geometry of the background. 

The purpose of this paper is to highlight the fact that consistent truncations provide a fertile ground for the discovery of non-trivial Killing and conformal Killing tensors. Indeed, we will exhibit several consistent truncations whose metrics have insufficient isometries to make separability manifest, and yet have fully separable massless \HJ equations.  Such metrics therefore have at least one non-trivial conformal Killing tensor, as was discussed recently in  \cite{Chervonyi:2015ima}.

Conversely, holography and the study of microstate geometries has led to physically interesting metrics that are generated by non-trivial solutions of consistent truncations. Such metrics are often  very complicated and seem rather intractable.  The fact that some of them have conformal Killing vectors, or conformal Killing tensors, means that these geometries can be probed using scalar waves and geodesics far more easily than one would expect. 

A consistent truncation in supergravity is the embedding of a lower-dimensional supergravity theory into a higher-dimensional one.  To say that the truncation is ``consistent," means that if one solves the equations of motion in the  lower-dimensional theory, then the result also leads to a solution of the  higher-dimensional theory, via an ``uplift." For this to occur, the space-time manifold, $\mathcal{M}$, of the $D$-dimensional theory must be a fibration with a compact $n$-dimensional Riemannian manifold, $\mathcal{F}$, over the spacetime manifold, $\mathcal{B}$, of the $d$-dimensional theory (and so $D=d+n$). 

The simplest examples arise when $\mathcal{F}$ has a transitive symmetry group, $G$, and one can easily obtain a consistent truncation by restricting to the fields that are singlets under the action of $G$. This includes the ``trivial'' torus compactifications. However, there are now well-known consistent truncations using spheres in which one allows fields to have very specific, non-trivial  (and non-singlet) dependence on the spherical harmonics. Such consistent truncations include the M-theory on  $S^7$  \cite{deWit:1984nz,deWit:1986iy} and on $S^4$ \cite{Nastase:1999cb,Nastase:1999kf}, respectively,  type  IIB supergravity on $S^5$ \cite{Khavaev:1998fb,Pilch:2000ue,Cvetic:2000nc,Lee:2014mla,Baguet:2015sma},  massive type IIA on $S^6$ \cite{Guarino:2015jca,Guarino:2015vca}, certain other sphere reductions \cite{Cvetic:1999xp,Cvetic:2000dm,Nastase:2000tu}, and particular $S^3$  compactifications of six-dimensional supergravities \cite{Samtleben:2019zrh}.  Indeed, it was the surprising separability of the massless wave equation for the ``superstrata'' solutions of six-dimensional supergravity \cite{Bena:2017upb,Walker:2019ntz}, which were subsequently shown to be part of an $S^{3}$ truncation in \cite{Mayerson:2020tcl}, that stimulated this more detailed study of separability in consistent truncations.

A consistent truncation requires ``uplift" formulae that precisely define how the dynamics of the fields on $\cB$ are encoded into the dynamics of the fields on $\mathcal{M}$. Once a solution of the theory on $\cB$ is found, inserting it into the uplift ansatz must then, necessarily, give a solution to the equations of motion on $\cM$. Since we are interested in the geodesics and the \HJ equation \eqref{HJ-eqs},   we are only going to need the metric uplift formula, that is, how the  vectors, scalars and metric on $\cB$ uplift to the complete metric on $\cM$.  Fortunately there is a universal approach to obtaining such an uplift formula based on the Kaluza-Klein Ansatz using the techniques first developed in \cite{deWit:1984nz} and then used to construct the consistent truncations above. 

The general uplift formula for the metric will be discussed  in Section~\ref{sec:uplift}. For now, we note that, in the  absence of Kaluza-Klein (KK) vector fields, the full metric takes the  form of a warped product:
\begin{align}
ds_{\cM}^{2} \equiv G_{MN} \, dx^{M}\, dx^{N} = \Delta ^{- \frac{2}{d-2}} \,   ds_{\cB}^{2} + ds_{\cF}^{2}  \,. \label{metricDecomp}
\end{align}
where  $ds_{\cB}^{2}=  g_{\mu\nu}(x)dx^\mu dx^\nu $ is the metric on the $d$-dimensional base, $\cB$. A mixing between the base and the fiber in the metric \eqref{metricDecomp} arises from a nontrivial  dependence of the internal metric, $ds_\cF^2=g_{mn}(x,y)dy^mdy^n$, and of the warp factor, $\Delta(x,y)$.
From the perspective of the lower-dimensional theory, these deformations of the internal metric, and warp factor, encode non-trivial scalar fields.  Typically, if all these scalar fields vanish then the internal metric, $d\sc_\cF= \go_{mn}(y)dy^mdy^n$, is ``round'' in that it has a transitive isometry, and then the scalar fields  on $\cals B$ are viewed as inducing  a deformation of  the ``round metric,''   on $\cals F$. 

More precisely, for a consistent truncation on a sphere $S^n$, the internal metric is given by the universal formula for its inverse:
\begin{equation}\label{InvMet}
\Delta ^{  -\frac{2}{d-2}}\,g^{mn}\eql {1\over 4}\,M_{ABCD}(x) \,K^{m\,AB}(y)K^{n\,CD}(y)\,,
\end{equation}
where $K^{AB} $ are the Killing vectors corresponding to the $\so(n+1)$ isometry of the round sphere metric and  $M_{ABCD} $  is an $\so(n+1)$ tensor determined by the scalar fields. In particular, for the vanishing scalars,
$M_{ABCD}=\delta^{AB}_{CD}$, $\Delta=1$, and \eqref{InvMet} reproduces the round sphere metric, $\go_{mn}$. Requiring this round sphere metric to solve the supergravity equations defines the vacuum of the theory, and the choices $M_{ABCD}$ and  $\Delta$ for this vacuum sets the normalization of the Killing vectors, $K^{AB}$. 

We will show in this paper that there is a remarkable  synergy between consistent truncations on spheres and the separability of the  \HJ equation.  At its core, this synergy works because of the theorem that all  rank two Killing tensors on a ``round'' sphere, or more generally any constant curvature manifold, are linear combinations of  bilinears in the Killing vectors \cite{Kalnins1980a,
 sumitomo1981,Takeuchi1983,Thompson1986}.  This means that (\ref{InvMet}) can be decomposed into Killing tensors and thus the \HJ equation can be separated to some degree, depending on the details of those Killing tensors.  We will show how this works in detail, establishing when a complete separation can be achieved and when the \HJ equation is only partially separable.  There is a lot of literature on separability of the \HJ equations in general, but the new element in our work is the symbiosis with the form of the internal metrics in consistent truncations.

We will also introduce the idea of ``separating isometries'' as a powerful  tool as well as a classification technique.  The idea is to consider subgroups, $G$, of the isometry group $\overset{_{\phantom{.}\circ}}{G}{}$  of the ``round'' metric and consider the most general consistent truncations that have $G$ as an isometry.  We find that separability of such a generic consistent truncation (in the absence of KK vector fields) is entirely determined by $G$, and if the massless \HJ equation is fully separable we will refer to  $G$ as  a separating isometry. 

In the body of this paper, we will focus on the separation of the \HJ equation, and return to the wave equation in Appendix~\ref{App:sepwave}.  We simply note here that if one broadens the notion of separability to encompass finite dimensional representations of gauge symmetries, then such a separation of the wave equation can be extended to include non-Abelian KK fields, and this may be important to probing non-trivial supergravity solutions. However, outside Appendix~\ref{App:sepwave}, we will use the  standard definition of separability and apply it to the  \HJ equation. As we will discuss, this will limit us to Abelian  KK fields. The separation of the \HJ equation is, of course, also important to probing non-supergravity solutions because it is directly related to the integrability of the geodesic problem.

In Section \ref{sect:separability} we give a brief survey of the relevant ideas, and literature, on separability, and in Section \ref{sect:MetTrunc} we review the relevant parts of consistent truncations. In particular, we recall the universal uplift formula for the metric of consistent truncations on spheres, including  vector fields, and introduce the notion of ``partial'' separability of the \HJ equation. The corresponding discussion of partial separability of massless scalar wave equations can be found in Appendix~\ref{App:sepwave}.  This is followed, in Section \ref{sec:pexample}, by a pedagogical discussion of an explicit example coming from an $\SO(3)\times \SO(3)$ invariant truncation of type IIB supergravity, which captures most of the generic features of our problem. Then in Section~\ref{sec:SepSph} we summarize some pertinent technical results for separability of  \HJ equations on Riemannian manifolds, introduce the idea of separable isometries and then use this to classify isometries that lead to separable \HJ equation in generic consistent truncations on  $S^2$, $S^3$, $\ldots$, $S^7$. This is illustrated with several explicit examples arising from known solutions of M-theory and type IIB supergravity in Section~\ref{ss:sepexhol}. In Section~\ref{sect:MicroExamples} we go back to six-dimensional microstate geometries and demonstrate separability in a large family of superstrata. A distinct feature here is that non-trivial Kaluza-Klein vector fields participate in the separation of variables. Finally, we conclude with a discussion of our results and their significance in Section~\ref{sect:Conclusions}.

\section{A brief summary of separability}
\label{sect:separability}

Conditions for separability of the  geodesic \HJ equation,  \eqref{HJ-eqs}, on Riemannian and Lorentzian manifolds have been well studied with the fundamental results going back to St\"ackel \cite{Stackel1893}, Levi-Civita \cite{LeviCivita1904a} and Eisenhart \cite{10.2307/1968433,eisenhart1997riemannian}. Separability here means that \eqref{HJ-eqs} admits a complete solution of the form:
\begin{equation}\label{SepS}
S(x^1,\ldots, x^D; c_1,\ldots, c_D)\eql\sum_{M=1}^D S_M(x^M;c_1,\ldots,c_D)\,,\qquad \det \left(\frac{\partial^{2}S}{\partial z^{M}\partial c_{N}} \right) \neq 0\,.
\end{equation}
 One can also consider multiplicative separability of the wave equation,  \eqref{WaveEqn}, of the form:\footnote{There is no sum over the index $M$ in the completeness condition here, so that the determinant is of a $2D\times 2D$ matrix, see (4.1) of \cite{Benenti2002}.}
\begin{align}
\Psi(z) = \prod_{M=1}^D \Psi_{M}(x^{M},\tilde{c}_{1},\tilde{c}_{2},\cdots \tilde{c}_{D})\,, \qquad \det \begin{pmatrix}
\frac{\partial}{\partial \tilde{c}_{N}} \left(\frac{\Psi_{M}'}{\Psi_{M}} \right) \\
\frac{\partial}{\partial \tilde{c}_{N}} \left(\frac{\Psi_{M}''}{\Psi_{M}} \right)
\end{pmatrix} \neq 0 \,, \label{ProdSep}
\end{align}
for a separate set of constants $\tilde{c}_{M}$. It turns out that separability of the wave equation of the form \eqref{ProdSep} implies that the Hamilton-Jacobi equation  \eqref{HJ-eqs} will separate additively as in  \eqref{SepS}. 
However, the converse is not necessarily true. For orthogonal metrics it can be shown to occur when the Ricci-tensor has vanishing off diagonal terms, while a full consideration for non-orthogonal systems are far more involved.\footnote{See \cite{Kalnins1980a} and the references therein for a discussion of these issues.}

Following on the earlier work by Levi-Civita and Eisenhart, more recently,  Kalnins and Miller (see, the monograph \cite{KMWBook} and the references therein), Benenti (see, e.g.,  the review \cite{Benenti2016}), and others gave  a variety of methods for establishing separability of both the \HJ equations and the corresponding wave equations in a geometric and coordinate independent way with a central role played by non-trivial (conformal) Killing tensors.

A convenient framework for analyzing   symmetries of the \HJ equation \eqref{HJ-eqs} is to think of symmetric tensors on $\cM$ as  functions on the phase space,  $T^*\cals M$, using the map: 
\begin{equation}\label{}
K^{M_1\ldots M_k}\qquad \longleftrightarrow \qquad K \equiv K^{M_1\ldots M_k}\,p_{M_1}\ldots p_{M_k}\,.
\end{equation}
The geodesic Hamiltonian for the metric, $G_{MN}$, is then defined as
\begin{equation}\label{}
H_\cM\eql {1\over 2}\,G^{MN}\,p_Mp_N\,,
\end{equation}
and the (conformal) Killing vector/tensor equations,  \eqref{KillTens} and \eqref{confKtens1}, are equivalent to the following equations for the Poisson brackets,
\begin{equation}\label{}
\{H_\cM,\cals K\}\eql 0\,,\qquad \{H_\cM,\zeta\}\eql -\gamma\,H_\cM\,,\qquad \{H_\cM,\xi\}\eql -\eta\,H_\cM\,.
\end{equation}

\subsection{Separability and consistent truncation}
\label{ss:sepConcTrunc}

We now assume that the metric has the form given in \eqref{metricDecomp} and \eqref{InvMet}. It follows that the geodesic Hamiltonian, $H_\cM$, on $\cM$  can be written as
\begin{equation}\label{HonM}
H_\cM\eql \Omega(x ,y ) \left[\,H_\cB(x ;\pi )+H_\cF(x ,y ;p )\,\right]\,,
\end{equation}
where $H_\cB\,\equiv\, {1\over 2}\,g^{\mu\nu}\pi_\mu\pi_\nu$ is the geodesic Hamiltonian on the base and
\begin{equation}\label{}
H_\cF~\equiv~ {1\over 2}\, \Delta^{-{2\over d-2}} g^{mn}\,p_mp_n\eql {1\over 8}\,M_{ABCD}\,K^{AB}K^{CD}\,,
\end{equation}
is the  Hamiltonian on the fiber.\footnote{Note that we define the  fiber Hamiltonian  using the rescaled metric $\Delta^{-2/(d-2)}g_{mn}$ on $\cF$.} It is rather remarkable that the uplift formula for the metric \eqref{InvMet}
leads to a simple, factorized dependence of $H_\cM$ on the warp factor
$\Omega\eql \Delta^{2\over d-2}\,.$
This turns out crucial for our analysis in two respects.  

First, when considering the massless \HJ equation \eqref{HJ-eqs} with $m^2=0$, the dependence on the warp factor drops out and one is left with the dynamics described by a much simpler Hamiltonian $H_\cM+H_\cF$. In many physically relevant examples that we will discuss, such as holographic RG-flows or  microstate geometries, the \HJ equation for $H_\cB$  trivially separates in natural coordinates on $\cB$, and so the separability reduces to considering the fiber, $\cF$. The  Hamiltonian $H_\cF$ can be expanded
\begin{equation}\label{HFexp}
H_\cF\eql \sum_\omega g_\omega(x )\,\cals K_\omega(y ;p )\,,
\end{equation}
where $g_\omega(x )$ are linearly independent functions on the base and $\cals K_\omega(y ;p )$ are Killing tensors for the round metric on $S^n$. In order to separate the massless \HJ equation between the base and the fiber, one must be able to set those Killing tensors to constants.  This can be done consistently only when
$\{H_\cF,\cals K_\omega\}=0$, and, for generic functions,  $g_\omega(x)$,  this means that the Killing tensors in \eqref{HFexp} must be in involution on $S^n$:
\begin{equation}\label{Kinvol}
\{\cals K_\omega,\cals K_{\omega'}\}\eql 0\,.
\end{equation}
The full separation of the massless \HJ equation reduces then to a well studied problem \cite{Kalnins1986}, but with an interesting twist. Since the geodesic Hamiltonian for the round metric, $\go_{mn}$, arises in the expansion \eqref{HFexp}, one is looking for separating coordinates for the \HJ equation on the round sphere, $S^n$, that   simultanously provide consistent separation of all other terms in  \eqref{HFexp}.  

Explicit examples of consistent truncations in the literature usually involve further reduction to a sector of the lower-dimensional theory that is invariant under some symmetry group $G$. In the higher dimensional theory, this symmetry becomes an isometry group of the uplifted metric \eqref{metricDecomp} along the fiber, and restricts the tensors, $\cals K_\omega$, in \eqref{HFexp} to those invariant under $G$. This leads to a subtle interplay between the symmetry of a truncation and separability of the massless \HJ equation. It turns out that in all  examples where the massless  \HJ equation separates, it does so because of isometries of the uplifted metric, and one can classify those symmetries by rather straightforward group theoretic arguments that we will present in Section~\ref{sec:SepSph}.

Secondly, the factorized warp factor in \eqref{HonM} determines symmetries of the full \HJ equation.  Note that, given a Killing tensor, $\cals K(y ;p)$, with respect to the fiber Hamiltonian, it satisfies
\begin{equation}\label{}
\{H_\cM,\cals K\}\eql \{\ln \Omega,\cals K\}\,H_\cM\,,
\end{equation}
and hence typically becomes a {\it conformal} Killing tensor for the full metric.  Reversing the argument, conformal Killing tensors for a warped metric may be a telltale of a Killing tensor lurking in the background and a separable massless \HJ equation.

The bottom line is that the structure of the base manifold depends on the details of the low-dimensional physics.  In many situations the  metric on $\cB$ does not present an obstacle to separability and so we will focus on the role of the compactification manifold, $\cF$.  We therefore begin by reviewing some  of the pertinent literature. 

\subsection{Separability on Riemannian manifolds}
\label{ss:sepRmflds}

Let $M$ be an $n$-dimensional Riemannian manifold.\footnote{Much of this discussion also carries over to pseudo-Riemannian manifolds.} A classic result of Levi-Civita \cite{LeviCivita1904a} gives the necessary and sufficient conditions for separability of the  Hamilton-Jacobi equation for a ``natural'' Hamiltonian
\begin{equation}\label{HJgen}
H\eql {1\over 2}\, g^{MN}p_Mp_N+U(x)\,,
\end{equation}
in a given coordinate system, $x^M$. Those Levi-Civita separability conditions require that:
\begin{equation}\label{LCcond}
\begin{split}
\frac{\partial H}{\partial x^{M}} \frac{\partial H}{\partial x^{N}} \frac{\partial^{2} H}{\partial p_{M} \partial p_{N} }  + \frac{\partial H}{\partial p_{M}} &\frac{\partial H}{\partial p_{N}} \frac{\partial^{2} H}{\partial x^{M} \partial x^{N} } -\frac{\partial H}{\partial x^{M}} \frac{\partial H}{\partial p_{N}} \frac{\partial^{2} H}{\partial p_{M} \partial x^{N} } - \frac{\partial H}{\partial p_{M}} \frac{\partial H}{\partial x^{N}} \frac{\partial^{2} H}{\partial x^{M} \partial p_{N} } \eql 0\,,\end{split}
\end{equation}
where there is no sum on $M$ and $N$.  Thus \eqref{LCcond} must me identically satisfied at all points on $T^*M$ for all distinct values of $M$ and $N$, $M\not=N$. In particular,  the vanishing  of the quartic terms in the momenta in \eqref{LCcond}  implies that  the Hamilton-Jacobi equation for the geodesic Hamiltionian, given by the first term in \eqref{HJgen}, must be separable irrespective of the potential, $U(x)$.

Separability  is not simply a property of the Hamiltonian, but also a property of coordinates. In this respect,  the  Levi-Civita conditions \eqref{LCcond} provide a direct calculation by which one can verify separability of the Hamilton-Jacobi equation in given coordinates  without constructing the resulting system of ODEs. What is less obvious is that, starting with the Hamiltonian \eqref{HJgen} written in some arbitrary coordinates, one can use \eqref{LCcond} to obtain coordinate-independent conditions on $H$ under which the Hamilton-Jacobi equation is separable and eventually  find the system of separating coordinates.

%

For orthogonal systems, that is when the metric tensor is diagonal, it was shown by  Eisenhart \cite{10.2307/1968433,eisenhart1997riemannian}, and then for general metrics by  Kalnins and Miller \cite{Kalnins1980a,Kalnins1981}, that geodesic separation is related to the existence of Killing vectors and Killing tensors of rank two. Subsequent work led to a variety of general criteria for separability of \eqref{HJgen}, which can be found in the comprehensive reviews  \cite{KMWBook,Benenti2016}.

The important point is that separability of the \HJ equation implies  the existence of  a complete set of linear and quadratic conserved quantities for the geodesic equation. (This is what is meant by saying that the geodesic equations are completely integrable.) The latter concept is coordinate invariant.  However, given such a set of conserved quantities, one would like to use them to exhibit a set of separable coordinates for the \HJ equation.  Fortunately this is possible, provided certain additional conditions are satisfied. To that end,  we will use the following practical, coordinate independent  method for establishing separability of the geodesic \HJ equation, given  in   Section 7 of \cite{Benenti2002} with  references to original papers. 

\medskip
\begin{enumerate}[leftmargin=20pt,topsep=0pt] 
\renewcommand{\labelenumi}{\it \theenumi}
\renewcommand{\theenumi}{C.\arabic{enumi}}
\item  \label{C1}  \it There must exist $0\leq r\leq n$ commuting Killing  vectors, $K_\alpha$, and $m=n-r$ Killing tensors, $\cals K_\omega$, of rank two  such that the system $(K_\alpha,\cals K_\omega)$ is integrable. This means that the Killing tensors, $\cals K_\omega$,  are invariant under $K_\alpha$'s and have vanishing Poisson brackets among each other.
\end{enumerate}
\medskip
Using the metric to lower/raise indices of the Killing tensors, one can view them as endomorphism of $TM$ or $T^*M$. Let $\Delta\subset TM$ be the distribution spanned by the Killing vectors and $\Delta^\perp$ its orthogonal complement with respect to the scalar product defined by the metric.
\medskip
\begin{enumerate}[leftmargin=20pt,topsep=0pt] 
 \setcounter{enumi}{1}
\renewcommand{\labelenumi}{\it \theenumi}
\renewcommand{\theenumi}{C.\arabic{enumi}}
\item  \label{C2}
\it  As endomorphisms of $TM$, the Killing tensors, $\cals K_\omega$, must preserve $\Delta^\perp$ and,  as linear operators on $\Delta^\perp$,  must be linearly independent and commute at each point on $M$.
\end{enumerate}
\medskip

Condition \ref{C2} implies that  at each point on $M$ there are $m$ linearly independent mutually orthogonal  eigenvectors, $V_a$, $a=1,\ldots,m$,  of the Killing tensors  acting on $\Delta^\perp$. Then, using  (i), one  proves that those eigenvectors form a completely integrable distribution, which, by the Frobenius theorem, implies that there exists  a coordinate system, $(x^M)=(x^a,\xi^\alpha)$,  such that
\begin{equation}\label{}
V_a \eql v_a(x)\,\partial _{a} \quad \text{(no sum)}\,,\qquad K_\alpha\eql \partial_{\alpha}\,.
\end{equation}
Note that the eigenvectors, $V_a$, do not necessarily commute and the functions, $v_a(x)$ can depend on all the coordinates $x^a$.  The important point is that the commutators of the $V_a$ close into the $V_a$, and so define a foliation, {\it \`a la} Frobenius, that can be coordinatized by the $x^a$. 
The system $(x^M)=(x^\alpha,\xi^\alpha)$ then defines a set of separable coordinates.

The set of the Killing tensors, $\cals K_\omega$, $\omega=1,\ldots,m$,  trivially includes the metric itself, which we identify with $\cals K_m$.
In the separable coordinates, $(x^a,\xi^\alpha)$, the metric and  other Killing tensors   have the semi-diagonal standard form,
\begin{equation}\label{stdform}
\cals K_\omega \equiv \cals K^{MN}_\omega p_Mp_N\eql \sum_{a=1}^m \,\cals K_\omega{}^{aa}\, p_a^2+\sum_{\alpha\beta=1}^r \cals K_\omega^{\alpha\beta}\,p_\alpha p_\beta\,,
\end{equation}
where 
\begin{equation}\label{expandK}
\cals K_\omega^{aa}\eql\varphi_\omega{}^a\,,\qquad \cals K_\omega ^{\alpha\beta}\eql \varphi_\omega{}^a\phi_a^{\alpha\beta}\,.
\end{equation}
The $m\times m$ matrix $(\varphi_\omega{}^a)\eql (\varphi_a{}^\omega)^{-1}$ is the inverse of what is known as the St\"ackel matrix. Both $\varphi_a{}^\omega=\varphi_a{}^\omega(x^a)$ and $\phi_a^{\alpha\beta}=\phi_a^{\alpha\beta}(x^a)$ depend only on the coordinate, $x^a$, corresponding to the lower index,  and all quantities in \eqref{expandK} are independent of the $\xi^\alpha$.  

Finally, the Killing vectors and tensors provide $n$ independent  integrals  of motion
\begin{equation}\label{}
K_\alpha\eql c_\alpha\,,\qquad \cals K_\omega \eql c_\omega \,,
\end{equation}
and the corresponding separated solution of the  geodesic \HJ equation is found from the  
generalized St\"ackel system of ODEs, which we  write using the momenta, $p_M=\partial_MS$,
\begin{equation}\label{Stacksys}
p_\alpha=c_\alpha\,,\qquad  p_a^2+\phi_a^{\alpha\beta}c_\alpha c_\beta \eql \varphi_a{}^\omega c_\omega\,.
\end{equation}
This makes the relation between Killing vectors/tensors, conserved quantities and separability of the \HJ equation precise.

Finally, for completness, let us note that for the natural \HJ equation \eqref{HJgen} with a nontrivial potential, $U$, additional conditions for separability are
\begin{equation}\label{}
K_\alpha U\eql 0\qquad \text{and}\qquad d(\cals K_\omega dU)\eql 0\,,
\end{equation}
for all Killing vectors and tensors. In the second equation, one should recall that, by raising an index with the metric,  $\cals K_\omega$ acts as an endomorphism on $T^*M$ and so the term in the parentheses is to be viewed as a one-form.  The corresponding  modification of \eqref{Stacksys} can be found in \cite{Benenti2002}.

\section{Metrics of consistent sphere truncations}
\label{sect:MetTrunc}

In this section we summarize the results for the uplift of the metric for generic Kaluza-Klein reductions. We then define the notion of ``partial'' separability of the \HJ equation between  the compact manifold, $\cF$, and the manifold of the reduced theory, $\cB$, and derive  necessary conditions for when it holds.   We then specialize to consistent truncations on spheres and discuss the structure of the metric on the fiber.
 
\subsection{Uplifting the metric}
\label{sec:uplift}

The uplift formulae for consistent truncations can be extremely complicated, especially for the tensor gauge fields.\footnote{See, however, more recent progress in, for example,   \cite{deWit:2013ija,Hohm:2014qga,Godazgar:2015qia,Baguet:2015sma}.}  However, the uplift formula for the metric can be deduced using some relatively simple insights. 

To make a long story short, we start with the standard  Kaluza-Klein ansatz for the metric on $\cals M$,
\begin{equation}\label{KKansatz}
G_{MN}\eql  
\begin{pmatrix}
G_{\mu\nu}(x,y)+B_\mu{}^m(x) B_\nu{}^n(x,y) g_{mn}(x,y) & B_\mu{}^m(x,y) g_{mn}(x,y) \\
  g_{mn}(x,y) B_\nu{}^n(x,y) & g_{mn}(x,y)
\end{pmatrix} 
 \,,
\end{equation}
where we have indicated the dependence of different terms on the coordinates, $x^\mu$, $\mu=1,\ldots,d$, on the base space-time, $\cB$, and $y^m$, $m=1,\ldots,n$,  on the internal compact manifold, $\cF$. The metric, $G_{\mu\nu}(x,y)$,   is related to the actual metric, $ g_{\mu\nu}(x)$, in the lower-dimensional theory on $\cals B$ by the rescaling
\begin{equation}\label{resmet}
G_{\mu\nu}\eql \Delta^{-\frac{2}{d-2}}\, g_{\mu\nu}\,,
\end{equation}
where the warp factor, $\Delta(x,y)$, is given by 
\begin{equation}
 \Delta  ~=~\sqrt{ \frac{\det(g_{mn})}{\det(\go_{mn})}}   \,.
\label{Deltadefd}
\end{equation}
 With this conformal rescaling, the Einstein action on $\cM$ reduces to the Einstein action on $\cB$.  

Next, one implements the main premise of the Kaluza-Klein program that  gauge invariance of the vector fields   must descend from general coordinate invariance of the fibration, which implies:\footnote{For a careful check of the relation between diffeomorphism and gauge symmetry for the Kaluza-Klein Ansatz,  see  \cite{Nastase:1999cb}.}
\begin{equation}
B_\mu{}^m~=~    A^\cI{}_\mu (x) \,  K^{\cI \, m} (y)   \,,
\label{gaugeframes}
\end{equation}
where $A^\cI{}_\mu (x)$ are the Kaluza-Klein gauge fields, $K^{\cI \, m} (y)$ are the Killing vectors for the {\it round metric}, $\go_{mn}(y)$,  on $\cF$ and $\cI, \cJ, \ldots$ are adjoint gauge indices.

Finally, the correct formula for the inverse of the internal metric on $\cF$ in supergravity theories can be deduced by using (\ref{gaugeframes}) to compare the supersymmetry transformations of the metric on $\cM$ with  supersymmetry transformations of the gauge fields in the theory on $\cB$ \cite{deWit:1984nz,deWit:1986iy,Pilch:2000ue}. One finds that:
\begin{equation}
\Delta^{-\frac{2}{d-2}} \, g^{mn} ~=~  \cU_{\cI \cJ}(x) \,  K^{\cI \, m} (y)  \,  K^{\cJ \, n} (y)   \,,
\label{metuplift}
\end{equation}
where $ \cU_{\cI \cJ}(x)$ is a specific, and known,  matrix constructed from the scalar fields of the theory on $\cB$.  The normalization of this matrix depends on the normalization of the Killing vectors, but can be set using the fact that when the scalars vanish  one has $\cU_{\cI \cJ} \sim \delta_{\cI \cJ}$ and the metric must be that of the round compact manifold.

Putting this all together, consistent truncation for a background involving scalar fields and Kaluza-Klein vectors on $\cB$, requires that the inverse metric on $\cM$ must be given by:\footnote{Note that $\mu,\nu,\dots$ indices will always be raised and lowered with the metric $ g_{\mu\nu}$ in \eqref{resmet}. }
\begin{equation}
G^{MN} ~=~ 
\Delta^{\frac{2}{d-2}} \,   \begin{pmatrix}
 g^{\mu \nu}(x)   & -    A^{\cI \mu}  (x)  \, K^{\cI \, n} (y)  \\
 - A^{\cI \nu} (x)  \, K^{\cI \, m} (y)  & \Big(\, \cU_{\cI \cJ}(x) +    A^{\cI \sigma}  (x)  A^\cJ{}_\sigma  (x)  \, \Big) \,  K^{\cI \, m} (y)  \,  K^{\cJ \, n} (y)
\end{pmatrix} 
 \,.
\label{MetUplift}
\end{equation}
The inverse metric following from $ds_{\cF}^{2}$, is then given by (\ref{metuplift}),  and gives a part of the lower right entry of (\ref{MetUplift}) with the warp factor factored out. 

\subsection{The Hamilton-Jacobi equation and the ``partial'' separation}
\label{ss:sepHJE}

The \HJ equation \eqref{HJ-eqs} for the inverse metric \eqref{MetUplift} reads
\begin{equation}\label{HJeqsf}
\Delta^{2\over d-2}\,\Bigg[g^{\mu\nu}{\partial S\over\partial x^\mu}{\partial S\over\partial x^\nu}-2 A^{\cals I\,\mu}K^{\cals I\,m}{\partial S\over\partial x^\mu}{\partial S\over\partial y^m}+
\Big(\, \cU_{\cI \cJ}  +    A^{\cI \sigma}   A^\cJ{}_\sigma   \, \Big) \,  K^{\cI \, m}   K^{\cJ \, n}  \,
{\partial S\over\partial y^m}{\partial S\over\partial y^n}
\Bigg]\eql -m^2\,.
\end{equation}
We will say that   \eqref{HJeqsf} is ``partially separable'' if the principal function $S(x,y)$ can be taken as a sum
\begin{equation}\label{}
S(x,y)\eql S_\cB(x)+S_\cF(y)\,,
\end{equation}
of functions that depend on the coordinates along $\cB$ and $\cF$, respectively.  The question is under what conditions the equation \eqref{HJeqsf} can be then reduced to a \HJ equation along $\cB$ of the form,
\begin{equation}\label{effeqs}
g^{\mu\nu}(x){\partial S_\cB\over\partial x^\mu}{\partial S_\cB\over\partial x^\nu}+\Lambda^\mu(x){\partial S_\cB\over\partial x^\mu}+V(x)\eql 0\,.
\end{equation}

It is clear that for a warp factor, $\Delta(x,y)$, that is a generic function on $\cals M$,  we must  set $m^2=0$, that is, the separation will hold  only for the massless \HJ equation given by the expression in the square bracket in \eqref{HJeqsf} set to zero.\footnote{The exception being when there is only a single ``breathing mode," so that $\cU_{\cI \cJ}\propto \delta_{\cI \cJ}$ and $\Delta = \Delta(x)$. In this instance the full massive \HJ equation may be separable.} Next consider the dependence  on the vector fields, $\cals A_\mu{}^{\cals I}$. Although the sum over the indices $\cals I$ and $\cals J$ runs over the entire adjoint representation, the Kaluza-Klein vector fields may gauge only a  subgroup of the isometries of the round metric on $\cals F$. Let $K_i$ be the vector fields on $\cals F$ corresponding to generators of the gauged symmetry, which are some linear combinations of the Killing vectors, $K^{\cals I}$. Hence, to be more precise, we should use
\begin{equation}\label{}
\cals A_\mu{}^{\cals I}K^{\cals I\,m}\eql \cals A_\mu{}^{i}K_i{}^m\,,
\end{equation}
and rewrite  \eqref{HJeqsf} in terms of the sum on the right hand side. 
The rest of the analysis is easier to carry out by looking at  the corresponding Hamiltonian obtained by setting
\begin{equation}\label{}
\pi_\mu\eql {\partial S\over\partial x^\mu}\,,\qquad p_m\eql {\partial S\over\partial y^m}\,.
\end{equation}

The resulting Hamiltonian for the massless \HJ equation can be recast into the following form
\begin{equation}\label{massH}
H\eql H_\cB(x;\pi)-A^{\mu\,i}(x)\pi_\mu K_i+\sum_\omega g_\omega(x)\cals K_\omega(y;p)\,,
\end{equation}
where the functions $g_\omega(x)$ are linearly independent combinations  of the functions $\,\cU_{\cI \cJ}$ and  $A^{\cI \sigma}   A^\cJ{}_\sigma $ while $\cals K_\omega$ are linear combinations of bilinears in the Killing vectors, $K^{\cals I}$.
Observe that the entire dependence of $H$ on the coordinates and momenta along the fiber, $\cF$, is through $K_i$  and $\cals K_\omega$.  Separation of the \HJ equation between the base and the fiber, resulting in the effective equation \eqref{effeqs}, amounts to setting consistently both $K_i$ and $\cals K_\omega$ to constants. Consistency means that $K_i$ and $\cals K_\omega$ must be actual constants of motion, that is
\begin{equation}\label{PwithH}
\{H,K_i\}\eql 0\,,\qquad \{H,\cals K_\omega\}\eql 0\,.
\end{equation}
Assuming that the Kaluza-Klein vector fields are suitably generic, and assuming the  linear independence of the functions $g_\omega(x)$ in the construction,  we deduce from \eqref{PwithH} that for a consistent partial separation we must have
\begin{equation}\label{invK}
\{K_i,K_j\}\eql 0\,,\qquad \{K_i,\cals K_\omega\}\eql 0\,,\qquad \{\cals K_\omega,\cals K_{\omega'}\}\eql 0\,.
\end{equation}

The  equations in \eqref{PwithH} and \eqref{invK} mean that (i)  the subgroup gauged by the Kaluza-Klein vector fields must be an abelian group of isometries of the metric in $H$, that is the metric \eqref{MetUplift} without the warp factor, (ii)   $\cals K_\omega$ are Killing tensors for that metric and are  invariant under the gauged isometries, and (iii)  the Killing tensors, $\cals K_\omega$,  are in involution.

The conditions \eqref{PwithH} and \eqref{invK} for the partial separation of the massless \HJ equation do not in general imply separability along the fiber. As was observed in Section~\ref{ss:sepRmflds}, this requires some further conditions on the consistent truncation, and we will return to these in Section~\ref{sec:SepSph}.

Before concluding we note that the wave equation affords a broader and interesting notion of partial separability. In particular, one can allow non-abelian KK gauge symmetries if one also allows the wave functions to lie in finite-dimensional representations of the  gauge group. More details may be found in Appendix~\ref{App:sepwave}.

\subsection{Sphere truncations}
\label{ssec:sphtr}

We now set the Kaluza-Klein vector fields to zero and specialize the metric  to that of the  sphere, $S^n$,  embedded in $\bbR^{n+1}$ as the hypersurface, $Y^AY^A=1$, where $Y^A$, $A=1,\ldots,n+1$, are the Cartesian coordinates. We will not use the induced metric from $\bbR^{n+1}$ but the uplifted fiber metric  \eqref{metuplift}. In terms of the Cartesian  coordinates, the $\so(n+1)$ Killing vectors are given by:
\begin{equation}\label{KillVecs}
K^{AB}\eql Y^AP_B-Y^BP_A\,,
\end{equation}
where $P_A$ is the momentum conjugate to $Y^A$ in $T^*\bbR^{n+1}$.  Since the Killing vectors   \eqref{KillVecs} are  tangent to the sphere, a Poisson bracket of any function of those vectors on $T^*S^n$ can be evaluated directly in the ambient space. This is clearly much simpler than working in some explicit coordinates on the sphere, which will be defined only locally.   

The Hamiltonian, $H_\cF$, in \eqref{HonM} is defined by the internal  metric given by the left hand side in  \eqref{metuplift}. It is convenient to normalize it such that
\begin{equation}\label{HcF}
H_\cF \eql {1\over 8}\,M_{ABCD} K^{AB}K^{CD}\,,
\end{equation}
were $M_{ABCD}$ is an $\so(n+1)$ tensor that depends on the position on the base, $\cB$, but is constant along~$S^n$. The geodesic Hamiltonian for the round metric on the unit radius sphere is then given by
\begin{equation}\label{geodHamSn}
H_{S^n}\eql {1\over 4} \,K^{AB}K^{AB}\,.
\end{equation}
It is clear that each term in the Hamiltonian \eqref{HcF} has the vanishing Poisson bracket with $H_{S^n}$ and thus is a Killing tensor for the round metric.\footnote{The Killing tensors given by \eqref{HcF}, with constant $M_{ABCD}$, have been  studied extensively in the mathematical literature (see, e.g., \cite{Schoebel2015a} and the references therein).}

The tensor, $M_{ABCD}$, has obvious symmetries
\begin{equation}\label{trsyms}
M_{ABCD}\eql -M_{BACD}\eql -M_{ABDC}\qquad \text{and}\qquad M_{ABCD}=M_{CDAB}\,,
\end{equation}
that follow from  \eqref{HcF}. This, however, does not account for the symmetry within the pairs of positions and the pairs of momenta after we substitute \eqref{KillVecs} in  \eqref{HcF}. A particularly elegant way to remove that degeneracy is to require that 
\begin{equation}\label{Bianchi1}
M_{A[BCD]}\eql 0\,,
\end{equation}
which together with \eqref{trsyms} defines $M_{ABCD}$ as an algebraic Riemann tensor in $\bbR^{n+1}$. When combined with the fact that all  rank two Killing tensors on a sphere, or more generally any constant curvature manifold, are linear combinations of  bilinears in the Killing vectors \cite{Kalnins1980a,
 sumitomo1981,Takeuchi1983,Thompson1986}, one obtains the 1-1 correspondence between the Killing tensors on $S^n$ and the algebraic Riemann tensors on the ambient space \cite{McLenaghan2004}.

We also note, that, like the Riemann tensor, if one assumes the symmetries (\ref{trsyms}), then (\ref{Bianchi1}) is equivalent to:
\begin{equation}\label{Bianchi2}
M_{[ABCD]}\eql 0\,.
\end{equation}

As an $\so(n+1)$ tensor, $M_{ABCD}$ decomposes into three components,
\begin{equation}\label{YTdec}
M_{ABCD}\eql M^\bullet_{ABCD}+ M^{\,\Yboxdim4pt\yng(2)}_{ABCD}+M^{\,\Yboxdim4pt\yng(2,2)}_{ABCD}\,,
\end{equation}
corresponding to the irreducible representations with dimensions\footnote{One may also note that the three representations of $\so(n+1)$  descend from a single irreducible representation of $\gl(n+1,\bbR)$ with the Young tableaux $\Yboxdim6pt\yng(2,2)$ \cite{McLenaghan2004}.}
\begin{equation}\label{}
\text{dim}\,\bullet\eql 1\,,\qquad \text{dim}~\Yboxdim8pt\yng(2) \eql {1\over 2}n(n+3)\,,\qquad \text{dim}~\Yboxdim8pt\yng(2,2) \eql {1\over 12} (n^2-4)(n+1)(n+3)\,.
\end{equation}
Adding these dimensions, we obtain the dimension of the space of  rank two Killing tensors on $S^n$ given by  the 
Delong \cite{delong1982killing}, Takeuchi \cite{Takeuchi1983}, Thompson \cite{Thompson1986} formula,
\begin{equation}\label{}
\text{dim}\,\cals K^2(S^n)\eql  {1\over n}\binom{n+2}{3}\binom{n+1}{2}\,.
\end{equation}

The singlet representation in \eqref{YTdec} corresponds, modulo the base dependent conformal factor,  to the geodesic Hamiltonian, $H_{S^n}$, for the round metric in \eqref{geodHamSn}. Using \eqref{KillVecs} and expanding the right hand side in \eqref{geodHamSn} we get
\begin{equation}\label{HXP}
H_{S^n}\eql {1\over 2}\,\big[(Y\cdot Y)(P\cdot P)-(Y\cdot P)(Y\cdot P)\big]\,,
\end{equation}
where the dot denotes the contraction in $\bbR^{n+1}$. Let $(y^m,p_n)$ be some coordinates and their conjugate momenta on $S^n$. Then $p_m=Y_m^A P_A$, where  $Y_m^A\equiv \partial_mY^A$. Using standard identities for the  harmonics, $Y^A$ and $Y_m^A$,  on a unit sphere:
\begin{equation}\label{harmid}
Y^AY^A\eql 1\,,\qquad Y^A_mY^A_n\eql\go_{mn}\,,\qquad \go^{mn}Y^A_mY^B_n\eql\delta^{AB}-Y^AY^B\,,
\end{equation}
one can show that the momenta $p_m$ and $P_A$ satisfy
\begin{equation}\label{}
P_A\eql \go^{mn}Y_m^A\,p_n+Y^A(Y\cdot P)\,.
\end{equation}
Substituting this in \eqref{HXP} and  using identities \eqref{harmid}, we verify that indeed
\begin{equation}\label{roundmetr}
H_{S^n}\eql {1\over 2}\,\go^{mn}\,p_mp_n\,.
\end{equation}

The Hamiltonian, $H_\cF$, corresponds  to a deformation of the round metric by a linear combination of the Killing tensors. As we have already discussed in Section~\ref{sect:separability}, it is natural to expand $H_\cF$ into linearly  independent functions on $\cB$, cf. \eqref{HFexp} and \eqref{massH},
\begin{equation}\label{Hexpand}
H_\cF\eql \sum_\omega g_\omega(x )\,\cals K_\omega(y;p)\,,
\end{equation}
where, for a separable \HJ equation, the round sphere  Killing tensors $\cals K_\omega$ must be in involution. This  severely restricts the possible terms in \eqref{Hexpand}. 

One method of imposing such a restriction in a controllable manner is to assume that the internal metric is invariant under some isometry $G\subset \SO(n+1)$. This is quite natural from the point of view of   consistent truncations where such isometries correspond to the scalar fields in the lower-dimensional supergravity being invariant under a subgroup of the gauge group. Let $K_i$, $i=1,\ldots,\text{dim}\,\fg$ be the Killing vectors corresponding to the generators, $T_i$, of the Lie algebra, $\fg\subset\so(n+1)$, of $G$. By the same argument that led to \eqref{Kinvol}, we must have
\begin{equation}\label{}
\{K_i,\cals K_\omega\}\eql 0\,,
\end{equation}
that is the allowed $\cals K_\omega$'s in \eqref{Hexpand} must be invariant under $\fg$.
As we will show in the next two sections, this restriction based on symmetry is quite powerful and leads to a large class of separable Hamiltonians.

\section{A pedagogical example}
\label{sec:pexample}

In this section we  illustrate some of the general discussion above with a simple example of an uplift of the $\SO(3)\times \SO(3)$ invariant sector of the maximal gauged  supergravity in five dimensions to type IIB supergravity on $S^5$. The uplifted metric in this truncation was obtained in \cite{Bobev:2020fon} using \eqref{metuplift} and shown to reproduce the one for half-BPS Janus solution derived in \cite{DHoker:2007zhm} directly in type IIB supergravity. An almost identical analysis  holds for the uplift of the $\SO(4)\times\SO(4)$ invariant sector of $\cals N=8$, four-dimensional supergravity to M-theory. See, \cite{Pope:2003jp,  DHoker:2009lky,Bobev:2013yra} for details of solutions constructed using uplift formulae.

\subsection{The metric and the massless \HJ equation}
\label{ss:SO3SO3}

The ten-dimensional metric derived using \eqref{metuplift} is given by\footnote{The coupling constant $g$ is related to the radius of the internal manifold. In the following we set $g=1$.} 
\begin{equation}
ds_{10}^{2} = \left( X_{1} X_{2}\right)^{1/4} \left[ds^2_{1,4} + \frac{4}{g^{2}} \left( d\theta^{2}+ \frac{\cos^{2}\theta}{X_{1}}\, d\Omega_{2}^{2} + \frac{\sin^{2}\theta}{X_{2}}\, d\widetilde{\Omega}_{2}^{2} \right) \right] \label{SO3SO3met}\,.
\end{equation}
The first term in the square bracket is the metric  $ds_{1,4}=g_{\mu\nu}dx^\mu dx^\nu$  for a solution in five-dimensions and  the second term is the metric along the fiber, $S^5$. This metric has the same structure as in \eqref{metricDecomp}, except that we have conveniently factored out the warp factor, $\Delta^{-2/3} = (X_{1}X_{2})^{1/4}$, in both terms. The $S^5$ here is a fibration of two unit two-spheres with the metrics
\begin{align}
d\Omega_{2}^{2} = d\phi_{1}^{2} + \sin^{2}\phi_{1}\, d\phi_{2}^{2}\,, \qquad d\widetilde{\Omega}_{2}^{2} = d\xi_{1}^{2} + \sin^{2}\xi_{1}\, d\xi_{2}^{2}\,,
\end{align}
over the interval  $0\leq\theta\leq\pi/2$. The deformations of the round metric on $S^5$ are parametrized by two scalar fields, $\alpha(x)$ and $\chi(x)$, in the five-dimensional theory through the  functions
\begin{equation}\label{Xfncts}
X_{1}= \sin^{2}\theta +e^{4\alpha(x)} \cosh 4\chi(x) \cos^{2}\theta  \qquad \text{and} \qquad X_{2}=e^{-4\alpha(x)}\cosh 4\chi(x) \sin^{2}\theta + \cos^{2}\theta \,.
\end{equation}

From the form of the metric  \eqref{SO3SO3met}, separability of the \HJ equation is by no means obvious. Yet, it is quite straightforward to check that by taking
\begin{align}
S(x,y) = S_{x}(x) + S_\theta (\theta) +S_{\phi_{1}}(\phi_{1}) +S_{\phi_{2}}(\phi_{2}) +S_{\xi_{1}}(\xi_{1}) +S_{\xi_{2}}(\xi_{2})\,,
\end{align}
the \HJ equation along $S^5$ separates into the following system of ODEs:
\begin{equation}\label{killsep}
S'_{\phi_{2}}(\phi_{2})=c_{\phi_{2}} \,, \qquad S'_{\xi_{2}}(\xi_{2})=c_{\xi_{2}}\,,  
\end{equation}
for the cyclic coordinates, $\phi_2$ and $\xi_2$, respectively, and 
\begin{equation}\label{ktsep}
\begin{split}
S_{\phi_{1}}'(\phi_{1})^{2} + \csc^2\phi_1\,c_{\phi_{2}}^2-c_{\phi_{1}}  & \eql  0\,, \\[6 pt]
S_{\xi_{1}}'(\xi_{1})^{2} + \csc^2\xi_1\,c_{\xi_{2}}^2 - c_{\xi_{1}}  & \eql  0\,, \\[6 pt]
S'_{\theta}(\theta)^{2} + \sec^2\theta\,c_{\phi_{1}}   + \csc^2\theta\,c_{\xi_{1}} -c_{\theta}  & \eql 0 \,.
\end{split}
\end{equation}
The resulting effective \HJ equation in five-dimensions is
\begin{equation}\label{fiveHJ}
g^{\mu\nu}\,{\partial S_x\over \partial x^\mu}{\partial S_x\over \partial x^\nu}+{1\over 4}\,
\Big[ c_{\theta}  -c_{\phi_{1}} - c_{\xi_{1}}   + 
\cosh 4\chi\left(c_{\phi_{1}} e^{4\alpha}+c_{\xi_{1}} e^{-4\alpha}  \right) \Big]\eql 0\,,
\end{equation}
which proves  partial separability in this example. Along the way we have also shown that the massless \HJ equation on $\cM$ fully separates along the $S^5$. For the solutions in \cite{Bobev:2020fon,DHoker:2007zhm}, the five-dimensional metric is a deformation of the AdS$_5$ metric with the scalar fields and the components of the metric tensor depending only on a radial coordinate, $r$. The full   separability of  \eqref{fiveHJ} is then trivial.

From  the ODEs \eqref{killsep} and \eqref{ktsep}, we also see that the separation involves two Killing vectors, which can be identified with the Cartan generators of the $\so(3)\times\so(3)$ isometry, two Killing tensors and a conformal Killing tensor, corresponding to the ODEs in (\ref{ktsep}) respectively. Those five conserved vectors/tensors are in involution, something that is much easier to see by working in the ambient space.

\subsection{The ambient space perspective}
\label{ss:ambsp}

The two $\SO(3)$'s of the isometry act  in $\bbR^6$ as rotations on $(Y^1,Y^2,Y^3)$ and  $(Y^4,Y^5,Y^6)$, respectively, generated  by the Killing vectors
\begin{equation}\label{}
(K^{12},K^{23},K^{31})\qquad \text{and}\qquad (K^{45},K^{56},K^{64}).
\end{equation}
Using this  embedding of the  $\so(3)\times\so(3)$  isometry in $\so(6)$ given by
\begin{equation}\label{}
\bf 6\quad \to\quad (3,1)+(1,3)\,,
\end{equation}
the resulting branchings of the $\so(6)$ irreps in \eqref{YTdec} are:\footnote{Our group theory conventions are the same as in \cite{Slansky:1981yr}. To compute branching rules here and in Section~\ref{sec:SepSph}, we have made an extensive use of the Mathematica package LieArt \cite{Feger2020a}.}
\begin{equation}\label{decso33}
\begin{split}
{\bf 1}  \quad \to\quad & (\irrep{1},\irrep{1}) \,,\\
{\bf 20'}  \quad \to\quad & (\irrep{1},\irrep{1})+(\irrep{3},\irrep{3})+(\irrep{5},\irrep{1})+(\irrep{1},\irrep{5})\,,\\
{\bf 84}  \quad \to\quad & (\irrep{1},\irrep{1})+2(\irrep{3},\irrep{3})+(\irrep{5},\irrep{1})+(\irrep{1},\irrep{5})+(\irrep{5},\irrep{3})+(\irrep{3},\irrep{5})+(\irrep{5},\irrep{5})\,.
\end{split}
\end{equation}
The uplifted metric  \eqref{metuplift} is a linear combination of three invariant Killing tensors on $S^5$ corresponding to the singlets in \eqref{decso33}. A convenient basis for those invariants consists of  the  ``Casimir invariant'' of $\so(6)$ \begin{equation}\label{}
\cals C_2^{\so(6)}\eql {1\over 4}\sum_{A,B=1}^6 K^{AB}K^{AB}\,,
\end{equation}
which is the geodesic Hamiltonian for the round metric on $S^5$ in \eqref{geodHamSn}, and the  Casimir invariants
of the two $\so(3)$'s:
\begin{equation}\label{}
 \cC_2^{\so(3)_1}\eql {1\over 4} \sum_{A,B=1}^3 K^{AB}K^{AB}\,,\qquad  \cC_2^{\so(3)_2}\eql {1\over 4} \sum_{A,B=4}^6 K^{AB}K^{AB}\,.
\end{equation}

The most general Hamiltonian, $H_\cF$, along the fiber invariant under  the $\SO(3)\times\SO(3)$ isometry is given by  a linear combination of these Casimirs,
\begin{equation}\label{HFso3}
H_\cF\eql g_0(x) \,\cC_2^{\so(6)}+g_1(x)\, \cC_2^{\so(3)_1}+g_2(x)\, \cC_2^{\so(3)_2},
\end{equation}
where $g_\omega(x)$ are arbitrary functions on the base. The three Casimir tensors are clearly in involution, together with the two Cartan generators, $K_1=K^{12}$ and $K_2=K^{45}$, of $\so(3)\times\so(3)$. 

As we have  discussed  in Section~\ref{ss:sepRmflds}, the existence of this number of  Killing vectors/tensors in involution is a necessary condition for separability of the Hamiltonian \eqref{HFso3}. To demonstrate separability of $H_\cF$, let us introduce   explicit coordinates, $(\theta,\phi_{1,2},\xi_{1,2})$, on $S^5$, such that 
\begin{equation}
\begin{aligned}
\begin{split}
Y^{1} &=   \cos \theta \, \sin \phi_{1} \, \cos\phi_{2} \,, \\
Y^{2} &=   \cos \theta \, \sin \phi_{1} \, \sin \phi_{2} \,, \\
Y^{3} &=   \cos \theta \, \cos \phi_{1} \,,
\end{split}
\qquad \qquad
\begin{split}
Y^{4} &=   \sin \theta \, \sin \xi_{1} \, \cos \xi_{2} \,, \\
Y^{5} &=   \sin \theta \, \sin \xi_{1} \, \sin \xi_{2} \,, \\
Y^{6} &=   \sin \theta \, \cos \xi_{1} \,.
\end{split}
\end{aligned}
\end{equation}
In terms of these coordinates and canonically conjugate momenta, we  find
\begin{equation}\label{kkso3}
K_1\eql p_{\phi_2}\,,\qquad K_2\eql p_{\xi_2}\,,
\end{equation}
\begin{equation}\label{}
\cC_2^{\so(3)_1}\eql {p_{\phi_1}^2\over 2}+{1\over 2}\,\csc^2\phi_1\,p_{\phi_2}^2\,,\qquad 
\cC_2^{\so(3)_2}\eql {p_{\xi_1}^2\over 2}+{1\over 2}\,\csc^2\xi_1\,p_{\xi_2}^2\,,
\end{equation}
and
\begin{equation}\label{c2so6}
\cC_2^{\so(6)}\eql {p_\theta^2\over 2}+\sec^2\theta\,\cC_2^{\so(3)_1}+\csc^2\theta\,\cC_2^{\so(3)_2}\,.
\end{equation}
The nested structure of \eqref{kkso3}-\eqref{c2so6} implies that by setting
\begin{equation}\label{}
K_1\eql c_{\phi_2}\,,\qquad K_2\eql c_{\xi_2}\,,\qquad \cC_2^{\so(3)_1}\eql {c_{\phi_1} \over 2}\,,\qquad 
\cC_2^{\so(3)_2}\eql {c_{\xi_1} \over 2}\,,\qquad \cC_2^{\so(6)}\eql {c_\theta \over 2}\,,
\end{equation}
the massless \HJ equation with the fiber Hamiltonian \eqref{HFso3} is partially separable and fully separable along the fiber, with the separation constants $c_\theta,\ldots,c_{\xi_2}$.

The fiber metric in \eqref{SO3SO3met} is recovered by setting
\begin{equation}\label{fsso3}
g_0\eql {1\over 4}\,,\qquad g_1\eql {1\over 4}\,( e^{4\alpha}\,\cosh 4\chi-1)\,,\qquad g_2\eql  {1\over 4}\,( e^{-4\alpha}\,\cosh 4\chi-1)\,,
\end{equation}
and is just a representative of a whole family of uplifted metrics for which the \HJ equation is separable due to the $\so(3)\times\so(3)$ isometry.

This simple calculation illustrates how an isometry can be used to restrict the allowed terms in  \eqref {Hexpand} and makes it more clear how the complicated functions $X_1$ and $X_2$ given in  \eqref{Xfncts} arise upon inversion of the metric from a simple internal Hamiltonian \eqref{HFso3}.

\subsection{Separability and symmetry}
\label{ssec:sepandsymm}

It is also quite instructive to explore, in this explicit example, whether the isometry of the metric alone can account for separability. To this end consider \eqref{SO3SO3met} but with the most general $\SO(3)\times\SO(3)$ invariant metric on the $S^{5}$:
\begin{align}
ds_{10}^{2} = ds_{1,4}^2 + \frac{4}{g^{2}} \left( d\theta^{2}+ A_{1}(x,\theta)\, d\Omega_{2}^{2} + A_{2}(x,\theta)\, d\widetilde{\Omega}_{2}^{2} \right) \,.
\end{align} 
where $A_{1,2}(x,\theta)$ are arbitrary functions of the base manifold and $\theta$. We have also dropped the warp factor for convenience. 

Proceeding with the separation of the massless \HJ equation as in Section~\ref{ss:SO3SO3}, we end up with the partially separated equation,
\begin{equation}\label{parsepso3}
g^{\mu\nu}\,{\partial S_\cB\over\partial x^\mu}{\partial S_\cB\over\partial x^\nu} +{1\over 2} \left[c_{\theta}^{2}+c_{\phi_{1}}^{2}\left( \frac{1}{A_{1}(x,\theta)} - \frac{1}{\cos^{2}\theta}  \right) +c_{\xi_{1}}^{2}\left( \frac{1}{A_{2}(x,\theta)} - \frac{1}{\sin^{2}\theta}  \right)\right]\eql 0\,.
\end{equation}
So the non-trivial identities required for separation are:
\begin{equation}\label{SepId}
\frac{1}{A_{1}(x,\theta)} - \frac{1}{\cos^{2}\theta} \eql f_{1}(x) \,, \qquad 
\frac{1}{A_{2}(x,\theta)} - \frac{1}{\sin^{2}\theta} \eql f_{2}(x) \,, 
\end{equation}
where $f_{1,2}(x)$ are some functions on $\cB$. For the specific $\SO(3)\times \SO(3)$ invariant solution considered in Section~\ref{ss:SO3SO3}:
\begin{align}
A_{1}(x,\theta) &=\sec^2\theta\,  X_{1}(x,\theta) =  e^{4\alpha(x)}\cosh 4 \chi(x) +\tan^{2}\theta  \,,\\[6 pt]
A_{2}(x,\theta) &= \csc^2\theta\,X_{2}(x,\theta) =e^{-4\alpha(x)}\cosh 4 \chi(x) +\cot^{2}\theta   \,,
\end{align}
and the identities \eqref{SepId} are indeed satisfied. A priori there is no reason to expect $A_{1,2}(x,\theta)$ to be constrained in such a way to allow this non-trivial separation.

We see from this example that separability of the massless \HJ equation is not just a consequence of the underlying symmetry, but results from an interplay between the symmetry and the structure of the uplifted metric in \eqref{HcF}.

\section{Separating isometries on spheres}
\label{sec:SepSph}

When looking at a consistent truncations on the $n$-sphere, $S^n$, one sometimes tries to simplify the problem by looking at sectors that are invariant under a particular  sub-algebra, $\fg$, of the full algebra of isometries,  $\so(n+1)$. This approach also turns out to be a powerful tool in the investigation of separability within consistent truncations, and, as we will discuss, the  separability of the \HJ equation for a {\it generic} $\fg$-invariant sector depends solely upon  the choice of $\fg$.  Indeed, we define $\fg$ to be a {\it separating isometry} precisely when the {\it generic}  $\fg$-invariant sector of the consistent truncation on $S^n$ has a separable \HJ equation in the absence of reduction gauge vectors.

It is important to note why we are insisting on the separability of {\it generic} $\fg$-invariant sectors. This is because, in Section \ref{ss:sepHJE} we crucially needed generic functions,  $g_\omega(x)$, in (\ref{HFexp}) in order to arrive at (\ref{invK}),  and, in particular, show that  partial separability requires that the Killing tensors, $\cals K_\omega$, in \eqref{Hexpand}  must be in involution.  If some of the $g_\omega$'s vanish, or these functions are linearly dependent,  then one could discard, or combine, some of the  $K_\omega$'s into a new basis so that the  new $g_\omega$’s become linearly independent. The resulting new $K_\omega$'s might then be in involution with one another and lead to separability of the \HJ equation.  This situation would arise if we started with a Lie algebra, $\fg$, that was not a separating isometry, but we chose a point in the configuration space at which $\fg$ becomes enhanced to a larger symmetry, $\hat \fg$, that is a separating isometry. It is also quite possible that there are examples in which there are separable loci in the configuration space of a non-separable isometries, $\fg$, and yet these loci do not involve some enhancement of the isometry.   

To avoid the complications of such special cases, we insist upon generic $\fg$-invariant sectors and can therefor invoke  (\ref{invK}).  Having made this step, we will systematically examine sub-algebras, $\fg \subset \so(n+1)$, for $n=2,\ldots,7$, and determine which algebras lead to separating isometries, or not.  This will automatically lead to a partially separable massless \HJ equation.  We start by  explaining our approach and we tabulate  the group theoretic details in Appendix~\ref{App:tables}. We then catalog the range of possibilities, giving the Killing tensors and the separable coordinates.

\subsection{Separating isometries}
\label{ssec:sepisoms}

Our task  is to classify ``separating isometries,'' $\fg\subset \so(n+1)$, of the uplifted metric \eqref{Hexpand} on $S^n$ for which the massless \HJ equation for the Hamiltonian \eqref{massH}, with no Kaluza-Klein vector fields, is fully separable along the fiber and partially separable between the fiber and the base.

By  separability along the fiber we mean that $\cals K_\omega$  have common separating coordinates as in \eqref{stdform}.  Since the  Hamiltonian, $H_{S^n}$, for the round metric is obviously invariant under $\fg$, the results summarized in Section~\ref{ss:sepRmflds} are directly applicable to our problem, which may now  be restated as follows: 

{\it For a given $\fg\subset \so(n+1)$,  determine whether the set  of Killing tensors, $\cals K_\omega$,  invariant under $\fg$ can be developed into to a set of $n$ Killing vectors/tensors that satisfy conditions \ref{C1} and \ref{C2} for separability of the geodesic \HJ equation for the round metric on $S^n$.}

Let $\mathbb K_\fg$ be the set of Killing tensors on $S^n$ invariant under $\fg$.  Since we are looking at the $\fg$-invariant sector, the $\cals K_\omega$ appearing in  \eqref{Hexpand} must be a linear combinations of the $\mathbb K_\fg$.

An isometry $\fg$ of the uplifted metric may fail to be separating for one of three reasons:
\begin{itemize}[leftmargin=25pt,topsep=0pt]
\setlength\itemsep{-3pt}
\item  [(i)] The set $\mathbb K_\fg$ is too large and  the invariant Killing tensors are not all in involution.
\item [(ii)] There is no extension of $\mathbb K_\fg$ to a set of $n$ Killing vectors/tensors in involution.
\item [(iii)] Some of the Killing tensors are not integrable as endomorphisms of the tangent space.
\end{itemize}

Intuitively one can understand the first two options as follows.  If $\fg$ is ``too small,'' then the  $\cals K_\omega$'s appearing in a generic expansion  \eqref{Hexpand}  may simply be too numerous for all of the   $\cals K_\omega$'s to be in involution and so they cannot all be set to independent constants, which means that separation fails.   If $\fg$ is ``too large,'' then $\mathbb K_\fg$ is small and there will need to be additional, non-$\fg$-invariant Killing tensors for the \HJ equation to separate. Those may or may not exist. The third possibility is a failure of the Killing tensors being integrable in the sense of endomorphisms of the tangent space as defined by  condition \ref{C2} in  Section~\ref{ss:sepRmflds}. They may  not preserve the subspace orthogonal to the Killing vectors or may not commute. In the last instance, there can be an integrable system of geodesics and yet there are no separating coordinates for the Hamilton-Jacobi equation.  We will give an example of this in Section~\ref{sss:su22ing2}.

Given $\fg$, the invariant Killing tensors, $\cals K_\omega\in\mathbb K_\fg$,  are in one-one correspondence with the singlets in the branching of the three representation of $\so(n+1)$ in \eqref{YTdec} under $\fg$. Obviously, there will be always at least one such tensor, which is the round metric on $S^n$. Other trivially invariant Killing tensors come from various Casimir invariants, which are constructed as follows.

For a subalgebra $\frak h\subset \so(n+1)$ with the generators, $t_i=(t_i{}^A{}_B)$, in $\bbR^{n+1}$, the Killing vectors,
$K(t_i)$, satisfying 
\begin{equation}\label{}
\{K(t_i),Y^A\}\eql t_i{}^A{}_B Y^B\,,
\end{equation}
are 
\begin{equation}\label{}
K(t_i)\eql {1\over 2}\,t_i{}^A{}_B K^{AB}\,.
\end{equation}
The Casimir invariant  of $\fh$ is defined as 
\begin{equation}\label{}
\cC^\fh_2\equiv -\kappa^{ij} K(t_i)K(t_j)\,,
\end{equation}
where $\kappa_{ij}$ is the Cartan-Killing form on $\fh$. In particular, we have 
\begin{equation}\label{}
H_{S^n}\eql \cC^{\so(n+1)}_2\,.
\end{equation}

Suppose that $\fg$ is a direct sum algebra
\begin{equation}\label{decomg}
\fg\eql \fh_1\oplus \ldots\oplus \fh_s\oplus\frak u(1)_1\oplus\ldots\oplus \frak u(1)_\nu\,,
\end{equation}
where $\fh_i$ are simple, and is embedded in $\so(n+1)$ through a chain of maximal subalgebras, 
\begin{equation}\label{gchain}
\so(n+1)\equiv  \fg_1\supset \fg_2\supset\ldots\supset \fg_k\equiv\fg\,,
\end{equation}
of length $k$. Then  the Casimir invariants and the   bilinears in the Killing vectors of the $\frak u(1)$'s in \eqref{decomg},
\begin{equation}\label{ninvlist}
\cC_2^{\frak h_i}\,,\quad i=1,\ldots,s\,;\qquad \cC^{\fg_a}_2\,,\quad a=1,\ldots,k\,;\qquad K_\alpha K_\beta\,,\quad \alpha,\beta=1,\ldots,\nu\,,
\end{equation}
are all invariant under $\fg$ and are in involution. However, note  the actual number of the independent invariants in \eqref{ninvlist} may be less than $s+k+\nu(\nu+1)/2$ as we will see in some examples below.

An empirical observation is that when the total number of singlets under $\fg$ exceeds the number of the obvious  invariants \eqref{ninvlist}, not all Killing tensors in $\mathbb K_\fg$ will be in involution. This means that a separating isometry cannot be too small. Another observation is that if one of the algebras in the chain \eqref{gchain} is a separating isometry then all other algebras higher in the chain are separating isometries as well. Conversely, if some algebra in not separating, all algebras lower in the chain cannot be separating. This leads to a systematic procedure by which we have classified all chains of separating isometries for the consistent truncations on $n$-spheres, $S^n$, for $n=2,\ldots,7$, listed in the tables in Appendix~\ref{App:tables}.

We thus have an effective method of enumerating many of the elements of $\mathbb K_\fg$ and testing whether they can be in involution with one another.  At the other extreme, when  $\fg$ is ``large,''  $\mathbb K_\fg$ is ``small,'' we  are going to need to find non-$\fg$-invariant Killing tensors so as to complete the separation of the \HJ equation. There is also a useful systematic way to achieve this.  If one has a chain of inclusions:
\begin{equation}\label{gchain2}
 \fg_1\subset \fg_2\subset\ldots\subset \fg_k\equiv\fg\,,
\end{equation}
then the Casimirs $\cC^{\fg_a}$ are all in involution with one another.  As we will see, this can prove to be an effective way of generating the missing Killing tensors.

\subsection{Separating isometries on $S^n$, $n=1,\ldots,6$}
\label{sss:sumtbl}
In this, and the next, section, we will check explicitly separability for the isometries  listed in Tables~\ref{tblsph16} and \ref{tblsph}. This is most easily done by simply constructing explicitly separating coordinates, which is typically quite straightforward given the isometry. We will also discuss some interesting examples where separability fails  when the isometry is further decreased. We start with the families of separable chains that are present on all spheres.

\subsubsection{$\fg=\so(n+1)$}
\label{sss:sonp1}

For the maximal isometry, we have only one invariant Killing tensor,
\begin{equation}\label{}
\mathbb K_{\,\fg}\eql \left\langle\cC_2^{\so(n+1)} \,\right\rangle\,.
\end{equation}

Separating coordinates for the geodesic Hamiltonian on spheres have been fully classified by Kalnins and Miller\cite{Kalnins1986}. All those coordinates, up to an equivalence,  are  orthogonal, which means that the metric tensor is diagonal. The  equivalence  is a linear change of cyclic variables. An earlier,  partial classification in terms of polyspherical coordinates was given by Vilenkin \cite{vilenkin1978special}. In fact, these are the only coordinates that we will need  in our examples. For an ambient space characterization of separating coordinates on spheres, see \cite{Schoebel2015a}.

\subsubsection{$\fg=\so(n)$}
\label{sss:son}

The $\so(n)$ isometry acts on $Y^1,\ldots,Y^n$, and 
\begin{equation}\label{}
\mathbb K_{\,\fg}\eql \left\langle\cC_2^{\so(n+1)} ~,~  \cC_2^{\so(n)}\,\right\rangle\,.
\end{equation}
Setting
\begin{equation}\label{septheta}
Y^I \eql \cos\theta\,y^I\,,\quad I=1,\ldots,n\;\qquad Y^{n+1}\eql\sin\theta\,,
\end{equation}
where $y^Iy^I=1$ define $S^{n-1}$, one finds that
\begin{equation}\label{}
\cC_2^{\so(n+1)}\eql {p_\theta^2\over 2}+{1\over 2}\,\sec^2\theta\,\cC_2^{\so(n)}\,.
\end{equation}
We can now use  any separating coordinates for the geodesic Hamiltonian on $S^{n-1}$ to separate simultanously both Killing tensors.

\subsubsection{$\fg\eql \so(p)\oplus \so(q)$,\quad  $p+q=n+1$}
\label{sss:sopsoq}

The $\so(p)$ and $\so(q)$ isometries act on $Y^1\,,\ldots\,,, Y^p$ and $Y^{p+1}\,,\ldots\,,Y^{p+q}$, respectively. There are three independent invariant Killing tensors,
\begin{equation}\label{}
\mathbb K_{\,\fg}\eql \left\langle\cC_2^{\so(n+1)} ~,~  \cC_2^{\so(p)} ~,~ \cC_2^{\so(q)}\,\right\rangle\,.
\end{equation}
The two sets of ambient coordinates parametrize $S^p\times S^q$ with radii $\cos\theta$ and $\sin\theta$, respectively, fibered over an interval parametrized by $\theta$. Then
\begin{equation}\label{}
\cC_2^{\so(n+1)}\eql {p_\theta^2\over 2}+\sec^2\theta\,\cC_2^{\so(p)}+\csc^2\theta\,\cC_2^{\so(q)}\,,
\end{equation}
and all three invariants separate using any separating coordinates on $S^p$ and $S^q$.

\subsubsection{$\fg\eql \su(m)\oplus \frak u(1)$,\quad $2m=n+1$}
\label{sss:sumu1nodd}
We take as the generator of $\frak u(1)$ the block diagonal, antisymmetric matrix with $J_{12}=\ldots=J_{2m-1\,2m}=1$, which defines a complex structure in $\bbR^{2m}$. Its Killing vector is\footnote{Here and below, the $\frak u(1)$ generators may be differently normalized and/or may be linear combinations of the $\frak u(1)$ generators in Tables~\ref{tblsph16} and \ref{tblsph}, which are obtained by the rules of LieArt \cite{Feger2020a}.}
\begin{equation}\label{Jdef}
J_{2m}\eql\sum_{A=1}^m\,K^{2A-1\,2A}\,.
\end{equation}
The $\su(m)$ Killing vectors are in involution with $J_{2m}$. One can check that
\begin{equation}\label{sucastoso}
\cC^{\so(2m)}\eql 2\,\cC^{\su(m)}+{m-2\over 2m}\,J_{2m}^2\,,
\end{equation}
 We thus find that there are only two independent singlets and 
\begin{equation}\label{}
\mathbb K_{\,\fg}\eql \left\langle\cC_2^{\so(2m)} ~,~  J_{2m}^2\right\rangle\,.
\end{equation}
Note that this also shows that  the naive counting of invariants for this isometry exceeds the actual number of invariants that follow from the branching rules.

To find the extra Killing vectors/tensors we can use the embedding chain  method  (\ref{gchain2}), based upon $\SU (k) \times \U{1} \subset \SU (k+1)$, which leads to the Casimirs of sequentially embedded $\cC^{\su(k)}$'s, and to the Killing vectors of the $\rm U(1)$ factors.  Using (\ref{sucastoso}), we can recast  the  former in terms of $\cC_2^{\so(2k)}$ and so arrive at the following  natural set of Killing vectors/tensors that are in involution: 
\begin{equation}\label{caskil}
\cC_2^{\so(2m)}\,,\quad \cC_2^{\so(2m-2)}\,,~\ldots~,\quad \cC_2^{\so(4)}\,,\quad K^{12}\,,\quad K^{34}\,,\quad \ldots\,, \quad K^{2m-1\,2m}\,.
\end{equation}
These are the  $(m-1)$  Casimir tensors of $\so(2k)$'s acting on the first $2k$ Cartesian coordinates and $m$ Cartan generators of $\so(2m)$. The corresponding coordinates $(\theta_2,\ldots,\theta_m;\phi_1 ,\ldots,\phi_m)$ are defined by
\begin{equation}\label{}
X^{2j-1}+i\,X^{2j}\eql x^j\,e^{i\phi_j}\,,\qquad j=1,\ldots,m\,,
\end{equation}
where
\begin{equation}\label{}
x^{m}  \eql \cos\theta_{m}\,,\qquad 
x^{m-1}   \eql \sin\theta_{m}\,\cos\theta_{m-1}\,,\quad\ldots
\end{equation}
are given in terms of standard spherical angles on $S^{m-1}$. Then $\phi_1,\ldots,\phi_m$ are cyclic with
\begin{equation}\label{}
K^{2j-1\,2j}\eql p_ {\phi_j}\,,\qquad j=1,\ldots,m\,.
\end{equation}
In particular,
\begin{equation}\label{jjs}
J_{2m}\eql p_{\phi_1}+\ldots +p_{\phi_m}\,.
\end{equation}
The Casimirs of $\so(2k)$, $n=1,\ldots,N+1$, are given inductively by 
\begin{equation}\label{}
\begin{split}
C_2^{\so(2)} & \eql {p_{\phi_1}^2\over 2}\,,\\
&\vdots \\
C_2^{\so(2k)} & \eql {p_{\theta_n}^2\over 2}+{1\over 2}\sec^2\theta_n\,p_{\phi_n}^2+\csc^2\theta_n\,C_2^{\so(2k-2)}\,,\qquad k=2,\ldots m\,.
\end{split}
\end{equation}
The nested structure of these Casimirs is precisely what is needed for separability.

\subsubsection{$\fg\eql \su(m-1)\oplus \frak u(1)\oplus\frak u(1)$,\quad $2m=n+1$}
\label{sss:sumu1u1even}

This isometry is obtained by breaking $\su(m)$ in Section~\ref{sss:sumu1nodd} to $\su(m-1)\oplus \frak u(1)$. We may take the two $\frak u(1)$ Killing vectors as
\begin{equation}\label{}
J_{2m-2}\qquad \text{and}\qquad J_{2m}\,,
\end{equation}
with the range of summation in \eqref{Jdef} set accordingly. The five singlets correspond to
\begin{equation}\label{invforpw}
\mathbb K_{\,\fg}\eql \left\langle\cC_2^{\so(2m)} ~,~ \cC_2^{\su(m-1)}  ~,~ J_{2m-1}^2 ~,~J_{2m-2}J_{2m} ~,~J_{2m}^2\right\rangle\,.
\end{equation}
The identity \eqref{sucastoso} allows us to use $\cC_2^{\so(2m-2)}$ instead of $\cC_2^{\su(m-1)}$. Using the same coordinates as in Section~\ref{sss:sumu1nodd}, we find
\begin{equation}\label{}
J_{2m-2}\eql p_{\phi_1}+\ldots +p_{\phi_{m-1}}\,,
\end{equation}
with everything else remaining the same. This shows that the smaller isometry is still separating.

\subsubsection{$\fg\eql \su(m)$ or $\su(m-1)\oplus \frak u(1)$\,,\quad $2m=n+1$}
\label{sss:dropu1}

Reducing the isometry by dropping $\frak u(1)$ in $\frak u(m)$ in Section~\ref{sss:sumu1nodd} or in $\frak u(m-1)$ in Section~\ref{sss:sumu1u1even} does not change the invariants and the system remains separable in the same coordinates as before.

\subsubsection{$\fg\eql \su(m)\oplus u(1)$, \quad $2m=n$}
\label{sss:su2minn1}

 The invariants are
\begin{equation}\label{invsum}
\mathbb K_{\,\fg}\eql \left\langle\cC_2^{\so(2m+1)} ~,~ \cC_2^{\su(m)} ~,~ J_{2m}^2\right\rangle\,.
\end{equation}
The $\su(m)\oplus \frak u(1)$ lies inside $\so(2m)$ in Section~\ref{sss:son}. After using \eqref{septheta}, the separation is the same as in Section~\ref{sss:sumu1nodd}. 

\subsubsection{$\fg\eql \su(m)\subset \su(m)\oplus u(1)$, \quad $2m=n$,\quad $m\geq 3$}
\label{sss:su2minn1du1}

The isometry here is the reduction of the one in  Section~\ref{sss:su2minn1} by dropping the $\frak u(1)$. For  $m=2$, this produces too many invariants that fail to be in involution. However, for  $m\geq 3$, the invariants are the same as in  \eqref{invsum}, and one can separate the system using the same coordinates as in Section~\ref{sss:su2minn1}.

\subsubsection{$\fg=\fg_2\subset\so(7)$}
\label{sss:g2inso7}
Using an explicit representation of $\fg_2$ as an automorphism of the algebra of unit octonions, one can check that 
\begin{equation}\label{g2so7}
\cC^{\fg_2}_2\eql \cC^{\so(7)}\,.
\end{equation}
This explains why there is only one singlet and
\begin{equation}\label{}
\mathbb K_{\,\fg}\eql \left\langle\cC^{\so(7)}\right\rangle\,.
\end{equation}
Hence any separating coordinates on the round $S^6$ can be used here.

\subsection{Separating isometries on $S^7$}
\label{ss:siS7}

In this section we discuss selected isometries, $\fg\subset\so(8)$, on $S^7$ in Table~\ref{tblsph} in Appendix~\ref{App:tables}.

\subsubsection{$\fg=\su(2)\oplus\su(2)\oplus\su(2)\oplus\frak u(1)$ and $\su(2)\oplus\su(2)\oplus\frak u(1)\oplus\frak u(1)$}
\label{sss:su2su2su2u1}

We start with the smaller isometry which consists of $\su(2)_ L^{(1)}\times \frak u(1)_R^{(1)}$ acting on $Y^{1},\ldots,Y^4$ and $\su(2)_L^{(2)}\times \frak u(1)_R^{(2)}$ acting on $Y^{5},\ldots,Y^8$. The six singlets are 
\begin{equation}\label{}
\mathbb K_{\,\fg}\eql \Big\langle \cC_2^{\so(8)} ~,~ \cC_2^{\su(2)_L^{(1)}} ~,~ \cC_2^{\su(2)_L^{(2)}} ~,~ R^{(1)}_3R^{(1)}_3 
~,~ R^{(1)}_3R^{(2)}_3 ~,~ R^{(2)}_3R^{(2)}_3 
 \Big\rangle\,,
\end{equation}
where $R_3^{(i)}$ is a Killing vector  in $\su(2)_L^{(i)}$ for  $i=1,2$. The natural separating coordinates correspond to the Killing vectors of the Cartan subalgebra of $\fg$, which is the same as the Cartan subalgebra of $\so(8)$, and the three Casimir invariants above. Hence we set
\begin{equation}\label{}
\begin{split}
Y^1+i\,Y^2 & \eql \cos\theta\,\cos\zeta_1\,e^{i\phi_1}\,,\qquad Y^3+i\,Y^4 \eql \cos\theta\,\sin\zeta_1\,e^{i\phi_2}\,,\\[6 pt]
Y^5+i\,Y^6 & \eql \sin\theta\,\cos\zeta_2\,e^{i\psi_1}\,,\qquad Y^7+i\,Y^8 \eql \sin\theta\,\sin\zeta_2\,e^{i\psi_2}\,,
\end{split}
\end{equation}
which gives
\begin{equation}\label{}
K^{12}\eql p_{\phi_1}\,,\qquad K^{34}\eql p_{\phi_2}\,,\qquad K^{56}\eql p_{\psi_1}\,,\qquad K^{78}\eql p_{\psi_2}\,,
\end{equation}
\begin{equation}\label{}
\begin{split}
\cC_2^{\su(2)_L^{(1)}} & \eql{1\over 16}\,\left[\, p_{\zeta_1} +\sec^2\zeta_1\,p_{\phi_1}+\csc^2\zeta_1\,p_{\phi_2}\,\right]\,,\\
\cC_2^{\su(2)_L^{(2)}} & \eql{1\over 16}\,\left[\, p_{\zeta_2} +\sec^2\zeta_2\,p_{\psi_1}+\csc^2\zeta_2\,p_{\psi_2}\,\right]\,,
\end{split}
\end{equation}
and
\begin{equation}\label{}
\cC_2^{\so(8)} \eql {p_\theta^2\over 2}+8\,\sec^2\theta\,\cC_2^{\su(2)_L^{(1)}}+8\,\csc^2\theta\,\cC_2^{\su(2)_L^{(2)}}\,,
\end{equation}
and the separability is manifest.

We also note that there is yet another $\su(2)\oplus\su(2)\oplus\frak u(1)\oplus\frak u(1)$ subalgebra in $\so(8)$ defined by the embedding
\begin{equation}\label{}
\bfs 8\quad\to\quad ({\bf 2,2})(0,0)+({\bf 1,1})(1,1)+({\bf 1,1})(1,-1)+({\bf 1,1})(-1,1)+({\bf 1,1})(-1,-1)\,.
\end{equation}
However, it leads to seven invariants with not all of them in involution. Hence it is not a separating isometry.

\subsubsection{$\fg=\so(7)_{v,c}$}
\label{sss:so7vc}

The $\so(7)_{v,c}$ subalgebras can be constructed using the two spinor representations of $\so(8)$, or, equivalently, as subalgebras preserving  a selfdual tensor, $C_+^{ABCD}$, and an anti-selfdual tensor, $C_-^{ABCD}$, respectively (see, e.g.,  \cite{deWit:1984nz}). The $C_\pm^{ABCD}$ are tensors in $\bbR^8$  that also satisfy
\begin{equation}\label{}
C_\pm ^{ABEF}C_\pm^{CDEF}\eql {1\over 12}\,\delta^{AB}_{CD}-4\, C_\pm^{ABCD}\,.
\end{equation}
The two subalgebras are conjugate under a reflection in $\rm O(8)$. The singlet in each sector corresponds to
\begin{equation}\label{}
C_2^{\so(7)_{v,c}}\eql {3\over 4}\,C_2^{\so(8)}\,.
\end{equation}
and hence one can use any separating coordinates for the round metric on $S^7$.

\subsubsection{$\fg= \fg_2\subset\so(7)_{s,v,c}\subset\so(8)$}
\label{sss:g2inso8}

One can embed $\fg_2$ in $\so(8)$ via the three $\so(7)$'s. Although the three embeddings are equivalent, one is led naturally to different separating coordinates depending on which one is chosen.

For $\fg_2\subset\so(7)_s$, we can take it to be the same as in Section~\ref{sss:g2inso7}. In particular,  \eqref{g2so7} holds. Using the coordinates introduced in Section~\ref{sss:son}, we obtain separating coordinates that consist of $\theta$ in \eqref{septheta} and any separating coordinates on $S^6$.

Another realization of $\fg_2\subset\so(7)_{v,c}$ is by requiring invariance of both tensors $C_\pm^{ABCD}$. Starting with the two Cartan generators of $\fg_2$,
\begin{equation}\label{}
K^{12}-K^{56}\,,\qquad K^{12}-2 K^{34}+K^{56}\,,
\end{equation}
one finds that there is one more Killing vector on $S^7$ in involution with those two, such that the three together span the Cartan subalgebra of $\so(6)$ with the generators
\begin{equation}\label{Kvecsd}
 K^{12}\,,\qquad   K^{34}\,,\qquad   K^{56}\,.
\end{equation}
 It is then straightforward to determine that the four Killing tensors, 
\begin{equation}\label{Ktensd}
  C_2^{\so(4)}\,,\qquad   C_2^{\so(6)}\,,\qquad   C_2^{\frak g_2}\,,\qquad   C_2^{\so(8)}\,,
\end{equation}
are invariant under \eqref{Kvecsd} and in involution with each other. In fact, the seven Killing vectors/tensors in \eqref{Kvecsd} and \eqref{Ktensd} satisfy both conditions \ref{C1} and \ref{C2}, which proves that the \HJ equation in the $\fg_2$ invariant sector is separable. 

The corresponding separating coordinates can be defined by
\begin{equation}\label{theXs}
\begin{split}
X^1+i\,X^2 & \eql Y^1\,\sin\theta_2\sin\theta_3\,e^{i\phi_1}\,,\\
X^3+i\,X^4 & \eql Y^1\,\cos\theta_2\sin\theta_3\,e^{i\phi_2}\,,\\
X^5+i\,X^6 & \eql Y^1\,\cos\theta_3\,e^{i\phi_3}\,,\\
X^7&\eql Y^2\,,\\
X^8&\eql Y^3\,,
\end{split}
\end{equation}
where 
\begin{equation}\label{theYs}
\begin{split}
Y^1 & \eql \cos\zeta_1\cos\zeta_2\,,\\
Y^2 & \eql {1\over \sqrt 2}\left(\sin\zeta_1-\cos\zeta_1\sin \zeta_2\right)\,,\\
Y^2 & \eql {1\over \sqrt 2}\left(\sin\zeta_1+\cos\zeta_1\sin\zeta_2\right)\,.
\end{split}
\end{equation}
These coordinates provide an example of polyspherical coordinates \cite{vilenkin1978special,Kalnins1986} on $S^7$, where $(\theta_2,\theta_3,\phi_1,\phi_2,\phi_3)$   
parametrize an  $S^5$ fibered over an $S^2$ with coordinates $(\zeta_1,\zeta_2)$. The three Killing vectors \eqref{Kvecsd} are
\begin{equation}\label{}
K^{12}\eql p_{\phi_1}\,,\qquad K^{34}\eql p_{\phi_2}\,,\qquad K^{56}\eql p_{\phi_3}\,,
\end{equation}
while the Killing tensors are given by
\begin{equation}\label{}
\begin{split}
C_2^{\so(4)} & \eql {p_{\theta_2}^2\over 2}+{1\over 2}\,\sec^2\theta_2\,p_{\phi_2}^2+\csc^2\theta_2\,{p_{\phi_1}^2\over 2}\,,\\[6 pt]
C_2^{\so(6)} & \eql  {p_{\theta_3}^2\over 2}+{1\over 2}\,\sec^2\theta_3\,p_{\phi_3}^2+\csc^2\theta_3\,C_2^{\so(4)}\,,\\[6 pt]
C_2^{\frak g_2} & \eql {p_{\zeta_2}^2\over 3}+{2\over 3}\,\sec^2\zeta_2\,C_2^{\so(6)}\,,\\[6 pt]
C_2^{\so(8)} & \eql {p_{\zeta_1}^2\over 2}+{3\over 2}\sec^2\zeta_1\,C_2^{\frak g_2}\,.
\end{split}
\end{equation}
Hence, setting the momenta and the Casimirs to constants, we obtain a separating system \eqref{Stacksys} for the \HJ equation.

\subsection{Examples of non-separating isometries}
\label{ss:nonsepis}

We will now discuss some examples of isometries where the separation fails. 

\subsubsection{$\fg=\so(p)\oplus\so(q)\oplus\so(r)$,\quad $p+q+r=n+1$}
\label{sss:notsepqr}
The three rotations act block diagonally  $\bbR^p\times \bbR^q\times\bbR^r$.  In addition to the Casimirs, $\cC_2^{\so(p)}$, $\cC_2^{\so(q)}$ and $\cC_2^{\so(r)}$ one also has 
\begin{equation}\label{}
\cC_2^{\so(p)\oplus\so(q)} \eql {1\over 4}\, \sum_{A,B=1}^{p+q}  \, K^{AB} \, K^{AB}\,,
\end{equation}
and the other two for $\so(p)\oplus\so(r)$ and $\so(q)\oplus\so(r)$ given by the similar formulae. One can check that these invariants have nonvanishing Poisson bracket among each other and hence this isometry is not separating.

This example is prototypical when $\fg$ is a sum of more than two diagonal subalgebras. This leads to a large number of invariants some of which fail to be in involution. We have omitted all those symmetries from the tables in Appendix~\ref{App:tables}.

\subsubsection{$\fg=\su(2)_M\oplus \fh$}
\label{sss:notsepqr}

The $\su(2)_M$ here is the maximal subalgebra of a larger symmetry, $\tilde\fg$.  The simplest example is  $\tilde\fg=\so(5)$, where  the generators of  $\su(2)_M$  are given by the Killing vectors
\begin{equation}\label{gensu25}
\begin{split}
K(t_1) & \eql K^{12}+\sqrt{\frac{3}{2}}\, K^{23}+\sqrt{\frac{3}{2}}\, K^{34}-K^{45}\,,\\[6 pt]
K(t_2) & \eql -K^{14}+\sqrt{\frac{3}{2}}\, K^{23}-K^{25}-\sqrt{\frac{3}{2}}\, K^{34}\,,\\[12 pt]
K(t_3) & \eql 2 K^{15}+K^{24}\,.
\end{split}
\end{equation}
The two singlets are
\begin{equation}\label{su2mxinv}
\mathbb K_{\,\fg}\eql \left\langle\cC_2^{\so(5)} ~,~ \cC_2^{\su(2)_\text{M}}\right\rangle\,.
\end{equation}

To determine separability one can  proceed as follows:\footnote{See, e.g., Section 5 in \cite{Kalnins1981}.}  First compute all Killing vectors and Killing tensors that are in involution with the invariants \eqref{su2mxinv} and then check whether there exists a set of four Killing vectors/tensors in involution that extends \eqref{su2mxinv}.

A direct calculation shows that the only Killing vectors that are in involution with  $\cC_2^{\su(2)_\text{M}}$ are those in \eqref{gensu25}. The seven  Killing tensors   in involution with the ones in \eqref{su2mxinv} are $\cC_2^{\so(5)}$ and six bilinears in the Killing vectors \eqref{gensu25}. It is then impossible to choose from this set the required four Killing vectors/tensors in involution and hence the system does not separate.

The same happens in the other non-separable examples in  Tables~\ref{tblsph16} and \ref{tblsph}  involving  maximal $\su(2)$'s,  where it is  impossible to find the required set  Killing vectors/tensors satisfying conditions \ref{C1} and/or   \ref{C2} in Section~\ref{ss:sepRmflds}.

\subsubsection{$\fg=\su(2)\oplus\su(2)\subset\fg_2\subset\so(7)$}
\label{sss:su22ing2}

In this example, separability fails in an interesting way. The embedding $\su(2)\oplus\su(2)\subset\so(7)$ is given by the branching 
\begin{equation}\label{}
\bf 7\quad\to\quad (2,2)+ (3,1)\,.
\end{equation}
Denote the two $\su(2)$'s as $\su(2)_L$ and $\su(2)_R$, respectively. They act  on $Y^1,\ldots,Y^4$ as $\so(4)\simeq\su(2)_L\oplus\su(2)_R$ and only $\su(2)_L$ acts as $\so(3)$ on $(Y^5,Y^6,Y^7)$. The four invariants are
\begin{equation}\label{KTsu22}
\mathbb K_{\,\fg}\eql \left\langle\cC_2^{\so(7)} ~,~ \cC_2^{\su(2)_L}~,~ \cC_2^{\su(2)_R}~,~ \cC_2^{\so(3)} \right\rangle\,.
\end{equation}
The space of Killing tensors that are in involution with these invariants is 23-dimensional and is spanned by $\cC_2^{\so(7)}$, $\cC_2^{\so(3)}$ and 21 bilinears in the Killing vectors of $\fg$. They are also in involution with any two commuting Cartan generators of $\fg$. So, one can easily extend \eqref{KTsu22} to a set of six Killing vectors/tensors in involution. This satisfies condition \ref{C1} for separability. However, using any coordinates on $S^6$, one can check that as endomorphisms of $TS^6$, the Killing tensors \eqref{KTsu22} commute only on a subspace of dimension two. Hence to satisfy condition \ref{C2} we would need a four-dimensional subspace in $TS^6$ spanned by commuting Killing vectors, which is impossible. 

This truncation provides an explicit example where there is a maximal set of Killing vectors and Killing tensors that are in involution, and so one has an integrable system of geodesics, yet there are no separating coordinates for the Hamilton-Jacobi equation.

\subsubsection{$\fg=\su(3)\subset\so(8)$}
\label{sss:su3}

The two singlets span
\begin{equation}\label{KTsu3}
\mathbb K_{\,\fg}\eql \left\langle\cC_2^{\so(8)} ~,~ \cC_2^{\su(3)} \right\rangle\,.
\end{equation}
The Killing tensors that are in involution with those in (\ref{KTsu3}) are $\cC_2^{\so(8)}$ and bilinears in the Killing vectors of $\su(3)$.  Using any convenient coordinates on $S^7$ one finds that no linear combination of those bilinears commutes, as endomorphisms, with $ \cC_2^{\su(3)}$, and that the latter does not preserve the subspaces perpendicular to either one or two commuting Killing vectors. This shows that the condition \ref{C2} is always violated and thus \eqref{KTsu3} are not simultaneously separable.

\subsubsection{$\fg=\sp(4)\oplus\su(2)\subset\so(8)$}
\label{sss:sp4su2}

There are only two singlets as one finds that
\begin{equation}\label{}
\cC_2^{\so(8)}\eql 4\,\cC_2^{\sp(4)}-2\,\cC_2^{\su(2)}\,.
\end{equation}
Then, similarly as in Section~\ref{sss:su3},  there are no Killing tensors that are simultaneously in involution and commute with $\cC_2^{\su(2)}$, which in turn does not preserve the orthogonal complement of the Killing vectors corresponding to the Cartan subalgebra of the isometry.

\section{More separable examples from holography}
\label{ss:sepexhol}

In this section we present some further examples of solutions belonging to consistent truncations of type IIB supergravity and M-theory on $S^5$ and $S^7$, respectively, and show that the massless \HJ equation is fully separable. Just as in the simpler example in Section~\ref{sec:pexample}, separability is by no means obvious given the explicit form of the metrics in local coordinates. Those metrics and additional  explicit discussion of separability can be found in Appendix~\ref{app:ExpEx}.

\subsection{RG-flows in type IIB supergravity}
\label{ss:IIBflows}

The two RG-flows in this section are obtained by uplifting supersymmetric domain wall solutions in the $\cals N=8$ gauged supergravity in five dimensions to solutions of type IIB supergravity in ten dimensions. The uplifted metric along $S^5$ is then given by \cite{Khavaev:1998fb,Pilch:2000ue}
\begin{equation}\label{}
M_{ABCD}\eql\text{const}\times \widetilde \cV_{ABab}\,\widetilde \cV_{CDcd}\,\Omega^{ac}\,\Omega^{bd}\,,
\end{equation}
where $\widetilde \cV_{ABab}$ are matrix elements of the scalar 27-bein that parametrizes the coset space $\rm E_{6(6)}/USp(8)$, and $\Omega^{ab}$ is the $\rm USp(8)$-invariant metric.

\subsubsection{$\cals N=1$ flow}
\label{ss:N1flowinIIB}
This solution is a holographic dual of the  RG-flow of $\Neql{4}$ Yang Mills  down to an $\Neql{1}$ supersymmetric\footnote{We count unbroken supersymmetries on the field theory side of the gauge/gravity duality.}  ``Leigh-Strassler''  conformal fixed point  in the infra-red driven by giving a mass to one chiral multiplet \cite{Freedman:1999gp}.  In five-dimensional supergravity the flow is defined by two scalar fields, $\alpha(r)$ and $\chi(r)$, and a metric function, $A(r)$. The five-dimensional metric is given by the usual Poincar\'e slicing:\footnote{Note that IIB supergravity is typically written using a ``mostly minus'' signature.}
\begin{equation}
ds_{1,4}^2 ~=~ e^{2 A(r)}\big(\, \eta_{\mu \nu} \, dx^\mu \, d x^\nu \big) ~-~ dr^2  \,, \qquad \mu \eql 0\,,\ldots\,,3  \,.
\label{5dPoincare}
 \end{equation}
The internal metric in terms of ambient coordinates is given by (3.4) in \cite{Pilch:2000fu}:
\begin{equation}\label{pwmetric}
ds_2^2(\alpha,\chi)\eql {a^2\over 2}{\sech\chi\over \xi}\,(dY^AQ^{-1}_{AB}dY^B)+{a^2\over 2}{\sinh\chi\tanh\chi\over\xi^3}(Y^AJ_{AB}dY^B)^2\,,
\end{equation}
and the warp factor
\begin{equation}\label{}
\Delta^{-{2\over 3}}\eql \xi\,\cosh\chi\,.
\end{equation}
Here $Q$ is a diagonal matrix with
$Q_{11}=\ldots=Q_{44}=e^{-2\alpha}$ and $Q_{55}=Q_{66}=e^{4\alpha}$, $J$ is an antisymmetric matrix with $J_{12}=J_{34}=J_{56}=1$, and $\xi^2=Y^AQ_{AB}Y^A$. The constant $a$ is related to the radius, $L$, of the AdS$_5$ metric \eqref{5dPoincare}, when $\alpha=\chi=0$, by $a=\sqrt 2\, L$. The gauge coupling constant is $g=2/L$.

While the full truncation for this flow has an $\su(2)\times \frak u(1)$ symmetry, the isometry of the metric \eqref{pwmetric} is enhanced to $\su(2)\times u(1)_1\times u(1)_2$, where $\su(2)\times\frak u(1)=\su(2)_L\times \frak u(1)_R\subset\so(4)$ acting on $Y^1,\ldots,Y^4$, while the second $\frak u(1)$ is the rotation in the $Y^{5,6}$-plane. This means that this metric is of the type in Section~\ref{sss:sumu1u1even} and should have a separable \HJ equation. 

One can check that the fiber Hamiltonian, $H_\cF$, corresponding to \eqref{pwmetric}, is
\begin{equation}\label{PWham}
\begin{split}
H_\cF  \eql {1\over a^2}\cosh^2 \chi\, \Big[  (Q_{AB}Y^AY^B)& (Q_{CD}P_CP_D) -(Q_{AB}Y^AP_B)(Q_{CD}Y^CP_D)\Big]\\ 
& \qquad -{1\over a^2}\sinh^2\chi\, \big[(QJ)_{AB}Y^AP_B\big]\big[(QJ)_{CD}Y^CP_D\big]\,,
\end{split}
\end{equation}
and indeed decomposes into the invariant Killing tensors, \eqref{invforpw}, as follows ($a^2=1$):
\begin{equation}\label{HFeffect}
\begin{split}
H_\cF & \eql 2\rho^2\cosh^2\chi\,  C_2^{\so(6)}+2\rho^{-4}(1-\rho^6)\cosh^2\chi\,  C_2^{\so(4)}  -
\rho^{-4}\sinh^2\chi\,J_4^2 \\[3 pt] & \quad-2 \rho^2\,\sinh^2\chi\,J_4 R
 \quad -{1\over 2}\, \rho^2(1-2\rho^6+\cosh 2\chi)\,R^2\,,
\end{split}
\end{equation}
where,
\begin{equation}\label{}
R\equiv J_6-J_4\eql K^{56}\,.
\end{equation}
  Setting
\begin{equation}\label{N12unv}
 C_2^{\so(6)}\eql c_6\,,\qquad C_2^{\so(4)}\eql c_4\,,\qquad J_4\eql c_J\,,\qquad R\eql c_R\,,
\end{equation}
in \eqref{HFeffect}, where $c_6$, $c_4$, $c_J$, and $c_R$ are constants, we obtain the effective potential, $U(r)$,  for the \HJ equation \eqref{effeqs} on the base. For the metric \eqref{5dPoincare}, this equation is then separable.

It follows from \eqref{sucastoso} for $m=2$ that
\begin{equation}\label{CasamirSO4proptoSU2}
\cC_2^{\so(4)}\eql 2\,\cC_2^{\su(2)_{L,R}}\,.
\end{equation}
The $\SU(2)_L$ symmetry acts transitively on $S^3$.  Hence the constant $c_4$ represents the energy of   motion on $S^3$.

We refer the reader to Appendix~\ref{app:N1flowinIIB} for an elementary derivation of the effective  \HJ equation on $\cB$ using the original local coordinates   in \cite{Pilch:2000fu}.

\subsubsection{$\cals N=2$ flow}
\label{ss:N2flowinIIB}

This solution represents  flows of $\Neql{4}$ Yang Mills down to into the infra-red $\Neql{2}$ supersymmetric theory in which mass is given to an $\Neql{2}$ hypermultiplet.  In five-dimensional supergravity the flow  is defined by two scalar fields, $\alpha$ and $\chi$. The five-dimensional metric is given by the  Poincar\'e slicing (\ref{5dPoincare}).  The five-dimensional flow and its IIB uplift were obtained in \cite{Pilch:2000ue}.\footnote{For the uplifted metric, see also \cite{Brandhuber:2000ct}.} 

The uplifted metric along $S^5$ has the same $\rm \SU(2)\times U(1)\times U(1)$ isometry as the $\cals N=1$ flow in the previous subsection. Hence the \HJ equation is partially separable, with the Hamiltonian, $H_\cF$, in the basis of invariants \eqref{N12unv} given by
\begin{equation}\label{N2HF}
\begin{split}
H_\cF\eql 2 \rho ^2 \cosh ( 2 \chi)  \,\cC^{\so(6)} & +\rho^{-4}\,\big[2-2 \rho ^6 \cosh (2 \chi )\big]\,\cC_2^{\so(4)}
\\[6 pt] &+\rho^{-4}\sinh^2(2\chi)\,J_4^2+\rho^2\,\big[(\rho^6-\cosh(2\chi)\big]\,R^2\,.
\end{split}
\end{equation}
One may note that \eqref{N2HF}, unlike \eqref{HFeffect}, involves only four out of five invariant Killing tensors. This is a reflection of a simpler analytic structure of this flow. We refer to Appendix~\ref{app:N2flowinIIB} for  details of the metric and an explicit  separation of the \HJ equation in local coordinates used in  \cite{Pilch:2000ue}.

\subsection{RG-flows/Janus solutions in M-theory}
\label{ss:Mthex}

The two solutions in this section are obtained by an uplift from the $\cals N=8$, $d=4$ gauged supergravity \cite{Wit1982} to M-theory on $S^7$. 
The uplifted metric is obtained using \cite{deWit:1984nz}
\begin{equation}\label{}
M_{ABCD}\eql \text{const}\times  (U^{ij}{}_{AB}+V^{ijAB})(U_{ij}{}^{CD}+V_{ijCD})\,,
\end{equation}
where $U^{ij}{}_{AB}\,,\ldots\,, V_{ijCD}$ are matrix elements of the scalar 56-bein parametrizing the  $\rm E_{7(7)}/SU(8)$ coset in the so-called $\SL(8,\bbR)$ basis \cite{Cremmer:1979up}. Using $\rm E_{7(7)}$ identities, one can also invert \eqref{InvMet} to obtain an explicit expression for the internal metric as a pull-back of a deformed metric in $\bbR^8$ \cite{Varela:2015ywx,Kruger:2016agp}. 

\subsubsection{ An $\cals N=1$ RG-flow in M-theory}
\label{sss:CPWflow}

This example represents a flow of ABJM theory \cite{Aharony:2008ug} down to a non-trivial infra-red conformal fixed point \cite{Warner:1983vz,Nicolai:1985hs,Benna:2008zy,Klebanov:2008vq}.  In an  $\SU(3)\times \rm U(1)$-invariant truncation  of the $\cals N=8$, $d=4$ gauged supergravity,  the flow \cite{Ahn:2000aq,Ahn:2000mf} is defined by two scalar fields, $\alpha$ and $\chi$. The four-dimensional metric is given by the usual Poincar\'e slicing and,  just as for the analogous RG-flow in five dimensions in Section~\ref{ss:N1flowinIIB}, the uplifted metric has an enhanced $\SU(3)\times \rm U(1)\times U(1)$ isometry along $S^7$. This means that it falls into the class of separable metrics discussed in Section~\ref{sss:sumu1u1even}. Indeed, the fiber Hamiltonian, $H_\cF$, has exactly the same form as in \eqref{PWham}, when written in the ambient space, $T^*\bbR^8$, but now with 
$
Q= \diag\left(\rho^{-1},\rho^{-1},\rho^{-1},\rho^{-1},\rho^{-1},\rho^{-1},\rho^3,\rho^3\right)\,,
$
and the complex structure matrix $J_{12}=\ldots= J_{78}=1$  \cite{Corrado:2001nv}. In terms of the invariants \eqref{invforpw}, it is given by
\begin{equation}\label{HFeffect}
\begin{split}
H_\cF & \eql 2\rho^2\cosh^2\chi\,  C_2^{\so(8)}+2\rho^{-2}(\rho^4-1)\cosh^2\chi\,  C_2^{\su(3)}  +
 {\rho^{-2}\over 6}\big[ 3+(\rho^4-4)\cosh^2\chi\, \big]\,J_6^2 \\ & \quad- \rho^2\,\sinh^2\chi\,J_6 R
  +{\rho ^2 \over 2}\, \left(  \rho ^4-\cosh^2\chi\right)\,R^2\,,
\end{split}
\end{equation}
where,
\begin{equation}\label{}
R~\equiv ~ J_8-J_6\eql K^{78}\,.
\end{equation}
Further details for this example can be found in Appendix~\ref{ss:CPWapp}.

\subsubsection{$\rm SU(3)\times U(1)\times U(1)$ invariant RG-flows/Janus solutions}
\label{sss:Jansu3}

This solution was obtained in \cite{Pilch:2015dwa} by an uplift of the corresponding Janus solution \cite{{Bobev:2013yra}} in an  $\rm \SU(3)\times U(1)\times U(1)$  invariant truncation of the  $\cals N=8$, $d=4$ gauged supergravity. It involves a single complex scalar field,
\begin{equation}\label{}
z\eql \tanh\lambda\,e^{i\zeta}\,,
\end{equation}
that parametrizes the unit Poincar\'e disk.
The  invariance of the truncation implies that the uplifted metric has at least the same isometry and thus the fiber Hamiltonian, $H_\cF$, on $S^7$ should be a linear combinations of the invariants given in   Section~\ref{sss:sumu1u1even}.

Starting with the explicit form of the uplifted metric given in \eqref{SU3U1U1Spheremet}, one can verify  that 
\begin{equation}\label{HcFSU3U1U1}
H_\cF\eql g_1\, \cC_2^{\so(8)}+g_2\, \cC_2^{\su(3)}+g_3\, J_6^2+g_4\,J_6 R+g_5\,R^2\,,
\end{equation}
where $R=K^{78}$ and
\begin{equation}\label{}
\begin{split}
g_1 & \eql \cosh( 2\lambda)+\sinh(2\lambda)\,\cos\zeta\,,\\
g_2 & \eql -4\sinh(2\lambda)\,\cos\zeta \,,\\
g_3 & \eql \frac{1}{12} \sinh (2 \lambda ) \Big [4 \cos  \zeta + 3 \sin ^2 \zeta   \left(\sinh (4 \lambda )-2 \cos
   \zeta  \sinh ^2(2 \lambda )\right)\Big]\,,\\
g_4 & \eql \frac{1}{2} \sin  \zeta   \sinh (2 \lambda ) \Big[\sin (2 \zeta ) \sinh ^2(2 \lambda
   )+\sin (\zeta ) \sinh (4 \lambda )\Big]\,,\\
 g_5 & \eql \frac{1}{2} \,\Big[\big(\cosh (2 \lambda )+ \sinh (2 \lambda )\cos  \zeta \big)^2-1\Big] \big(\cosh (2 \lambda )+ \sinh (2 \lambda )\cos  \zeta \big)\,.\\
\end{split}
\end{equation}
Given the form of these functions it is not surprising that the corresponding metric obtained from $H_\cF$ should be rather involved. Yet the underlying symmetry assures that the \HJ equation is separable.
We refer the reader to Appendix~\ref{ss:SU3U1U1} for an explicit treatment of that separability in local coordinates.

\subsection{Coulomb branch flows}
\label{ss:Coulomb}

The simplest non-trivial examples of uplifted metrics correspond to the Coulomb flows in maximal supergravities, for which some of the earlier examples can be found in \cite{Freedman:1999gk,Cvetic:1999xx}. The scalar fields that define those flows are given by an $(n+1)\times (n+1)$ symmetric, unimodular matrix, $T_{AB}$, parameterizing  the coset space
\begin{equation}\label{}
{\SL(n+1,\bbR)\over\SO(n+1)}\,.
\end{equation}
The tensor, $M_{ABCD}$, is then of a Ricci type and the internal Hamiltonian is simply
\begin{equation}\label{Coulmet}
H_{\cals F} \eql {1\over 4}\,T_{AC}T_{BD}K^{AB}K^{CD}\,.
\end{equation}
The corresponding metric along the fiber is then \cite{Nastase:2000tu} (see, also \cite{deWit:1984nz,Nastase:1999cb,Cvetic:1999xp})
\begin{equation}\label{theCmet}
ds_\cF^2\eql \Delta^\beta\,{T_{AB}^{-1}\,dY^AdY^B} \,,\qquad \Delta^{2-\beta n}\eql T_{AB}Y^AY^B\,,\end{equation}
where
\begin{equation}\label{}
\beta\eql {2\over n-1}{d-1\over d-2}\,,
\end{equation}
and the metric is that of  a conformally rescaled ellipsoid. 

In specific examples \cite{Freedman:1999gk,Cvetic:1999xx,Khavaev:2000gb,Gowdigere:2005wq}, one takes $T_{AB}$, to be diagonal,
\begin{equation}\label{Tdiag}
(T_{AB})\eql \diag(a_1,a_2,\ldots,a_{n+1})\,,\qquad a_1a_2\ldots a_{n+1}\eql 1\,,
\end{equation}
so that $H_\cF$ has the same form as in \eqref{PWham} with $J=0$. The isometry of the metric \eqref{theCmet} depends on how many identical eigenvalues arises in \eqref{Tdiag} and the separability of the corresponding \HJ equation follows from the discussion in  Sections~\ref{sss:son}, \ref{sss:sopsoq} and \ref{sss:notsepqr}.  In particular, the \HJ equation is partially separable if there are only two distinct eigenvalues in \eqref{Tdiag}, but fails to be separable if there are more than two as for instance for the flows discussed in \cite{Cvetic:2000tb,Gowdigere:2005wq}. The reason is that $k$ different eigenvalues break the isometry to a product $\SO(n_1)\times \ldots\times \SO(n_k)$, $n_1+\ldots+n_k=n+1$ as in Section~\ref{sss:notsepqr} and the resulting invariant  Killing tensors are not in involution.  This lack of separability might seem counterintuitive as the \HJ equation for  the metric \eqref{theCmet} on an ellipsoid with {\it constant} matrix, $T_{AB}$, does separate in the elliptic Jacobi coordinates\footnote{See, e.g., \cite{KMWBook}} for any values of $a_1,\ldots,a_{n+1}$. The point is that in our problem those eigenvalues are functions on the base, $a_i=a_i(x)$, and lead to a non-trivial mixing between the base and the fiber when $k>2$.

\section{Microstate geometry examples with gauge-vectors  }
\label{sect:MicroExamples}

In this section we present the $(1,m,n)$ multi-mode superstrata solutions. These are solutions to $D=6$, $\mathcal{N}=(1,0)$ supergravity, coupled to two anti-self dual tensor multiplets and can be consistently truncated on a three-sphere, while retaining non-trivial gauge vectors, see \cite{Mayerson:2020tcl} for details. The appearance of gauge vectors modifies the analysis of separability relative to Section \ref{sec:SepSph}.

\subsection{Considerations for separability with gauge-vectors}

Recall that the massless Hamilton-Jacobi equation, for a truncation with non-trivial gauge vectors $A^{\mu\,i}(x)$, can be cast in the form of (\ref{massH}): 
\begin{equation}\label{massHb}
H\eql H_\cB(x;\pi)-A^{\mu\,i}(x)\pi_\mu K_i+\sum_\omega g_\omega(x)\cals K_\omega(y;p)\,,
\end{equation}
where 
\begin{align}
\sum_\omega g_\omega(x)\cals K_\omega(y;p)= \frac{1}{2}\Big(\, \cU_{\cI \cJ}  +    A^{\cI \sigma}   A^\cJ{}_\sigma   \, \Big) \,  K^{\cI \, m}   K^{\cJ \, n} \,.
\end{align}

In order to discuss when (\ref{massHb}) will separate for a deformed spherical fiber with non-trivial gauge-vectors, we can apply much of the discussion of Section \ref{sec:SepSph}, but with a couple of twists. To do so it will be convenient to:
\begin{itemize}
\item Define $\widetilde{H}_{\cF}$ and $\widetilde{M}_{ABCD}$ implicitly by:
\begin{equation}
\widetilde{H}_{\cF}={1\over 8}\,\widetilde{M}_{ABCD} K^{AB}K^{CD} = \frac{1}{2}\Big(\, \cU_{\cI \cJ}  +    A^{\cI \sigma}   A^\cJ{}_\sigma   \, \Big) \,  K^{\cI }   K^{\cJ}\,.
\end{equation}
Note that if the gauge-vectors vanish the identification $(\widetilde{H}_{\cF},\widetilde{M}_{ABCD})\to  (H_{\cF},M_{ABCD})$ is made with our previous discussion.
\item Suppose again that the uplifted geometry is invariant under
\begin{equation}\label{decomg2}
\fg\eql \fh_1\oplus \ldots\oplus \fh_s\oplus\frak u(1)_1\oplus\ldots\oplus \frak u(1)_\nu\,,
\end{equation}
where $\fh_i$ are simple, and $\mathfrak{g}$ is  embedded in $\so(n+1)$ through a chain of maximal subalgebras, 
\begin{equation}\label{gchain2}
\so(n+1)\equiv  \fg_1\supset \fg_2\supset\ldots\supset \fg_k\equiv\fg\,.
\end{equation} 
\item Define $\tilde{\fg} \subset \fg$ as the Lie algebra for the group under which the reduction gauge vectors $A^{\mu\,i}(x)$ transform and in which they are valued. 
\end{itemize}

Assuming that the the set of functions $(A^{\mu\,i}(x)\pi_\mu,g_\omega(x) )$ on $\cB$ are linearly independent, then separation for a given $\mathfrak{g}$ will generically be impeded unless:
\begin{itemize}
\item [(i)] The Lie algebra, $\fg$, corresponds to one of the ``separating isometries" discussed in section \ref{sec:SepSph} (or summarized in Appendix \ref{App:tables}) which separates the corresponding $H_{\cF}$ with vanishing gauge vectors. Since $\widetilde{M}_{ABCD}$ has the same index structure as $M_{ABCD}$ they transform in the same representation of $\fg$,  hence $\widetilde{H}_{\cF}$ will separate when the corresponding $H_{\cF}$ separates for the same $\fg$. 
\item [(ii)] The Lie algebra $\tilde{\fg} $, is of the form:
\begin{equation}\label{decomg3}
\tilde{\fg} \eql \tilde{\fh}_1\oplus \ldots\oplus \tilde{\fh}_s\oplus\frak u(1)_1\oplus\ldots\oplus \frak u(1)_\nu\,,
\end{equation}
where the $\tilde{\fh}_i$ are abelian subalgebras of the corresponding $\fh_{i}$. This is required so that condition (i) discussed below (\ref{invK}) is satisfied, ensuring the $K_i$ are in involution and can be consistently set to constants. 
\item [(iii)] The Killing vectors, $K_{i}$, generating the subalgebra $\tilde{\fg}\subset \fg$, are in involution with the set of Killing tensors/vectors required to separate the the $\widetilde{H}_{\cF}$ part of the Hamiltonian, i.e. those required to separate the corresponding $H_{\cF}$ discussed in Section \ref{sec:SepSph}. 
\end{itemize}

This discussion makes it clear that for separation with non-trivial gauge fields, one should start by imposing one of the ``separating isometries,"  $\fg$, from Section \ref{sec:SepSph} on $H_{\widetilde{\cF}}$ and restricting the gauge vectors to some abelian subalgebra $\tilde{\fg}\subset \fg$. We defer the full analysis of the necessary conditions on $\tilde{\fg}$ to future work, but note that there may be conspiracies between the functions $(A^{\mu\,i}(x)\pi_\mu,g_\omega(x) )$ on $\cB$ which would alter the above discussion.

\subsection{The $(1,m,n)$ superstrata}
\label{SS:1mnFamily}

The $(1,m,n)$ superstrata, \cite{Mayerson:2020tcl}, belong to a three sphere reduction of $D=6$, $\mathcal{N}=(1,0)$ supergravity, coupled to two anti-self dual tensor multiplets. The solutions are microstate geometries for the D1/D5/P black hole.

We will use the standard parametrization $(\theta,\phi_{1},\phi_{2})$ of the three-sphere which is embedded in the ambient $\mathbb{R}^{4}$ as
\begin{align}
Y^{1} =  \sin \theta \sin \phi_{1} \,, \qquad Y^{2}=\sin \theta \cos \phi_{1} \,, \qquad Y^{3} =  \cos \theta \sin \phi_{2} \,, \qquad Y^{4}= \cos\theta \cos \phi_{2} \,. \label{superstrataAmbCoord}
\end{align}
In addition, it will be convenient to identify the generators of an $\mathfrak{su}(2)\oplus  \mathfrak{u}(1)$ subalgebra of $\mathfrak{so}(4)$:
\begin{equation}\label{}
\begin{split}
L_{1}=K^{12}+K^{34} \,, \qquad L_{2}&=K^{13}-K^{24}\,, \qquad L_{3}=K^{14}+K^{23}\,, \\
R_{1} &= K^{12} - K^{34} \,,
\end{split}
\end{equation}
where $K^{AB}$ are the standard Killing vectors which generate $\mathfrak{so}(4)$ in the coordinates (\ref{superstrataAmbCoord}) and identify the Casimir:\footnote{There is no need to consider $\cC_2^{\su(2)}$ separately since it is proportional to $\cC_2^{\so(4)}$, see (\ref{CasamirSO4proptoSU2}) and (\ref{sucastoso}).}
\begin{align}
\cC_2^{\so(4)}= \frac{1}{4}\sum_{A,B=1}^{4} K^{AB}K^{AB}   \,.
\end{align}

Introducing the light cone coordinates: 
\begin{align}
u= \frac{1}{\sqrt{2}}(t-y) \qquad \text{and} \qquad v= u= \frac{1}{\sqrt{2}}(t+y)\,,
\end{align}
where $y\sim y+ 2\pi R_{y}$ is the periodic direction along which of the branes of underlying D1/D5/P system intersect, then the coordinates on $\cB$ are specified by $(u,v,r)$ where $r$ is a radial direction. The $(1,m,n)$ superstrata solutions can then be parametrized in terms of the positive real constant $a$, the inverse radius\footnote{This radius is proportional to the D1/D5 charges via $
g_{0}=(Q_{1}Q_{5})^{-1/4}\,.$}
of the three-sphere at infinity, $g_{0}^{-1}$, and by the specification of pair of holomorphic functions  
\begin{align} \label{F0F1def}
F_{0}=\sum_{n=1}^{\infty}b_{n}\xi^{n} \qquad \text{and} \qquad F_{1}=\sum_{n=1}^{\infty}d_{n}\xi^{n} \,,
\end{align}
where the $b_{n}$ and $c_{n}$ are real constants and $\xi$ is the complex variable,
\begin{align} \label{xidef}
\xi ~\equiv~\frac{r}{\sqrt{a^{2}+r^2}} \, e^{ i\frac{\sqrt{2}}{R_{y}} v } \,.
\end{align} 

It will be convenient to introduce the scalars:
\begin{align}
\chi_{A} = - \frac{ag_{0}^{2}R_{y}}{\sqrt{ 2(a^{2}+r^{2}) }}  \left( iF_{0} ,F_{0},-ie^{i\frac{\sqrt{2}}{R_{y}} v } F_{1}, e^{i\frac{\sqrt{2}}{R_{y}} v }F_{1}  \right) + \text{c.c.} \,, \label{chidef}
\end{align}
and the combination
\begin{align}
\Omega^{2} \equiv \frac{1 }{2R_{y}^{2}g_{0}^{4}} \left( 4- \chi_{A}\chi_{A} \right) = \frac{2}{g_{0}^{4}R_{y}^{2}} - \frac{a^{2}}{a^{2}+r^{2}}\left(\abs{F_{0}}^{2}+\abs{F_{1}}^{2} \right)\,. \label{Omega21mn}
\end{align}

The full $(1,m,n)$ superstrata can then be fully specified by:
\begin{equation}
\begin{aligned}
H_\cB(x;\pi)= - \frac{\pi_{1}^{2}}{a^{4}g_{0}^{6}R_{y}^{2}}+\frac{1}{2a^{4}R_{y}^{4}r^{2}(a^{2}+r^{2})\Omega^{2}} & \left( \frac{4r^{4}\pi_{1}^{2}}{g_{0}^{10}} + \frac{4a^{4}R_{y}^{2}r^{2}\pi_{1}(\pi_{1}-\pi_{2})}{g_{0}^{6}} \right. \\
& \qquad \left. + \frac{a^{8}R_{y}^{4}(\pi_{1}-\pi_{2})^{2}+2a^{4}R_{y}^{2}r^{2}(a^{2}+r^{2})^{2}\pi_{3}^{2}}{g_{0}^{2}} \right) \,,
\end{aligned} \label{1mnHB}
\end{equation}
\begin{equation}
\begin{aligned}
-A^{\mu i}(x)\pi_{\mu} K_{i} &= \frac{1}{\sqrt{2}g_{0}^{2}R_{y}\Omega^{2}} \left[ \frac{a^{2}}{r^{2}}(\pi_{1}-\pi_{2})(R_{1}+L_{1})+ \frac{a^{2}}{a^{2}+r^{2}}\left(\left(1-\frac{2}{a^{2}g_{0}^{4}R_{y}^{2}}\right)\pi_{1}-\pi_{2} \right)(R_{1}-L_{1}) \right] \\
& \qquad  +\frac{\pi_{1}}{\sqrt{2} a^{2}g_{0}^{6}R_{y}^{3}\Omega^{2}} \left[ (\chi_{1}\chi_{3}-\chi_{2}\chi_{4})L_{3} -(\chi_{1}\chi_{4}+\chi_{2}\chi_{3})L_{2} \right] \\
& \qquad \qquad  +\frac{\pi_{1}}{2^{3/2}a^{2}g_{0}^{6}R_{y}^{3}\Omega^{2}} \left[ (\chi_{3}^{2}+\chi_{4}^{2})(R_{1}+L_{1}) + \left( \chi_{1}^{2}+\chi_{2}^{2}  \right)(R_{1}-L_{1})\right]\,,
\end{aligned} \label{1mnGaugeVecs}
\end{equation}
and  
\begin{equation}
\begin{aligned}
\sum_\omega g_\omega(x)\cals K_\omega(y;p) &=   \frac{1}{4g_{0}^{2}R_{y}^{2}\Omega^{2}} \left[4\cC_2^{\so(4)}+ \frac{a^{2}}{r^{2}} (L_{1}+R_{1})^{2} - \frac{a^{2}}{a^{2}+r^{2}} (L_{1}-R_{1})^{2}\right]\,,
\end{aligned} \label{1mnFiber}
\end{equation}

There are two possible impediments to separability to note at this juncture:
\begin{itemize}
\item The gauge vectors identified in (\ref{1mnGaugeVecs}) belong to the non-abelian $\mathfrak{su}(2)\oplus \mathfrak{u}(1)\subset \mathfrak{so}(4)$, so the $K_{i}$ will not be in involution.
\item The base part of the massless Hamilton-Jacobi equation, (\ref{1mnHB}), may not separate on $\cB$, since $\Omega^{2}$ may mix all three base coordinates $(u,v,r)$ in a highly no-trivial manner.
\end{itemize}

We will show in the next subsection that each of these points can be overcome by appropriately restricting $F_{0}$ and $F_{1}$, leading to partially and fully separable massless \HJ equations for subfamilies of the $(1,m,n)$ superstrata.  
 
\subsection{Separable $\text{U}(1)\times \text{U}(1)$ subsectors}
In order to produce a partially separable solution between the base and fiber, points one and two in the list of the previous subsection must be addressed. This is simply done by restricting the isometry group and gauge vectors to a $\text{U}(1)\times \text{U}(1)$ subgroup. There are two distinct ways of accomplishing this, by either setting $F_{1}=1$ or $F_{0}=0$, giving the $(1,0,n)$ and $(1,1,n)$ subfamilies respectively. Denoting the generators of $\text{U}(1)\times \text{U}(1)$ by $(L,R)$, they can be identified in our three-sphere conventions as:
\begin{align}
L= K^{12} \qquad \text{and} \qquad  R=K^{34} \,.
\end{align}
The corresponding Casimirs are then given by:
\begin{align}
\cC_2^{\mathfrak{u}(1)_{L}}=L^{2} \qquad \text{and} \qquad  \cC_2^{\mathfrak{u}(1)_{R}}=R^{2}\,.
\end{align}

The most general $\text{U}(1)\times \text{U}(1)$ invariant $\sum_\omega g_\omega(x)\cals K_\omega(y;p)$ in concert with the most general expansion of $-A^{\mu i}(x)\pi_{\mu} K_{i}$ for $U(1)\times U(1)$ gauge vectors then takes the form:
\begin{align}
-A^{\mu i}(x)\pi_{\mu} K_{i} &=  h_{L}(x)L+h_{R}(x)R  \,, \label{aGenU1U1}\\
 \sum_\omega g_\omega(x)\cals K_\omega(y;p)&=   g_{1}(x) \cC_2^{\so(4)}  +g_{2}(x) \cC_2^{\mathfrak{u}(1)_{L}} + g_{3}(x) \cC_2^{\mathfrak{u}(1)_{R}}\,, \label{FiberGenU1U1}
\end{align}
where $h_{L,R}(x)$ and $g_{\omega}(x)$ are arbitrary functions of the base. 

Since all of the $(K_{i},K_{\omega})$ appearing in (\ref{aGenU1U1})-(\ref{FiberGenU1U1}) are in involution and are sufficient in number, the fiber is guaranteed to be fully separable and partially separate from the base in the massless \HJ equation. This is purely a consequence of the $\text{U}(1)\times \text{U}(1)$ isometry imposed on $\sum_\omega g_\omega(x)\cals K_\omega(y;p)= \frac{1}{2}\Big(\, \cU_{\cI \cJ}  +    A^{\cI \sigma}   A^\cJ{}_\sigma   \, \Big) \,  K^{\cI \, m}   K^{\cJ \, n}$ and restricting the gauge group to the abelian $\text{U}(1)\times \text{U}(1)$ group.

\subsubsection{The $(1,0,n)$ superstrata}
\label{SS:Superstrata10n}

The $(1,0,n)$ subfamily of superstrata is given by setting $F_{1}=0$, or equivalent $\chi_3=\chi_{4}=0$, in the $(1,m,n)$ family of Section \ref{SS:1mnFamily}, giving:
\begin{equation}
\begin{aligned}
H_\cB(x;\pi)= - \frac{\pi_{1}^{2}}{a^{4}g_{0}^{6}R_{y}^{2}}+\frac{1}{2a^{4}R_{y}^{4}r^{2}(a^{2}+r^{2})\Omega_{(1,0,n)}^{2}} & \left( \frac{4r^{4}\pi_{1}^{2}}{g_{0}^{10}} + \frac{4a^{4}R_{y}^{2}r^{2}\pi_{1}(\pi_{1}-\pi_{2})}{g_{0}^{6}} \right. \\
&  \left. + \frac{a^{8}R_{y}^{4}(\pi_{1}-\pi_{2})^{2}+2a^{4}R_{y}^{2}r^{2}(a^{2}+r^{2})^{2}\pi_{3}^{2}}{g_{0}^{2}} \right) \,,
\end{aligned} \label{10nHB}
\end{equation}
\begin{equation}
\begin{aligned}
-A^{\mu i}(x)\pi_{\mu} K_{i} &= \frac{\sqrt{2}}{g_{0}^{2}R_{y}\Omega_{(1,0,n)}^{2}} \left[\frac{a^{2}}{r^{2}}(\pi_{1}-\pi_{2})L+ \frac{a^{2}}{a^{2}+r^{2}}\left(\frac{2\pi_{1}}{a^{2}g_{0}^{4}R_{y}^{2}}+\pi_{2}-\pi_{1} \right)R \right]\\
& \qquad \qquad \qquad \qquad \qquad \qquad \qquad \qquad + \frac{\sqrt{2}}{a^{2}g_{0}^{2}R_{y}}\left(  1- \frac{2}{g_{0}^{4}R_{y}^{2}\Omega_{(1,0,n)}^{2}}\right)\pi_{1} R \,,
\end{aligned} \label{10nGaugeVecs}
\end{equation}
and
\begin{equation}
\begin{aligned}
\sum_\omega g_\omega(x)\cals K_\omega(y;p) &=  \frac{1}{g_{0}^{2}R_{y}^{2}\Omega_{(1,0,n)}^{2}}  \left[ \cC_2^{\so(4)}    +   \frac{a^{2}}{r^{2}} \cC_2^{\mathfrak{u}(1)_{L}} - \frac{a^{2}}{a^{2}+r^{2}} \cC_2^{\mathfrak{u}(1)_{R}}\right]\,,
\end{aligned} \label{10nFiber}
\end{equation}
where 
\begin{align}
\Omega_{(1,0,n)}^{2} = \frac{1 }{2R_{y}^{2}g_{0}^{4}} \left( 4- \chi_{1}^{2} - \chi_{2}^{2} \right) =\frac{2}{g_{0}^{4}R_{y}^{2}} - \frac{a^{2}}{a^{2}+r^{2}}\abs{F_{0}}^{2} \,. \label{Omega210n}
\end{align}

From these expressions we can see that the fiber will separate internally and partially from the base by setting:
\begin{align}
L=c_{L} \,, \qquad R=c_{R} \,, \qquad \cC_2^{\so(4)} =c_{1}\,.
\end{align}

\subsubsection{The $(1,1,n)$ superstrata}
\label{SS:Superstrata11n}

The $(1,1,n)$ subfamily of superstrata is given by setting $F_{0}=0$, or equivalent $\chi_1=\chi_{2}=0$, in the $(1,m,n)$ family of Section \ref{SS:1mnFamily}, giving:
\begin{equation}
\begin{aligned}
H_\cB(x;\pi)= - \frac{\pi_{1}^{2}}{a^{4}g_{0}^{6}R_{y}^{2}}+\frac{1}{2a^{4}R_{y}^{4}r^{2}(a^{2}+r^{2})\Omega_{(1,1,n)}^{2}} & \left( \frac{4r^{4}\pi_{1}^{2}}{g_{0}^{10}} + \frac{4a^{4}R_{y}^{2}r^{2}\pi_{1}(\pi_{1}-\pi_{2})}{g_{0}^{6}} \right. \\
&  \left. + \frac{a^{8}R_{y}^{4}(\pi_{1}-\pi_{2})^{2}+2a^{4}R_{y}^{2}r^{2}(a^{2}+r^{2})^{2}\pi_{3}^{2}}{g_{0}^{2}} \right) \,,
\end{aligned} \label{11nHB}
\end{equation}
\begin{equation}
\begin{aligned}
-A^{\mu i}(x)\pi_{\mu} K_{i} &= \frac{\sqrt{2}}{g_{0}^{2}R_{y}\Omega_{(1,1,n)}^{2}} \left[\frac{a^{2}}{r^{2}}(\pi_{1}-\pi_{2})L+ \frac{a^{2}}{a^{2}+r^{2}}\left(\frac{2\pi_{1}}{a^{2}g_{0}^{4}R_{y}^{2}}+\pi_{2}-\pi_{1} \right)R \right]\\
& \qquad \qquad \qquad \qquad \qquad \qquad \qquad \qquad - \frac{\sqrt{2}}{a^{2}g_{0}^{2}R_{y}}\left(  1- \frac{2}{g_{0}^{4}R_{y}^{2}\Omega_{(1,0,n)}^{2}}\right)\pi_{1} L \,,
\end{aligned} \label{11nGaugeVecs}
\end{equation}
and
\begin{equation}
\begin{aligned}
\sum_\omega g_\omega(x)\cals K_\omega(y;p) &=  \frac{1}{g_{0}^{2}R_{y}^{2}\Omega_{(1,1,n)}^{2}}  \left[ \cC_2^{\so(4)}    +   \frac{a^{2}}{r^{2}} \cC_2^{\mathfrak{u}(1)_{L}} - \frac{a^{2}}{a^{2}+r^{2}} \cC_2^{\mathfrak{u}(1)_{R}}\right]\,,
\end{aligned} \label{11nFiber}
\end{equation}
where 
\begin{align}
\Omega_{(1,1,n)}^{2} = \frac{1 }{2R_{y}^{2}g_{0}^{4}} \left( 4- \chi_{3}^{2} - \chi_{4}^{2} \right) =\frac{2}{g_{0}^{4}R_{y}^{2}} - \frac{a^{2}}{a^{2}+r^{2}}\abs{F_{1}}^{2} \,. \label{Omega211n}
\end{align}

From these expressions we can see that the fiber will separate internally and partially from the base by setting:
\begin{align}
L=c_{L} \,, \qquad R=c_{R} \,, \qquad \cC_2^{\so(4)} =c_{1}\,.
\end{align}

\subsubsection{Separating the base}

Consider again the the base part of the Hamiltonian for the $(1,m,n)$ superstrata (\ref{1mnHB}):
\begin{equation}
\begin{aligned}
H_\cB(x;\pi)= - \frac{\pi_{1}^{2}}{a^{4}g_{0}^{6}R_{y}^{2}}+\frac{1}{2a^{4}R_{y}^{4}r^{2}(a^{2}+r^{2})\Omega^{2}} & \left( \frac{4r^{4}\pi_{1}^{2}}{g_{0}^{10}} + \frac{4a^{4}R_{y}^{2}r^{2}\pi_{1}(\pi_{1}-\pi_{2})}{g_{0}^{6}} \right. \\
& \qquad \left. + \frac{a^{8}R_{y}^{4}(\pi_{1}-\pi_{2})^{2}+2a^{4}R_{y}^{2}r^{2}(a^{2}+r^{2})^{2}\pi_{3}^{2}}{g_{0}^{2}} \right) \,.
\end{aligned} \label{1mnHB2}
\end{equation}
This part of the Hamiltonian is independent of and so separates trivially in $u$. The only dependence on $v$ comes though (\ref{Omega21mn}):
\begin{align}
\Omega^{2} = \frac{2}{g_{0}^{4}R_{y}^{2}} - \frac{a^{2}}{a^{2}+r^{2}}\left(\abs{\sum_{n=1}^{\infty}b_{n}\left(\frac{r}{\sqrt{a^{2}+r^2}} \, e^{ i\frac{\sqrt{2}}{R_{y}} v } \right)^{n}}^{2} + \abs{\sum_{n=1}^{\infty}d_{n}\left(\frac{r}{\sqrt{a^{2}+r^2}} \, e^{ i\frac{\sqrt{2}}{R_{y}} v } \right)^{n}}^{2}  \right) \,,
\end{align} 
which clearly obstructs separation between $(v,r)$ for generic $F_{0}=\sum_{n=1}^{\infty} b_{n}\xi^{n}$ and $F_{1}=\sum_{n=1}^{\infty} d_{n}\xi^{n}$. 

However if only a single term appears in each of the sums, say $b_{n_{1}}$ and $d_{n_{2}}$, then the dependence on $v$ will be removed:
\begin{align}
\Omega^{2}_{\text{single}} = \frac{2}{g_{0}^{4}R_{y}^{2}} - \frac{a^{2}}{a^{2}+r^{2}}\left[b_{n_{1}}^{2}\left(\frac{r^{2}}{a^{2}+r^2}  \right)^{n_{1}} + d_{n_{2}}^{2}\left(\frac{r^{2}}{a^{2}+r^2}  \right)^{n_{2}}   \right]\,,
\end{align} 
so that $H_\cB(x;\pi)$ is trivially completely separable on $\cB$. 

Thus, if one restricts to single $(1,0,n)$ or $(1,1,n)$ the six-dimensional massless \HJ equations will be fully separable. This separability has been extensively exploited in \cite{Bena:2018bbd,Bena:2019azk,Bena:2020yii,Martinec:2020cml}.

\section{Conclusions}
\label{sect:Conclusions}

In this paper we have explored the interplay between consistent truncation and separability of the \HJ equation in supergravity solutions.
This work was stimulated by the discovery of separability in single-mode superstrata in \cite{Bena:2018bbd,Bena:2019azk,Bena:2020yii,Martinec:2020cml}, however, we have now shown that this was far from an accident: the structure of the uplifted metric of any consistent truncation is well-adapted to possible separability of both the wave equation and the \HJ equation. This does not mean a generic consistent truncation will result in separability, and we have exhibited several, relatively simple examples where such separability fails. 

To analyze this situation we introduced the idea of a separating isometry, which implies that if a consistent truncation without reduction gauge vectors preserves this isometry, then the metric will necessarily have a separable \HJ equation. We then classified the separating isometries on sphere truncations, $S^n$, for $n=2, \dots, 7$. The surprise is that there are, in fact, many non-trivial separable isometries, and this is what we mean when we say that consistent truncations are ``well-adapted'' to separability.  When reduction  gauge vectors are present, in addition to the requirement of a separating isometry, the gauge group must be restricted to an abelian subgroup of the specific separating isometry in order to not impede separability.  

Obviously, specifying isometries will imply the existence of Killing vectors, and the associated conserved momenta, but the separable isometries on sphere are ``too small'' for the Killing vectors to fully separate the \HJ equation.  This means that there must be rank-two Killing tensors to complete the separation, and in this paper we have classified and constructed these rank-two Killing tensors using bilinears of the Killing vectors on the sphere.

While we have classified separability using generic consistent truncations that preserve a specified isometry, this does not mean we have a complete classification of separability, even for consistent truncations.  We used ``genericity'' of the background to show that the Killing tensors appearing in the expansion of the inverse metric must all be in involution with one another.  As we noted earlier, for non-separating isometries there can be specialized loci of the consistent truncation on which the \HJ equation becomes separable.  In particular, this can happen at points in the moduli space at which the actual symmetry is enhanced to a separating isometry.  However, there are, almost certainly less trivial examples that do not involve symmetry enhancement, or could be based on symmetry enhancement that involves discrete isometries.   We have not classified these possibilities. 

The important practical conclusion here is that, {\it if a consistent truncation has a separating isometry then the \HJ equation is {\it always} separable in the absence of reduction gauge vectors.} Further, if reduction gauge vectors are present, \textit{they must be restricted to an abelian subgroup of the separating isometry in order to not impede separability.} If a consistent truncation only has a non-separating isometry then the  \HJ equation  is generically non-separable, but there can still be specialized configurations in which the  \HJ equation becomes separable.

Combining the mathematics of separability with the metrics arising through consistent truncations is not only a very interesting formal problem, but it also has important impact on supergravity, string compactifications and holography.   

In any circumstance in which one creates a supergravity background to address a physics problem, one of the first priorities is to probe that background, and geodesics and scalar waves are the simplest of such probes. Analysis then leads to issues of separability and this determines how one tries to solve for the dynamics of probes, and the extent to which such analysis can be achieved.   Consistent truncations arise in situations in which the higher dimensional physics of string theory, as well as holography arising out of string theory, can be reduced to some simpler, lower-dimensional system.  Thus our analysis casts a broad light on the extent to which one can analyze probes of a rich families of supergravity backgrounds. Indeed, in Section \ref{ss:sepexhol} we catalogued families of important holographic backgrounds in which the \HJ equation is separable. 

Beyond this utility, this work has two immediate applications:
\begin{itemize}
\item[(i)] Given a supergravity solution, one can check whether the massless \HJ equation partially separates. If so it may be a sign the the solution is part of a consistent truncation which may subsequently be identified. A first step in this identification would be to check that the geometry conforms to the uplifted form of (\ref{MetUplift}). 
\item[(ii)] Given a pair of supergravity theories related by consistent truncation, the separating isometries (see the tables in Appendix \ref{App:tables}) may be used to constrain ans\"{a}tze for solutions which will necessarily have a partially separable massless \HJ equations between the fiber and base in the uplift. Such solutions will be more amenable to analytic investigation due to the additional symmetry. 
\end{itemize}

The first application was precisely the historical development that occurred with the $(1,m,n)$ superstrata, \cite{Bena:2017upb,Walker:2019ntz,Mayerson:2020tcl}, which motivated this work. Examples which give weight to the second application, include the microstate geometry probe calculations of \cite{Bena:2018bbd,Bena:2019azk,Bena:2020yii,Martinec:2020cml} and the calculation of the spin 2 spectrum of the solution of Section \ref{ss:N1flowinIIB} in \cite{Bobev:2020lsk}. This later calculation was used as a check against the results coming directly from exceptional field theory using the methods of \cite{Malek:2019eaz,Malek:2020yue}. It is interesting to note that the examples where purely analytic results for the spin 2 spectrum can be found in \cite{Klebanov:2009kp,Bobev:2020lsk,Guarino:2020flh}, possess separating isometries. This is suggestive that studying solutions with separating isometries within the framework of exceptional field theory may prove particularly useful.

There are also other interesting avenues of investigation that follow on from our work. We have not examined the separability of the scalar wave equation in any great detail. We noted in  Section~\ref{sect:separability} that separability of the wave equation is a stronger constraint than separability of the \HJ equation, and so the latter is necessary for the former. It would be interesting to generalize our discussion to the wave equation and see if the concept of a separating isometry has similar leverage in separating the wave equation. It is also natural to broaden the question to consider a notion of separability that encompasses charged scalars that transform in finite-dimensional representations of some non-abelian gauge symmetry, and how that gauge symmetry might interact with a separating isometry. In Appendix~\ref{App:sepwave} we make some first steps in looking at this broader notion of separability and, once again, show that consistent truncations are well-adapted to separability.

It is also interesting to consider separability in a broader context than just consistent truncations. \textit{Given a manifold which can be written as a fibration, is it possible to develop a set of conditions, that if satisfied by the fiber, lead to partial separability of the \HJ equation between the base and fiber?} In this work the fiber was a deformed sphere restricted as necessary for a consistent truncation within supergravity, while the conditions amounted to a preservation of separating isometries and a restriction of the gauge group to some abelian subgroup of the separating isometry group. However, in string theory,  compactifications on Calabi-Yau manifolds are ubiquitous, so it is potentially extremely useful to determine whether there is a comparable analysis of Calabi-Yau fibers which are adapted for separability. To begin addressing this problem one could consider simple K\"{a}hler\footnote{For an example of such a manifold where separability proved useful see the conifold solutions of \cite{Pufu:2010ie}.} or hyper-K\"{a}hler fibers, so that conditions can be readily imposed via K\"{a}hler potentials.

Thus we have shown that the observations of \cite{Bena:2018bbd,Bena:2019azk,Bena:2020yii,Martinec:2020cml} made in the context of microstate geometries in six dimensions was the tip of a much larger ice-berg that extends across all  consistent truncations and has a broad impact for holographic field theory.  Indeed, it means that solving the dynamics of probes in supergravity backgrounds could be a rather more tractable problem that one might have naively expected.

\section*{Acknowledgments}
\vspace{-2mm}
We are  grateful to Nikolay Bobev and Jesse van Muiden for discussions. The work of KP and NW was supported in part by the DOE grant DE-SC0011687. NW  and RW were supported by the ERC Grant 787320 - QBH Structure. RW was also supported by the Research Foundation - Flanders
(FWO). KP and RW are grateful to the IPhT Saclay for hospitality during the initial stage of this project.

\vspace{1cm}
\appendix

\section{Separating the scalar wave equation}
\label{App:sepwave}

As we remarked in Sections~\ref{sect:introduction} and  \ref{sect:separability}, the geometric optics approximation to the wave equation, and its realization through WKB methods, mean that the separability of the scalar wave equation implies the separability of the \HJ equation, but the converse is not true. This link is  further weakened if the scalars are coupled to electromagnetic fields. Moreover, if, instead of considering a single scalar, we broaden the notion of separability to include   finite families of  scalars that transform in a non-trivial representations of a non-abelian gauge symmetry, then one really has lost the link with the  \HJ equation.

Charged scalar waves are also very interesting probes of supergravity solutions and this is an issue well worth exploring in the future:  what can we say about consistent truncations, and the broader notion of separability for the wave equation?  In this appendix we make a short excursion into  these ideas and once again find that there is a very powerful synergy between the metrics of consistent truncation and the broader notion of separability.

The massive scalar wave equation on $\cM$ is given by:
\begin{equation}
\nabla^M \nabla_M  \Psi  ~=~  \frac{1}{\sqrt{\abs{G}}} \,\partial_{M} \,\Big( \sqrt{\abs{G}} \, G^{MN}\partial_{N}\Psi \Big) ~=~ -m^{2}\Psi\,.
\label{Masswave}
\end{equation}

The Kaluza-Klein ansatz, \eqref{KKansatz}, for the metric can be written in terms of frames as 
\begin{equation}
E_M{}^A  ~=~ 
 \begin{pmatrix}
E_\mu{}^\alpha & B_\mu{}^m \, e_m{}^a \\
0 & e_m{}^a
\end{pmatrix} \,,
\end{equation} 
where the lower left block is set to zero using the off-diagonal Lorentz rotation. The indices $\alpha$ and $a$ are flat along $\cB$ and $\cF$, respectively. The rescaling   of the frames corresponding to \eqref{resmet} is
\begin{equation}
E_\mu{}^\alpha ~=~ \Delta^{-\frac{1}{d-2}} \,  e_\mu{}^\alpha    \,.
\label{Bframes}
\end{equation}
so that
\begin{equation}
 \Delta  ~\equiv~ \frac{\det(e_m{}^a)}{\det(\eo_m{}^a)}~=~\sqrt{ \frac{\det(g_{mn})}{\det(\go_{mn})}}   \,,
\label{Deltadef}
\end{equation}
where $\eo_m{}^a$ is a frame for the round metric $\go_{mn}$. From 
 (\ref{Bframes}) and (\ref{Deltadef}) we see that, 
\begin{equation}
\sqrt{\abs{G}} ~=~  \Delta^{-\frac{2}{d-2}} \,\det ( e_\mu{}^\alpha)  \,  \det (\eo_m{}^a) 
~=~  \Delta^{-\frac{2}{d-2}} \, \sqrt{ - \det (  g_{\mu \nu})  \,  \det (\go_{mn}) }\,,
\end{equation}
and hence 
\begin{equation}
\begin{aligned}
\Delta^{-\frac{2}{d-2}} \,  \nabla^M \nabla_M  \Psi  ~=~&  \frac{1} {\sqrt{|  g |}} \,\partial_\mu \Big [ \,\sqrt{|  g|} \,  g^{\mu \nu} \,\big(\partial_\nu - A^\cI{}_\nu  \, K^{\cI \, m}   \, \partial_m \big) \Psi  \,\Big]  \\ 
& + \frac{1} { \sqrt{\go} } \,\partial_m \bigg [ \, \sqrt{\go}  \,  \,\big(  -  g^{\mu \nu} \,A^\cI{}_\nu  \, K^{\cI \, m} \,\partial_\mu \Psi   \big)  + \big(\, \cU_{\cI \cJ} +  g^{\rho \sigma} A^\cI{}_\rho    A^\cJ{}_\sigma   \, \big) \,  K^{\cI \, m}   \,  K^{\cJ \, n}  \,\partial_n \Psi \,\bigg] \,,
\end{aligned}   
\label{sepLap0}
\end{equation}
where $\go$ is the determinant of the round metric on $\cF$.   This reduces to:
\begin{equation}
 \nabla^M \nabla_M  \Psi  ~=~  \Delta^{\frac{2}{d-2}} \,   \Big[\,  g^{\mu \nu} \,\big(\nabla_\mu - A^\cI{}_\mu  \, \cL_\cI \big)\,\big( \nabla_\nu - A^\cJ{}_\nu  \, \cL_\cJ \big) \,\Psi  ~+~ \cU_{\cI \cJ} \, \cL_\cI \, \cL_\cJ\, \Psi  \, \Big] \,, 
 \label{sepLap1}
\end{equation}
where, $g$ and $\nabla_{\mu}$ are the determinant of the metric and the covariant derivative on $\cB$ respectively. The expression:
\begin{equation}
\cL_\cI \, F   ~\equiv~  \,  K^{\cI \, m}  \,\partial_m  \,F  \,,
 \label{Lieder}
\end{equation}
denotes the Lie derivative.   To establish (\ref{sepLap1}) one must use the Killing equation for $K^{\cI \, m}$ on the ``round" compact manifold to remove the divergences of $K^{\cI \, m}$.

Observe that the operator 
\begin{equation}
\cU_{\cI \cJ}(x)  \, \cL_\cI \, \cL_\cJ\, \,,
 \label{SphereLap1}
\end{equation}
is a deformation of the Laplacian of the ``round'' metric on $\cF$ and  can be simplified in terms of the eigenmodes of this operator.  
Indeed, suppose  the scalar, $\Psi(x,y)$, can be written in terms of a set of harmonics that form a representation of the isometries on the round sphere:
\begin{equation}
\Psi(x,y)    ~=~  \sum_j  \, \psi_j (x) \, Y^j(y)\,.
 \label{harmdecomp}
\end{equation}
Then there is a constant representation matrix, $ T_\cI{}^i{}_j$, such that 
\begin{equation}
\cL_\cI \, Y^i (y)   ~= ~  T_\cI{}^i{}_j  \,Y^j(y)   \,,
 \label{Lierep}
\end{equation}
and (\ref{sepLap1}) reduces to 
\begin{equation}
\begin{aligned}
 &\Delta^{-\frac{2}{d-2}}\nabla^M \nabla_M  \Psi  \\
 &~~=   \, \Big[\,  g^{\mu \nu} \,\big(\delta^j_k \,\nabla_\mu - A^\cI{}_\mu  \, T_\cI{}^j{}_k   \big)\,\big(\delta^i_j \,  \nabla_\nu - A^\cJ{}_\nu  \, T_\cJ{}^i{}_j  \,  \big) \,\psi_i(x)   ~+~ \cU_{\cI \cJ} \, T_\cI{}^j{}_k  \,T_\cJ{}^i{}_j  \, \psi_i(x)  \, \Big] \,  Y^k(y) \,.
 \label{sepLap2}
 \end{aligned}
\end{equation}

There are several things to note at this juncture.  First, modulo the warp factor, the wave operator has separated and reduced to an operator entirely on $\cB$.  
The dependence on $\cF$ has been  reduced to group representations of harmonics on the round metric.  This is therefore a ``consistent truncation'' for any choice of harmonics, $Y^j(y)$.    This is what we mean when we refer to a ``broader notion'' of separability for the wave equation.   The standard (stronger) form of separability involves a single function on $\cB$ and a single function on $\cF$. Here the separation of variables involves a finite space of modes. 

If the background vector fields, $A^\cI{}_\mu$, are restricted to abelian $\rm U(1)$ subgroups of the isometries, $A^{\hat \cI}{}_\mu$, with charges $q_{\hat \cI}$. Then one can construct a $Y(y)$, giving a one-dimensional eigenspace under these $\rm U(1)$'s, with:
\begin{align}
\cL_{\hat{\cI}}Y(y)= iq_{\hat{\cI}} Y(y)\,. \label{U1GaugeExp}
\end{align}
When this occurs, the first part of the operator in  (\ref{sepLap1}), reduces to:
\begin{equation}
 \Big[\,  g^{\mu \nu} \,\big(\,\nabla_\mu - i q_{\hat \cI} \,A^{\hat \cI}{}_\mu \, \big)\,\big(\,  \nabla_\nu - i q_{\hat \cJ}  \, A^{\hat \cJ}{}_\nu  \,  \big) \,\psi(x)    \, \Big] \,  Y(y) \,.
 \label{sepLap3}
\end{equation}
This part of the Laplacian has now separated in the more traditional  (strong) sense: $ \Psi(x,y) = \psi(x)  Y(y)$.  

As with the \HJ equation,  the massive wave equation, (\ref{Masswave}), is only separable if $ \Delta^{-\frac{2}{d-2}}$ separates into the sum of a function of $x$ and a function of $y$. Such a separation is possible if $\cU_{\cI \cJ}\propto \delta_{\cI\cJ}$, but generically it will only be the massless wave equation that will be separable. 

Finally, we note that the broader notion of separability described here overlaps strongly with the idea of consistent truncation:  in both instances some higher-dimensional dynamics can be reduced exactly to dynamics in lower dimensions. It is thus an obvious question as to whether the broader concept of separability explored here goes beyond  the idea of consistent truncation.  The answer is obviously yes: a consistent truncation typically involves a precisely specified spectrum of higher-dimensional fields with a very limited set of representations of the isometry group involved in the dimensional reduction.  Here we are demonstrating broader separability for a scalar field in any representation and in any consistent truncation.  For example, eleven-dimensional supergravity doesn't even  have a scalar field, but the results presented here show that an eleven-dimensional scalar, in any representation of the underlying reduction isometry, is ``broadly separable'' in either of the standard sphere reductions.     There is thus an interesting set of open questions about the ``broader separability'' of the dynamics of fields that may not be part of the consistent truncation, and yet may be very useful probes of those backgrounds.

\section{Summary tables}
\label{App:tables}
\def\DG{{}}

In Tables~\ref{tblsph16} and  \ref{tblsph} in this appendix we list chains of isometries on $n$-spheres, $S^n$, $n=2,\ldots,7$, that  guarantee separability of the \HJ equation for the uplifted metric \eqref{KKansatz} as discussed in Section~\ref{sec:SepSph}. We also  present some examples of isometries where this separability fails by violating one of the general conditions \ref{C1} or \ref{C2} of Section~\ref{ss:sepRmflds}. 

The outline of the tables is as follows. For each sphere, $S^n$, the lie algebra, $\fg$, of an isometry imposed on the uplifted metric is given through a chain of maximal subalgebras of $\so(n+1)$, with each subsequent step indicated by an indentation. For example, on $S^3$, both $\so(3)$ and $\su(2)\oplus \frak u(1)$ are maximal subalgebras of $\so(4)$, while $\frak u(1)\oplus\frak u(1)$ is a maximal subalgebra of  $\su(2)\oplus \frak u(1)$. A particular embedding of $\fg$ in $\so(n+1)$ can be also determined directly from the branching of the vector representation, $\Yboxdim6pt\yng(1)\,$, of $\so(n+1)$ given in the third column. In the next three columns we give the number of singlets of $\fg$ in the branching of the three irreps of $\so(n+1)$ that comprise the metric, see \eqref{YTdec}. The last but one column indicates whether the \HJ equation is separable in the sense of Section~\ref{sec:SepSph}, while the last column gives a reference to a subsection where the separability or the lack thereof is discussed in more detail.

The lists of separating isometries in Tables~\ref{tblsph16} and  \ref{tblsph} are complete in the sense that by imposing a smaller isometry onto the uplifted metric will generically lead to a \HJ equation that is  not partially separable. This does not mean that there are no \HJ equations with that isometry that are separable, but only that the isometry itself does not guarantee separability for all  metric functions, $M_{ABCD}(x)$, in \eqref{InvMet} with that invariance.

Smaller subalgebras may arise in multiple chains of maximal subalgebras. To simplify the tables we usually list them only once. Similarly, we usually omit duplicates of subalgebras containing $\frak u(1)$ factors that differ either by the normalization or more generally linear combinations of the $\frak u(1)$ charges, see for example the $\su(2)\oplus\frak u(1)\oplus \frak u(1)$ on $S^5$ in Table~\ref{tblsph16} or $\su(3)\oplus\frak u(1)\oplus \frak u(1)$ in Table~\ref{tblsph}. They are equivalent as far as our problem is concerned.

Additional subtlety is present  on $S^7$, where an isometry subalgebra, $\fg$, may be embedded into $\so(8)$ acting on $\bbR^8$ through one of its eight-dimensional irrepses: $\bfs 8_s$, $\bfs 8_v$ or $\bfs 8_c$.\footnote{We use  the same conventions as in \cite{Slansky:1981yr,Feger2020a}, with the branchings $\bfs 8_s\to\bfs 1+\bfs 7$, $\bfs 8_{v,c}\to \bfs 8$ under the standard $\so(7)\subset\so(8)$.} Since the branchings of the tensor products of $\bfs 8_v$ and of the products of $\bfs 8_c$ are related by a reflection in $\bbR^8$, they are equivalent as far as separability is concerned. In the first column in Table~\ref{tblsph}, we have indicated the triality of the  embedding of $\fg$ into $\so(8)$. For the three non-separable examples indicated by the $*$, the analysis in Section~\ref{sss:sp4su2} is not completely exhaustive.

\begin{table}
\renewcommand{\arraystretch}{1.1}
\begin{center}
\resizebox{0.9\textwidth}{!}{%
\begin{tabular}{   @{\extracolsep{5 pt}}  
c l l c c c c c}
\toprule
\noalign{\smallskip}
 & \qquad  $\fg$ & $\Yboxdim6pt\yng(1)$  &  $\Yboxdim6pt\yng(2,2)$ &  $\Yboxdim6pt\yng(2)$ & $\bullet$ & SEP & Sec. \\
\noalign{\smallskip}
\midrule
\noalign{\smallskip}
$S^2$ & \DG $\so(3)$ & $\bfs 3$ & 0 & 0 & 1 & Y & \ref{sss:sonp1}\\ 
& \quad \DG $\frak u(1)$ & $(2)+(0)+(-2)$ & $-$ & 1 & 1 & Y & \ref{sss:son} \\
\noalign{\smallskip}
\midrule
\noalign{\smallskip}
$S^3$ & \DG $\so(4)$ & $\bfs 4$ & 0 & 0 & 1 & Y & \ref{sss:sonp1}\\
&\quad \DG  $\so(3)$ & $\bfs 1+\bfs 3$ &  0 & 1 &1   & Y & \ref{sss:son} \\ 
& \DG \quad $\su(2)\oplus \frak u(1)$ & $\bfs 2(1)+\bfs 2(-1)$ & 1 & 0 & 1 & Y & \ref{sss:sumu1nodd}\\ 
	& \DG \qquad $\frak u(1)\oplus \frak u(1)$ & $(1,1)+(1,-1)+(-1,1)+(-1,-1)$ & 2 & 1 & 1 & Y & \ref{sss:sumu1u1even} \\
\noalign{\smallskip}
\midrule
\noalign{\smallskip}
$S^4$ & \DG $\so(5)$ & $\bfs 5$ & 0 & 0 & 1 & Y & \ref{sss:sonp1}\\
&\DG \quad $\so(4)$ & $\bf 1+4$ & 0 & 1 & 1 & Y  & \ref{sss:son}\\
		&\DG \qquad $\su(2)\oplus \frak u(1)$ & $\bfs 1(0)+\bfs 2(1)+\bfs 2(-1)$ & 1 & 1 & 1 & Y & \ref{sss:su2minn1} \\ 
 	& \DG \quad $\so(3)\oplus \frak \so(2)$ & $ \irrep{1} (2)+ \irrep{1} (-2)+ \irrep{3} (0)$ & 1 & 1 & 1 & Y & \ref{sss:sopsoq} \\ 
	&   \quad  $\so(3)$ & $\bfs 5$   & 1 & 0 & 1  & N & \ref{sss:notsepqr}\\
\noalign{\smallskip}
\midrule
\noalign{\smallskip}
$S^5$ & \DG $\so(6) $ & $\bfs 6$ & 0 & 0 & 1 & Y & \ref{sss:sonp1}\\
& \DG \quad $\su(3)\oplus \frak u(1)$ & $\bfs 3(-2)+\overline{\bfs 3}(2)$ & 1 & 0 & 1  & Y & \ref{sss:sumu1nodd} \\ 
		&\DG  \qquad $\su(3)$ & $\bfs 3 +\overline{\bfs 3} $  & 1 & 0 & 1 & Y & \ref{sss:dropu1}\\
		&\DG $\qquad  \su(2)\oplus\frak u(1)\oplus\frak u(1)$ & 
		$ \irrep{1}(2, 2)+ \irrep{1} (-2,-2)+ \irrep{2} (1,-2)+ \irrep{2} (-1,2)$ & 3 & 1 & 1 & Y & \ref{sss:sumu1u1even}\\ 
			&\DG $\quad \qquad  \su(2)\oplus\frak u(1)$ & 
		$ \irrep{1}(  2)+ \irrep{1} (-2)+ \irrep{2} (-2)+ \irrep{2} (2)$ & 3 & 1 & 1 & Y & \ref{sss:dropu1}\\ 
	
		&\qquad  $  \so(3)\oplus \frak u(1)$ & $\irrep{3} (-2)+ \irrep{3}(2)$ & 2 & 0 & 1 & N & \ref{sss:notsepqr}\\ 
	& \DG \quad $\su(2)\oplus\su(2)\oplus \frak u(1)$  & 
	$ {\bf (1,1)}(2)+{\bf (1,1)}(-2)+{\bf (2,2)}(0)$ & 1 & 1 & 1  & Y & \ref{sss:sopsoq}\\ 
		& \DG $\qquad \su(2)\oplus\frak u(1)\oplus\frak u(1)$ &
	$\irrep{1} (0,2)+ \irrep{1} (0,-2)+ \irrep{2} (1,0)+ \irrep{2} (-1,0)$ & 3 & 1 & 1 & Y & \ref{sss:sumu1u1even} \\ 
	& \DG\quad  $\so(5)$ & $\bfs 1+\bfs 5$  & 0 & 1 & 1  & Y & \ref{sss:son}\\ 
	& \qquad   $\su(2)$ &  $\bfs 1+\bfs 5$  & 1 & 1 & 1  & N & \ref{sss:notsepqr}\\
	& \DG \quad $ \so(3)\oplus \so(3)$ & ${\bf (3,1)+(1,3)}$ & 1 & 1 & 1 & Y & \ref{sss:sopsoq} \\
\noalign{\smallskip}
\midrule
\noalign{\smallskip}
$S^6$ & \DG $\so(7)$ & $\bfs 7$  & 0 & 0 & 1 & Y & \ref{sss:sonp1}\\ 
& \DG \quad $\so(6)$ & $\bfs 1+\bfs 6$ & 0 & 1 & 1  & Y & \ref{sss:son} \\ 
		& \DG $\qquad \su(3)\oplus\frak u(1)$ & $ \irrep{1}(0)+ \irrep{3}(-2)+ \irrepbar{3}(2)$ & 1 & 1 & 1 & Y & \ref{sss:su2minn1} \\ 
 		& \DG $\quad \qquad \su(3)$ & $ \irrep{1} + \irrep{3} + \irrepbar{3} $ & 1 & 1 & 1 & Y & \ref{sss:su2minn1du1}\\ 
		& \quad\qquad $\so(3)\oplus\frak u(1)$ & $ \irrep{1}(0)+ \irrep{3}(-2)+ \irrep{3}(2)$ & 2 & 1 & 1  & N &\ref{sss:notsepqr}\\
	& \DG \quad $\su(2)\oplus \su(2)\oplus \su(2)$ & $\bf (2,2,1)+(1,1,3) $ & 1 & 1 &  1  & Y & \ref{sss:sopsoq}\\ 
		&\DG  $\qquad  \su(2)\oplus \su(2)\oplus \frak u(1) $ & 
		$(\irrep{2},\irrep{1})(1)+(\irrep{2},\irrep{1})(-1)+(\irrep{1},\irrep{3})(0)$ & 2 & 1 & 1  & Y & \ref{sss:su2minn1du1}\\ 
	& \DG \quad $\so(5)\oplus \so(2)$	& $\irrep{1} (2)+ \irrep{1} (-2)+ \irrep{5} (0)$ & 1 & 1 & 1  & Y & \ref{sss:sopsoq}
	\\ 
	&  \quad  $\frak g_2$ & $\bfs 7$ & 0 & 0 & 1 & Y & \ref{sss:g2inso7}\\
 	&   $\qquad \su(3)$ & $ \irrep{1} + \irrep{3} + \irrepbar{3} $ & 1 & 1 & 1 & Y & \ref{sss:su2minn1du1}\\ 
	& \qquad  $\su(2) $ & $\bfs 7$  & 1 & 0 &  1 & N & \ref{sss:notsepqr}\\
	& $\qquad \su(2)\oplus \su(2) $ & $(\irrep{2},\irrep{2})+(\irrep{3},\irrep{1})$ & 2 & 1 & 1 & N & \ref{sss:su22ing2}\\
\noalign{\smallskip}
\bottomrule
\end{tabular}
}
\medskip
\caption{\label{tblsph16} Separating isometries on $n$-spheres, $S^n$, $n=2,\ldots ,6$. }
\end{center}
\end{table}

\vfill\eject


\begin{table}[h]
\renewcommand{\arraystretch}{1.1}
\begin{center}
\resizebox{0.99\textwidth}{!}{%
\begin{tabular}{@{\extracolsep{5 pt}} c l l c c c c c}
\toprule
\noalign{\smallskip}
 & \hspace{43 pt}$\frak g$ &  $\Yboxdim6pt\yng(1)$ &  $\Yboxdim6pt\yng(2,2)$ &  $\Yboxdim6pt\yng(2)$ & $\bullet$  & SEP & Section\\
\noalign{\smallskip}
\midrule
$S^7$ & $\so(8)$ & $\bfs 8$ & 0 & 0 & 1 & Y & \ref{sss:sonp1}\\
$ s,v,c$ & \DG $\quad\su(2)\oplus \su(2)\oplus \su(2)\oplus \su(2)$ & $\bf (2,1,2,1)+(1,2,1,2)$ & 1 & 1 & 1 & Y & \ref{sss:sopsoq} \\ 
		& \DG \qquad $ \su(2)\oplus \su(2)\oplus \su(2)\oplus \frak u(1)$ & 
		${\bf(1,2,1)}(1)+{\bf (1,2,1)}(-1)+{\bf(2,1,2)}(0) $ & 2 & 1 & 1 & Y & \ref{sss:su2su2su2u1}\\
		& \DG $\qquad\quad\su(2)\oplus \su(2)\oplus\frak u(1) \oplus\frak u(1)$ & $ (\irrep{2},\irrep{1})(1,0)+(\irrep{2},\irrep{1})(-1,0)+(\irrep{1},\irrep{2})(0,1)+(\irrep{1},\irrep{2})(0,-1)$ & 4 & 1 & 1 & Y & \ref{sss:su2su2su2u1} \\ 
		
%
$ s$	& \DG \quad $\so(6)\oplus \so(2) $ & $\bfs 1(2)+\bfs 1(-2)+\bfs 6(0)$ & 1 & 1 & 1  & Y & \ref{sss:sopsoq}\\ 
		& \DG $\qquad \su(3)\oplus \frak u(1)\oplus \frak u(1)$ 
		& $ \irrep{1}(0,2)+\irrep{1}(0,-2)+\irrep{3}(2,0)+\irrepbar{3}(-2,0)$ & 3 & 1 & 1  & Y & \ref{sss:sumu1u1even}\\
$v,c $  & \quad $\su(4)\oplus\frak u(1)$ &  $  \irrep{4}(1)+ \irrepbar{4}(-1)$ & 1 & 0 & 1 & Y & \ref{sss:sumu1nodd}\\
 & \qquad $\su(3)\oplus\frak u(1)\oplus\frak u(1)$ & $ \irrep{1} (3,-1)+ \irrep{1} (-3,1)+ \irrep{3} (1,1)+ \irrepbar{3} (-1,-1)$ & 3 & 1 & 1 & Y & \ref{sss:sumu1u1even}
 	 \\
$s,v,c$	& $ \quad   \su(3)$ & $\bfs 8$ & 1 & 0 & 1 & N & \ref{sss:su3}  \\
	
$s$	&\DG \quad  $\so(7)$  & $\bf 1+7$ &  0 & 1 & 1  & Y & \ref{sss:son}\\
%

$v,c$	&\DG \quad  $\so(7)$  & $\bf 8$ & 0 & 0 & 1 & Y & \ref{sss:so7vc} \\
	& \qquad $\su(4)$ & $\irrep{4}+\irrepbar{4}$ & 1 & 0 &  1 & Y & \ref{sss:dropu1} \\
	& \quad\qquad $\su(3)\oplus\frak u(1)$ & $(\irrep{1})(3)+(\irrep{1})(-3)+(\irrep{3})(1)+(\irrepbar{3})(-1)$  
	& 3 & 1 & 1 & Y & \ref{sss:dropu1}\\
	& \qquad $\su(2)\oplus\su(2)\oplus\su(2)$ & $(\irrep{2},\irrep{1},\irrep{2})+(\irrep{1},\irrep{2},\irrep{2})$ & 2 & 1 & 1 & N$^*$ & \ref{sss:sp4su2}\\
	& \qquad $\sp(4)\oplus \frak u(1)$ & $ \irrep{4} (1)+ \irrep{4} (-1)$ & 2 & 0 & 1 & N$^*$ & \ref{sss:sp4su2} \\
	
$s,v,c$		& $\qquad  \DG \frak g_2$ & $\bfs 1+\bfs 7$ & 0 & 1 & 1  & Y & \ref{sss:g2inso8}\\

$s$	& \quad  $\so(5)\oplus \so(3)$ & $(\irrep{1},\irrep{3})+(\irrep{5},\irrep{1})$ & 1 & 1 & 1   & Y & \ref{sss:sopsoq}\\$v,c$	& \quad $\sp(4)\oplus\su(2) $ & $\bf (4,2)$ & 1 & 0 & 1 & N$^*$ & \ref{sss:sp4su2}\\

\bottomrule
\end{tabular}
}
\medskip
\caption{\label{tblsph} Examples of (non-)separating isometries on $S^7$. }
\end{center}
\end{table}
\normalsize

\section{Explicit examples}
\label{app:ExpEx}

In this appendix we show explicitly how the massless \HJ equation (\ref{HJ-eqs}):
\begin{align} \label{MHJEApp}
G^{MN}\,{\partial S\over\partial x^M}\,{\partial S\over\partial x^N}\eql 0 \,,
\end{align}
separates for the metrics introduced in Sections \ref{ss:sepexhol} and \ref{sect:MicroExamples}. 

\subsection{The $\cals N=1$ flow in Section~\ref{ss:N1flowinIIB}}
\label{app:N1flowinIIB}

The metric for the uplift of the $\cals N=1$ holographic RG-flow in Section~\ref{ss:N1flowinIIB} in terms of local coordinates in \cite{Pilch:2000fu} reads:
\begin{equation}\label{Neq1metricApp}
ds_{10}^2\eql {X^{1/2}\over \rho} \cosh \chi\,\left(\,ds_{1,4}^2-{a^2\over 2}\,\widetilde {ds}{}_5^2\right)\,,
\end{equation}
where 
\begin{equation}\label{}
ds_{1,4}^2 ~=~ e^{2 A(r)}\big(\, \eta_{i j} \, dx^i \, d x^j \big) ~-~ dr^2  \,, \qquad i,j \eql 0\,,\ldots\,,3  \,,
\end{equation}
and 
\begin{equation}\label{metr55}
\widetilde {ds}{}_5^2\eql {\sech^2\chi\over \rho^2\,}\,\Big[d\theta^2+{\rho^2\over 4\,X}(\sigma_1^2+\sigma_2^2)+{\sin^2(2\theta)\over 4\,X^2}\Big(d\phi-{\rho^6\over 2}\,\sigma_3\Big)^2\,\Big]+{\rho^4\over X^2}\Big(\sin^2\theta\,d\phi+{1\over 2}\,\cos^2\theta\,\sigma_3\Big)^2\,.
\end{equation}
The scalar fields, $\alpha(r)$ and $\chi(r)$, depend only on the radial coordinate, $r$,  and 
\begin{equation}\label{}
X(r,\theta)\eql \cos^2\theta +\rho(r)^6\sin^2\theta\,,\qquad \rho=e^\alpha(r)\,.
\end{equation}
The $\sigma_j$ are the standard $\SU(2)$-invariant forms, satisfying $d\sigma_i=\sigma_j\wedge\sigma_k$.  Explicitly, one can take Euler angles, $(\varphi_1, \varphi_2, \varphi_3)$, on $\rm SU(2)$ and write: 
\begin{equation}
\begin{aligned}
\sigma_1 ~=~ & \cos (\varphi_3)\,  d\varphi_1 +  \sin ( \varphi_1 )\, \sin ( \varphi_3)\, d\varphi_2 \,, \\
\sigma_2 ~=~ & \sin (\varphi_3)\,  d\varphi_1 -  \sin ( \varphi_1 )\, \cos ( \varphi_3)\, d\varphi_2\,, \\
\sigma_3 ~=~ & d\varphi_3 + \cos ( \varphi_1 )\,   d\varphi_2  \,.
\end{aligned}
\label{sigdefn}
\end{equation}
The metric  has an $\rm SU(2) \times U(1) \times U(1)$ isometry, with the two $\rm U(1)$'s given by a $\phi$-translation and a $\varphi_3$-translation, which is a rotation between $\sigma_1$ and $\sigma_2$.   It was shown in \cite{Corrado:2001nv} that the metric is that of a deformed  Hopf fibration over stretched $\mathbb C\mathbb P^2$, with the Hopf fiber given by 
$\omega\eql \sin^2\theta\,d\phi+{1\over 2}\,\cos^2\theta\,\sigma_3$.

Assuming a separable ansatz of the form:
\begin{equation}\label{}
S \eql S_{x}(x)+ S_\theta (\theta) +S_{\varphi_{1}}(\varphi_{1}) +S_{\varphi_{2}}(\varphi_{2}) +S_{\varphi_{3}}(\varphi_{3})+S_{\phi}(\phi)\,,
\end{equation}
then the massless Hamilton Jacobi equation (\ref{MHJEApp}), reduces to the following system of differential equations.

Three trivial ODEs along the fiber:
\begin{align}
S'_{\varphi_{2}} = c_{\varphi_{2}} \,, \qquad S'_{\varphi_{3}} = c_{\varphi_{3}}\,, \qquad S'_{\phi} = c_{\phi}\,,
\end{align}
corresponding to the three commuting Killing vectors of the isometries along the $\varphi_2$, $\varphi_3$ and $\phi$ coordinates of the metric (\ref{Neq1metricApp}).

Two non-trivial ODEs along the fiber:
\begin{align}
 S'_{\varphi_{1}}(\varphi_{1})^{2} +c_{\varphi_{3}}^{2}+ \left(\frac{c_{\varphi_{2}}-c_{\varphi_{3}}\cos \varphi_{1}}{\sin \varphi_{1}}  \right)^{2}-c_{\varphi_{1}} &=0  \,,\\
S'_{\theta}(\theta)^{2}+ \frac{c_{\phi}^{2}}{\sin^{2}\theta} +\frac{4 c_{\varphi_{1}}}{\cos^{2}\theta} -c_{\theta} &= 0\,,
\end{align}
corresponding to (conformal) Killing tensors.

The remaining differential equation purely on the base:
\begin{equation}
\begin{aligned}
 g^{\mu\nu} \frac{\partial S_{x}}{\partial x^{\mu}} \frac{\partial S_{x}}{\partial x^{\nu}}  &= -{g^2\over 8\rho^4} \Big[-4(c_{\varphi_{1}}+c_{\varphi_{3}}^{2}) +(c_{\phi}^{2}-c_{\theta} +4c_{\varphi_{1}} - 4 c_{\phi}c_{\varphi_{3}} )\rho^{6} -2 c_{\phi}^{2}\rho^{12} \\
& \qquad\qquad  +\left[4c_{\varphi_{3}}^{2}+ \left(c_{\phi}^{2}-c_{\theta}+4c_{\phi}c_{\varphi_{3}} \right)\rho^{6} +4c_{\varphi_{1}}(\rho^{6}-1)  \right] \cosh 2\chi\Big]\,,
\end{aligned}
\end{equation}
which agrees with \eqref{HFeffect}.

\subsection{The $\cals N=2$ flow in Section~\ref{ss:N2flowinIIB}}
\label{app:N2flowinIIB}

The type IIB supergravity uplift  of the metric  \cite{Pilch:2000ue} for the $\cN=2$ flow is given by: 
\begin{equation}
ds_{10}^2  ~=~   \Omega^2 \, ds_{1,4}^2 ~-~ ds_5^2\,, \label{N4MTmet}
\end{equation}
where $ds_{1,4}^2$ is given in (\ref{5dPoincare}) and the internal metric is given by\footnote{We have rescaled the $\sigma_j$ in \cite{Pilch:2000ue} by a factor of two and interchanged $\sigma_{1}\leftrightarrow \sigma_{3}$, so as to bring them into line with (\ref{sigdefn})}: 
\begin{equation}
ds_5^2  ~=~ 
{ L^2 \, (\cosh(2 \chi)\, X_1X_2)^{1/4}\over\rho^3} \left(
\frac{d \theta^2}{\cosh(2 \chi)}  ~+~  \frac{1}{4}\, \rho^6\cos^2\theta\,\Big({\sigma_3^2\over \cosh(2 \chi) \,X_2}
+{\sigma_1^2+\sigma_2^2\over X_1}\Big)~+~\sin^2\theta\,  {d\phi^2\over X_2} \right)
\end{equation}
The radius, $L$, of AdS$_5$ metric, (\ref{5dPoincare}), for vanishing scalars is related to gauge coupling constant, $L=\sqrt2/g$. The functions $X_{1,2}$ are defined by 
\begin{equation}
\begin{split}
X_1(r,\theta)&~\equiv~ \cos^2\theta + \rho(r)^6 \cosh(2\chi(r)) \, \sin^2\theta\,, \\
X_2(r,\theta)&~\equiv~ \cosh(2\chi(r))\cos^2\theta+ \rho(r)^6\sin^2\theta \,,
\end{split} 
\label{X1X2Omdefn}
\end{equation}
and the warp factor is
\begin{equation}\label{}
\Omega^2   ~\equiv~   {(\cosh(2 \chi)\,X_1X_2)^{1/4}\over \rho} \,.
\end{equation}

The metric $ds_5^2$ has an $\rm SU(2) \times U(1) \times U(1)$ isometry. The first $\rm U(1)$ is defined by $\phi$-translations and the second by rotating $\sigma_{1}$ and $\sigma_{2}$.  The symmetry action has co-dimension one on the compactification manifold, and this is manifest in the explicit dependence on the coordinate $\theta$.

%


Assuming a separable ansatz of the form:
\begin{equation}
\begin{split}
S(x,y) = S_{x}(x) + S_\theta (\theta) +S_{\varphi_{1}}(\varphi_{1}) +S_{\varphi_{2}}(\varphi_{2})+S_{\varphi_{3}}(\varphi_{3})+S_{\phi}(\phi)   \,,
\end{split}
\end{equation}
then the massless Hamilton Jacobi equation (\ref{MHJEApp}), reduces to the following system of differential equations.

Three trivial ODEs along the fiber:
\begin{align}
S'_{\varphi_{2}}(\varphi_{2})= c_{\varphi_{2}} \,, \qquad S'_{\varphi_{3}}(\varphi_{3})= c_{\varphi_{3}} \,, \qquad S'_{\phi}(\phi)= c_{\phi} 
\end{align}
corresponding to the three commuting Killing vectors of the isometries along the $\varphi_2$, $\varphi_3$ and $\phi$ coordinates of the metric (\ref{Neq1metricApp}).

Two non-trivial ODEs along the fiber:
\begin{align}
S_{\varphi_{1}}'(\varphi_{1})^{2} + \frac{c_{\varphi_{2}}^{2} - c_{\varphi_{3}}^{2} + 2 c_{\varphi_{2}}c_{\varphi_{3}} \cos \varphi_{1}}{\sin^{2}\varphi_{1}} -c_{\varphi_{1}}&= 0 \,, \\
S_{\theta}'(\theta)^{2} + \frac{c_{\phi}^{2}}{\sin^{2}\theta} + \frac{4 c_{\varphi_{1}}}{\cos^{2}\theta} - c_{\theta} &= 0  \,.
\end{align}
corresponding to (conformal) Killing tensors.

The remaining differential equation purely on the base:
\begin{equation}
 g^{\mu\nu} \frac{\partial S_{x}}{\partial x^{\mu}} \frac{\partial S_{x}}{\partial x^{\nu}}   =  \frac{1}{L^{2}\rho^{4}}\left[ 4 c_{\varphi_{1}} - 2 c_{\varphi_{3}}^{2} +c_{\phi}^{2}\rho^{12} + \left(c_{\theta} - c_{\phi}^{2} - 4c_{\varphi_{1}} \right)\rho^{6}\cosh 2\chi + 2 c_{\varphi_{3}}^{2}\cosh 4\chi \right] \,.
\end{equation}
which agrees with \eqref{N2HF}.   

\subsection{The $\cals N=1$ flow in Section~\ref{sss:CPWflow}}
\label{ss:CPWapp}

This example represents a flow of ABJM theory down to a non-trivial infra-red conformal fixed point \cite{Warner:1983vz,Nicolai:1985hs}.  In four-dimensional supergravity the flow \cite{Ahn:2000aq,Ahn:2000mf} is defined by two scalar fields, $\alpha$ and $\chi$. The four-dimensional metric is given by the usual Poincar\'e slicing;
\begin{equation}
ds_{1,3}^2 ~=~ dr^2 ~+~ e^{2 A(r)}\big(\, \eta_{i j} \, dx^i \, d x^j \big)  \,, \qquad i,j \eql 0\,,1\,,2  \,.
 \end{equation}

The M-theory uplift  of this flow is given in \cite{Corrado:2001nv}. The metric is given by\footnote{The coordinates employed here differ to those appearing in (4.23) of \cite{Corrado:2001nv} by $\phi\to -\psi$ and $\psi \to -(\phi +\psi)$.}:
\begin{equation} \label{metricNeq1621App}
ds_{11}^{2} = \frac{X^{2/3}}{\rho^{4/3}} \cosh^{4/3}\chi  \left( ds_{1,3}^{2} +ds_{7}^{2} \right)\,,
\end{equation}
with
\begin{equation}
\begin{aligned}
ds_{7}^{2} &= \frac{L^{2}\sech^{2}\chi}{\rho^{4}} \left\lbrace d\mu^{2} + \frac{\rho^{8}\cos^{2}\mu}{X}\,  \left[ ds_{\mathbb{CP}^{2}}^{2} + \frac{\rho^{8}}{ X}\sin^{2}\mu \, \left(  d\psi - \rho^{-8}d\phi - \frac{1}{2}\sin^{2}\theta \,  \sigma_{3}\right)^{2} \right.  \right.\\
& \qquad \qquad\qquad\qquad \qquad   \qquad \qquad\left. \left.  + \frac{\cosh^{2}\chi}{X}  \cos^{2}\mu  \left( d\psi +\tan^{2}\mu \, d\phi - \frac{1}{2}\sin^{2}\theta  \, \sigma_{3}\right)^{2} \right]  \right\rbrace
\end{aligned}\,,
\end{equation}
where the radii of the ``round'' $AdS_4$ and $S^7$ are $L/2$ and $L$ respectively. The functions, $\rho$ and $X$, are defined by: 
\begin{equation}
\rho \equiv e^{\alpha}\,, \qquad  X(r,\mu) ~\equiv~ \cos^2\mu+\rho(r)^8\sin^2\mu\,.
\end{equation}
The $\sigma_j$ are the standard $\text{SU}(2)$-invariant one-forms (\ref{sigdefn}) and  
\begin{align} \label{CP2metric}
ds_{\mathbb{CP}^{2}}^{2} = d\theta^{2} + \frac{1}{4}\sin^{2}\theta \, \left( \sigma_{1}^{2}+\sigma_{2}^{2}+\cos^{2}\theta \, \sigma_{3}^{2} \right)\,,
\end{align}
is the standard metric for $\mathbb{CP}^{2}$.

The metric, (\ref{metricNeq1621App}) has an $\text{SU}(3) \times \text{U}(1) \times \text{U}(1)$ isometry.  The $\text{SU}(3)$ action is transitive on the stretched $S^5$ defined by the Hopf fiber of $(d\phi - \coeff{1}{2}\sin^2\theta \, \sigma_3)$ over $\IC\IP^2$.  The first $\text{U}(1)$ is defined by $\psi$-translations and the second is a rotation between $\sigma_{1}$ and $\sigma_{2}$.  The symmetry action has co-dimension one on the compactification manifold, and this is manifest in the explicit dependence on the coordinate $\mu$.

Applying the discussion of Section \ref{ss:sepHJE}, one can now investigate the separability of the massless Hamilton-Jacobi equation. 

Assuming a separable ansatz of the form:
\begin{align}
S(x,y) = S_{x}(x)+S_{\mu}(\mu) + S_{\theta}(\theta) + S_{\varphi_{1}}(\varphi_{1}) + S_{\varphi_{2}}(\varphi_{2}) + S_{\varphi_{3}}(\varphi_{3}) + S_{\phi}(\phi)+S_{\psi}(\psi) \,,
\end{align}
then the massless Hamilton Jacobi equation (\ref{MHJEApp}), reduces to the following system of differential equations.

Four trivial ODEs along the fiber:
\begin{align}
S'_{\varphi_{2}}(\varphi_{2})= c_{\varphi_{2}}\,, \qquad S'_{\varphi_{3}}(\varphi_{3})= c_{\varphi_{3}}\,, \qquad
 S'_{\phi}(\phi)=c_{\phi}\,, \qquad S'_{\psi}(\psi)= c_{\psi}\,,   
\end{align}
corresponding to the four commuting Killing vectors of the isometries along the $\varphi_2$, $\varphi_3$, $\phi$ and $\psi$ coordinates of the metric (\ref{metricNeq1621App}).

Three non-trivial ODEs along the fiber:
\begin{align}
S'_{\varphi_{1}}(\varphi_{1})^{2} + \frac{c_{\varphi_{2}}^{2}+c_{\varphi_{3}}^{2} -2 c_{\varphi_{2}}c_{\varphi_{3}} \cos \varphi_{1}  }{\sin^{2} \varphi_{1}} - c_{\varphi_{1}} &= 0 \,,\\
S'_{\theta}(\theta)^{2} + \frac{4 c_{\varphi_{1}}}{\sin^{2}\theta} + \left(\frac{2c_{\varphi_{3}}+c_{\psi}}{\cos\theta} \right)^{2} - c_{\theta} &= 0  \,, \\
S'_{\mu}(\mu)^{2}  + \left( \frac{c_{\phi}}{\sin\mu} \right)^{2} + \frac{c_{\theta}}{\cos^{2} \mu}  - c_{\mu} &= 0 \,. 
\end{align}
corresponding to (conformal) Killing tensors.

The remaining differential equation purely on the base:
\begin{equation}
\begin{aligned}
g^{\mu\nu} \frac{\partial S_{x}}{\partial x^{\mu}} \frac{\partial S_{x}}{\partial x^{\nu}} &= \frac{1}{2L^{2}} \Big\lbrace  \left[c_{\theta}(\rho^{8}-1)-c_{\psi}^{2}\right]\rho^{-4} - \left[ c_{\mu}+c_{\phi}(2c_{\psi}+2c_{\phi} \rho^{8} - c_{\phi}) \right]\rho^{4}  \\
& \qquad \qquad \qquad +  \left[ \frac{c_{\psi}^{2}-c_{\theta}}{\rho^{4}} + \left( c_{\theta} - c_{\mu} +c_{\phi}(c_{\phi}+2c_{\psi}) \right)\rho^{4} \right]\cosh 2\chi  \Big\rbrace  \,,
\end{aligned}
\end{equation}
which agrees with \eqref{HFeffect}.

\subsection{The solution in Section~\ref{sss:Jansu3}}
\label{ss:SU3U1U1}

The full metric for this example is in \cite{Pilch:2015dwa} and may be written as: 
\begin{align}
ds_{11}^{2} = X^{1/3}\Sigma^{2/3} ds_{1,3}^{2} +ds_{\widetilde{S}^{7}}^{2}\,, \label{SU3U1U1met}
\end{align}
where $ds_{\widetilde{S}^{7}}^{2}$ is a deformed seven-sphere, parametrized by the coordinates, $(\chi,\theta,\varphi_{1},\varphi_{2},\varphi_{3},\phi)$, with metric:
\begin{equation}
\begin{aligned}
ds_{\widetilde{S}^{7}}^{2} &= m_{7}^{-2}\left( \frac{\Sigma}{X} \right)^{2/3} \left[ d\chi^{2} + \frac{X}{\Sigma}\cos^{2}\chi \left( ds_{\mathbb{CP}^{2}}^{2}+\frac{X}{\Sigma}\sin^{2}\chi \left(d\psi + \frac{1}{2}\sin^{2}\theta \, \sigma_{3} + \frac{\Xi}{X}d\phi \right)^{2} \right) \right. \\
& \qquad \qquad \qquad \qquad \qquad \qquad \qquad \qquad \qquad   + \left. \frac{1}{\Sigma^{2}}\left(d\phi+\cos^{2}\chi \left( d\psi + \frac{1}{2}\sin^{2}\theta \, \sigma_{3}\right) \right)^{2} \right] \,,
\end{aligned} \label{SU3U1U1Spheremet}
\end{equation}
\begin{align}
ds_{1,3}^{2} &= e^{2A(r)}ds_{AdS_{3}}^{2}+dr^{2}\,.,
\end{align}
and the $\mathbb{CP}^{2}$ metric is given by (\ref{CP2metric}). In these expressions $(X,\Xi,\Sigma)$ control the deformation of the seven-sphere through the four dimensional scalar fields, $\lambda(r)$ and $\zeta(r)$: 
\begin{equation}
\begin{aligned}
X(r) = \cosh 2\lambda & +\cos \zeta \sinh 2\lambda \,, \qquad   \Xi(r) = 2\cos \zeta \sinh 2\lambda \,, \\[6 pt]
\Sigma(r,\chi) &= \cosh 2\lambda - \cos \zeta \sinh 2\lambda \cos 2\chi \,.
\end{aligned}
\end{equation}

Assuming a separable ansatz of the form:
\begin{align}
S(x,y) = S_{x}(x) + S_{\chi}(\chi)+S_{\theta}(\theta)+S_{\varphi_{1}}(\varphi_{1}) + S_{\varphi_{2}}(\varphi_{2}) + S_{\varphi_{3}}(\varphi_{3}) +S_{\phi}(\phi) +S_{\psi}(\psi)  \,,
\end{align}
then the massless Hamilton Jacobi equation (\ref{MHJEApp}), reduces to the following system of differential equations.

Four trivial ODEs along the fiber:

\begin{align}
S'_{\varphi_{2}}(\varphi_{2})=c_{\varphi_{2}}\,, \qquad S'_{\varphi_{3}}(\varphi_{3})=c_{\varphi_{3}}\,, \qquad
 S'_{\phi}(\phi)=c_{\phi}\,, \qquad S'_{\psi}(\psi)=c_{\psi}\,,   
\end{align}
corresponding to the four commuting Killing vectors of the isometries along the $\varphi_2$, $\varphi_3$, $\phi$ and $\psi$ coordinates of the metric (\ref{SU3U1U1Spheremet}).

Three non-trivial ODEs along the fiber:
\begin{align}
  S'_{\varphi_{1}}(\varphi_{1})^{2} +\left(  \frac{ c_{\varphi_{2}}- c_{\varphi_{3}}\cos \varphi_{1} }{\sin \varphi_{1}} \right)^{2} - c_{\varphi_{1}} &= 0\,,\\
  S'_{\theta}(\theta)^{2} -c_{\psi}^{2} + 4 \left( \frac{c_{\varphi_{1}}+c_{\varphi_{3}}^{2}}{\sin^{2}\theta} \right) + \left( \frac{c_{\psi}-2c_{\varphi_{3}}}{\cos \theta} \right)^{2} -c_{\theta} &= 0 \,, \\
S'_{\chi}(\chi)^{2} -c_{\phi}^{2} + \left(\frac{c_{\phi}-c_{\psi}}{\sin \chi} \right)^{2}  + \left( \frac{c_{\theta}+c_{\psi}^{2}}{\cos^{2}\chi} \right) - c_{\chi} &= 0\,. 
\end{align}
corresponding to (conformal) Killing tensors.

The remaining differential equation purely on the base:
\begin{align}
 g^{\mu\nu} \frac{\partial S_{x}}{\partial x^{\mu}} \frac{\partial S_{x}}{\partial x^{\nu}} &= m_{7}^{2} \left\lbrace c_{\theta}\, \Xi(r) -X(r)\,\left[c_{\chi}+\left( c_{\phi}\, X(r)-c_{\psi}\, \Xi(r)\right)^{2} \right] \right\rbrace   \,.
\end{align}
which agrees with \eqref{HcFSU3U1U1}.

\subsection{The $(1,0,n)$ superstata solution in Section~\ref{SS:Superstrata10n}}
\label{ss:10nAppExample}

The full metric for the $(1,0,n)$ superstrata may be written as:
\begin{equation} \label{metric10nApp}
\begin{aligned} 
ds_{6}^{2} &= \frac{\Delta^{2}}{g_{0}^{2}\Omega^{2}} \left\lbrace -\frac{4\Sigma}{R_{y}^{2}}(du+dv)dv + \frac{2\Sigma}{a^{4}R_{y}^{4}}\left(R_{y}^{2}(a^{2}+r^{2})\Omega^{2} - \frac{2r^{2}}{g_{0}^{4}} \right)dv^{2} + \frac{g_{0}^{4}R_{y}^{2}\Omega^{4}}{2(a^{2}+r^{2})\Delta^{4}} dr^{2} \right. \\
& \left. \qquad- \frac{2\sqrt{2}a^{2}}{R_{y}}(du+dv)\left(\sin^{2}\theta \, d\phi_{1} - \cos^{2}\theta \, d\phi_{2}\right) - \frac{2\sqrt{2}}{a^{2}R_{y}}\left((a^{2}+r^{2})\Omega^{2} - \frac{2r^{2}}{g_{0}^{4}R_{y}^{2}} \right)\cos^{2}\theta \, dv \, d\phi_{2} \right. \\
& \qquad \qquad  \qquad\qquad\qquad\qquad\qquad \left. + \frac{g_{0}^{4}R_{y}^{2}\Omega^{4}}{2\Delta^{4}} d\theta^{2} + \frac{2}{g_{0}^{4}R_{y}^{2}}\sin^{2}\theta \, d\phi_{1}^{2} + \Omega^{2}\cos^{2}\theta  \, d\phi_{2}^{2}\right\rbrace \,,
\end{aligned}
\end{equation}
where 
\begin{equation}
\begin{aligned}
\Omega^{2} =\frac{1}{2g_{0}^{4}R_{y}^{2}}\left( 4-\chi_{1}^{2}-\chi_{2}^{2} \right) \,, \qquad 
 \Delta^{2} = \frac{4-\chi_{1}^{2}-\chi_{2}^{2} }{\sqrt{2}\sqrt{8-\chi_{1}^{2}-\chi_{2}^{2}+(\chi_{1}^{2}+\chi_{2}^{2})\cos 2\theta}}\,.
\end{aligned}
\end{equation}
and
\begin{equation}
\Sigma=r^{2}+a^{2}\cos^{2}\theta \,.
\end{equation}
In these expressions the details of the microstate geometry are specified by the scalars $(\chi_{1},\chi_{2})$, which always appear in the combination:
\begin{align}
\chi_{1}^{2}+\chi_{2}^{2} = 2g_{0}^{4}R_{y}^{2}\left(\frac{a^{2}}{a^{2}+r^{2}} \right) \abs{F_{0}}^{2}\,,
\end{align}
where $F_{0}$ is a tunable holomorphic function, depending only on the base mixing the $(u,v)$ coordinates (\ref{F0F1def}). 

Assuming a separable ansatz of the form:
\begin{align}
S(x,y) = S_{u}(u)+S_{vr}(v,r) +S_{\theta}(\theta)+S_{\phi_{1}}(\phi_{1}) + S_{\phi_{2}}(\phi_{2})   \,,
\end{align}
then the massless Hamilton Jacobi equation (\ref{MHJEApp}), reduces to the following system of differential equations.

One trivial ODE along the base and two along the fiber:
\begin{align}
S'_{u}(u)=c_{u}\,, \qquad S'_{\phi_{1}}(\phi_{1})=c_{\phi_{1}}\,, \qquad S'_{\phi_{2}}(\phi_{2})=c_{\phi_{2}}\,,   
\end{align}
corresponding to the three commuting Killing vectors of the isometries along the $u$, $\phi_{1}$ and $\phi_{2}$ coordinates of the metric (\ref{metric10nApp}).

A non-trivial ODE along the fiber:
\begin{align}
 S'_{\theta}(\theta)^{2} + \frac{c_{\phi_{1}}^{2}}{\sin^{2}\theta}+\frac{c_{\phi_{2}}^{2}}{\cos^{2}\theta} -c_{\theta} &= 0 \,, 
\end{align}
corresponding to a conformal Killing tensor.

The remaining differential equation purely on the base:
\begin{align}
g^{\mu\nu} \frac{\partial S}{\partial x^{\mu}} \frac{\partial S}{\partial x^{\nu}} -2 A^{\cI \mu} K^{\cI m} \frac{\partial S}{\partial x^{\mu}} \frac{\partial S}{\partial y^{m}} = - \frac{2}{R_{y}^{2}g_{0}^{2}\Omega^{2}}\left(c_{\theta} - \frac{a^{2}}{a^{2}+r^{2}}c_{\phi_{1}}^{2} + \frac{a^{2}}{r^{2}}c_{\phi_{2}}^{2}  \right)   \,,
\end{align}
where 
\begin{equation}
\begin{aligned}
-2 A^{\cI \mu} K^{\cI m}& = \frac{2\sqrt{2}}{R_{y}g_{0}^{2}\Omega^{2}} \frac{a^{2}}{r^{2}} \begin{pmatrix}
0 & \frac{r^{2}}{a^{2}+r^{2}} & -1 \\
0 & - \frac{r^{2}}{a^{2}+r^{2}} & 1 \\
0 & 0 & 0
\end{pmatrix} \\
& \qquad \qquad  + \frac{2\sqrt{2}}{a^{2}g_{0}^{2}R_{y}}\left( \frac{2}{g_{0}^{4}R_{y}^{2}\Omega^{2}} \left(\frac{r^{2}}{a^{2}+r^{2}}\right) -1 \right) \begin{pmatrix}
0 & 1 & 0 \\ 0 & 0 & 0\\ 0 & 0 & 0
\end{pmatrix}
\end{aligned}
\end{equation}
and $g^{\mu\nu}$ is the inverse of the reduced metric on the base given by:
\begin{align}
ds_{3}^{2} &= - \frac{a^{4}R_{y}^{2}g_{0}^{6}}{2} \left(  du +dv + \frac{2r^{2}}{a^{4}R_{y}^{2}g_{0}^{4}}\, dv\right)^{2} + \frac{g_{0}^{2}}{2} \Omega^{2} \left[ \frac{R_{y}^{2}}{a^{2}+r^{2}}\, dr^{2} + \frac{2r^{2}}{a^{2}}\left(1+\frac{r^{2}}{a^{2}} \right)\, dv^{2} \right]\,,
\end{align}
note that this metric is in the form of a time-like K\"{a}hler fibration, see \cite{Mayerson:2020tcl,Houppe:2020oqp}.

\subsection{The $(1,1,n)$ superstata solution in Section~\ref{SS:Superstrata11n}}
\label{ss:11nAppExample}

The full metric for the $(1,0,n)$ superstrata may be written as:
\begin{equation} \label{metric10nApp}
\begin{aligned} 
ds_{6}^{2} &= \frac{\Delta^{2}}{g_{0}^{2}\Omega^{2}} \left\lbrace -\frac{4\Sigma}{R_{y}^{2}}(du+dv)dv + \frac{2\Sigma}{a^{4}R_{y}^{4}}\left(R_{y}^{2}r^{2}\Omega^{2} + \frac{2}{g_{0}^{4}}(a^{2}-r^{2}) \right)dv^{2} + \frac{g_{0}^{4}R_{y}^{2}\Omega^{4}}{2(a^{2}+r^{2})\Delta^{4}} dr^{2} \right. \\
& \left. \qquad \qquad \qquad - \frac{2\sqrt{2}a^{2}}{R_{y}}(du+dv)\left(\sin^{2}\theta \, d\phi_{1} - \cos^{2}\theta \, d\phi_{2}\right) \right. \\
& \qquad \qquad  \qquad  \qquad   \left.- \frac{2\sqrt{2}}{a^{2}R_{y}^{3}} \left[ \frac{2}{g_{0}^{4}} \left(r^{2}\sin^{2}\theta \, d\phi_{1} +a^{2}\cos^{2}\theta \, d\phi_{2} \right) - R_{y}^{2}r^{2}\Omega^{2}\sin^{2}\theta  \, d\phi_{1}\right]\, dv \right. \\
& \qquad \qquad  \qquad\qquad\qquad\qquad\qquad\qquad \left. + \frac{g_{0}^{4}R_{y}^{2}\Omega^{4}}{2\Delta^{4}} d\theta^{2} + \Omega^{2}\sin^{2}\theta \, d\phi_{1}^{2} + \frac{2}{g_{0}^{4}R_{y}^{2}} \cos^{2}\theta  \, d\phi_{2}^{2}\right\rbrace
\end{aligned}
\end{equation}
where 
\begin{equation}
\begin{aligned}
\Omega^{2} =\frac{1}{2g_{0}^{4}R_{y}^{2}}\left( 4-\chi_{3}^{2}-\chi_{3}^{2} \right) \,, \qquad 
 \Delta^{2} = \frac{4-\chi_{3}^{2}-\chi_{4}^{2} }{\sqrt{2}\sqrt{8-\chi_{3}^{2}-\chi_{4}^{2}+(\chi_{3}^{2}+\chi_{4}^{2})\cos 2\theta}}\,.
\end{aligned}
\end{equation}
and
\begin{equation}
\Sigma=r^{2}+a^{2}\cos^{2}\theta \,.
\end{equation}
In these expressions the details of the microstate geometry are specified by the scalars $(\chi_{3},\chi_{4})$, which always appear in the combination:
\begin{align}
\chi_{3}^{2}+\chi_{4}^{2} = 2g_{0}^{4}R_{y}^{2}\left(\frac{a^{2}}{a^{2}+r^{2}} \right) \abs{F_{1}}^{2}\,,
\end{align}
where $F_{1}$ is a tunable holomorphic function, depending only on the base mixing the $(u,v)$ coordinates (\ref{F0F1def}). 

Assuming a separable ansatz of the form:
\begin{align}
S(x,y) = S_{u}(u)+S_{vr}(v,r) +S_{\theta}(\theta)+S_{\phi_{1}}(\phi_{1}) + S_{\phi_{2}}(\phi_{2})   \,,
\end{align}
then the massless Hamilton Jacobi equation (\ref{MHJEApp}), reduces to the following system of differential equations.

One trivial ODE along the base and two along the fiber:
\begin{align}
S'_{u}(u)=c_{u}\,, \qquad S'_{\phi_{1}}(\phi_{1})=c_{\phi_{1}}\,, \qquad S'_{\phi_{2}}(\phi_{2})=c_{\phi_{2}}\,,   
\end{align}
corresponding to the three commuting Killing vectors of the isometries along the $u$, $\phi_{1}$ and $\phi_{2}$ coordinates of the metric (\ref{metric10nApp}).

A non-trivial ODE along the fiber:
\begin{align}
 S'_{\theta}(\theta)^{2} + \frac{c_{\phi_{1}}^{2}}{\sin^{2}\theta}+\frac{c_{\phi_{2}}^{2}}{\cos^{2}\theta} -c_{\theta} &= 0 \,, 
\end{align}
corresponding to a conformal Killing tensor.

The remaining differential equation purely on the base:
\begin{align}
g^{\mu\nu} \frac{\partial S}{\partial x^{\mu}} \frac{\partial S}{\partial x^{\nu}} -2 A^{\cI \mu} K^{\cI m} \frac{\partial S}{\partial x^{\mu}} \frac{\partial S}{\partial y^{m}} = - \frac{2}{R_{y}^{2}g_{0}^{2}\Omega^{2}}\left(c_{\theta} - \frac{a^{2}}{a^{2}+r^{2}}c_{\phi_{1}}^{2} + \frac{a^{2}}{r^{2}}c_{\phi_{2}}^{2}  \right)   \,,
\end{align}
where 
\begin{equation}
\begin{aligned}
-2 A^{\cI \mu} K^{\cI m} &= \frac{2\sqrt{2}}{R_{y}g_{0}^{2}\Omega^{2}} \frac{a^{2}}{r^{2}} \begin{pmatrix}
0 & \frac{r^{2}}{a^{2}+r^{2}} & -1 \\
0 & - \frac{r^{2}}{a^{2}+r^{2}} & 1 \\
0 & 0 & 0
\end{pmatrix}\\
& \qquad + \frac{4\sqrt{2}}{a^{2}g_{0}^{6}R_{y}^{3}\Omega^{2}} \begin{pmatrix}
0 & \frac{a^{2}}{a^{2}+r^{2}} & 1 \\
0 & 0 & 0 \\
0 & 0 & 0
\end{pmatrix}  + \frac{2\sqrt{2}}{a^{2}g_{0}^{2}R_{y}} \begin{pmatrix}
0 & 0 & 1 \\
0 & 0 & 0 \\
0 & 0 & 0
\end{pmatrix}\,,
\end{aligned}
\end{equation}
and $g^{\mu\nu}$ is the inverse of the reduced metric on the base given by:
\begin{align}
ds_{3}^{2} &= - \frac{a^{4}R_{y}^{2}g_{0}^{6}}{2} \left(  du +dv + \frac{2r^{2}}{a^{4}R_{y}^{2}g_{0}^{4}}\, dv\right)^{2} + \frac{g_{0}^{2}}{2} \Omega^{2} \left[ \frac{R_{y}^{2}}{a^{2}+r^{2}}\, dr^{2} + \frac{2r^{2}}{a^{2}}\left(1+\frac{r^{2}}{a^{2}} \right)\, dv^{2} \right]\,,
\end{align}
note that this metric is, just like that for the $(1,0,n)$ geometry, is in the form of a time-like K\"{a}hler fibration, see \cite{Mayerson:2020tcl,Houppe:2020oqp}.

\begin{adjustwidth}{-1mm}{-1mm} 
\bibliographystyle{JHEP}      
\bibliography{Separability}       

\providecommand{\href}[2]{#2}\begingroup\raggedright\begin{thebibliography}{10}

\bibitem{Carter:1968rr}
B.~Carter, \emph{{Global structure of the Kerr family of gravitational
  fields}}, \href{https://doi.org/10.1103/PhysRev.174.1559}{\emph{Phys. Rev.}
  {\bfseries 174} (1968) 1559}.

\bibitem{Carter:1968ks}
B.~Carter, \emph{{Hamilton-Jacobi and Schrodinger separable solutions of
  Einstein's equations}}, {\emph{Commun. Math. Phys.} {\bfseries 10} (1968)
  280}.

\bibitem{Chervonyi:2015ima}
Y.~Chervonyi and O.~Lunin, \emph{{Killing(-Yano) Tensors in String Theory}},
  \href{https://doi.org/10.1007/JHEP09(2015)182}{\emph{JHEP} {\bfseries 09}
  (2015) 182} [\href{https://arxiv.org/abs/1505.06154}{{\ttfamily
  1505.06154}}].

\bibitem{deWit:1984nz}
B.~de~Wit, H.~Nicolai and N.~P. Warner, \emph{{The Embedding of Gauged $N=8$
  Supergravity Into $d=11$ Supergravity}},
  \href{https://doi.org/10.1016/0550-3213(85)90128-2}{\emph{Nucl. Phys.}
  {\bfseries B255} (1985) 29}.

\bibitem{deWit:1986iy}
B.~de~Wit and H.~Nicolai, \emph{The consistency of the s**7 truncation in d=11
  supergravity},
  \href{https://doi.org/10.1016/0550-3213(87)90253-7}{\emph{Nucl.Phys.}
  {\bfseries B281} (1987) 211}.

\bibitem{Nastase:1999cb}
H.~Nastase, D.~Vaman and P.~van Nieuwenhuizen, \emph{{Consistent nonlinear K K
  reduction of 11-d supergravity on AdS(7) x S(4) and selfduality in odd
  dimensions}},
  \href{https://doi.org/10.1016/S0370-2693(99)01266-6}{\emph{Phys. Lett. B}
  {\bfseries 469} (1999) 96}
  [\href{https://arxiv.org/abs/hep-th/9905075}{{\ttfamily hep-th/9905075}}].

\bibitem{Nastase:1999kf}
H.~Nastase, D.~Vaman and P.~van Nieuwenhuizen, \emph{{Consistency of the AdS(7)
  x S(4) reduction and the origin of selfduality in odd dimensions}},
  \href{https://doi.org/10.1016/S0550-3213(00)00193-0}{\emph{Nucl. Phys. B}
  {\bfseries 581} (2000) 179}
  [\href{https://arxiv.org/abs/hep-th/9911238}{{\ttfamily hep-th/9911238}}].

\bibitem{Khavaev:1998fb}
A.~Khavaev, K.~Pilch and N.~P. Warner, \emph{{New vacua of gauged N=8
  supergravity in five-dimensions}},
  \href{https://doi.org/10.1016/S0370-2693(00)00795-4}{\emph{Phys. Lett. B}
  {\bfseries 487} (2000) 14}
  [\href{https://arxiv.org/abs/hep-th/9812035}{{\ttfamily hep-th/9812035}}].

\bibitem{Pilch:2000ue}
K.~Pilch and N.~P. Warner, \emph{{N=2 supersymmetric RG flows and the IIB
  dilaton}}, \href{https://doi.org/10.1016/S0550-3213(00)00656-8}{\emph{Nucl.
  Phys.} {\bfseries B594} (2001) 209}
  [\href{https://arxiv.org/abs/hep-th/0004063}{{\ttfamily hep-th/0004063}}].

\bibitem{Cvetic:2000nc}
M.~Cvetic, H.~Lu, C.~Pope, A.~Sadrzadeh and T.~A. Tran, \emph{{Consistent SO(6)
  reduction of type IIB supergravity on S**5}},
  \href{https://doi.org/10.1016/S0550-3213(00)00372-2}{\emph{Nucl. Phys. B}
  {\bfseries 586} (2000) 275}
  [\href{https://arxiv.org/abs/hep-th/0003103}{{\ttfamily hep-th/0003103}}].

\bibitem{Lee:2014mla}
K.~Lee, C.~Strickland-Constable and D.~Waldram, \emph{{Spheres, generalised
  parallelisability and consistent truncations}},
  \href{https://doi.org/10.1002/prop.201700048}{\emph{Fortsch. Phys.}
  {\bfseries 65} (2017) 1700048}
  [\href{https://arxiv.org/abs/1401.3360}{{\ttfamily 1401.3360}}].

\bibitem{Baguet:2015sma}
A.~Baguet, O.~Hohm and H.~Samtleben, \emph{Consistent type iib reductions to
  maximal 5d supergravity},  \href{https://arxiv.org/abs/1506.01385}{{\ttfamily
  1506.01385}}.

\bibitem{Guarino:2015jca}
A.~Guarino, D.~L. Jafferis and O.~Varela, \emph{{String Theory Origin of Dyonic
  N=8 Supergravity and Its Chern-Simons Duals}},
  \href{https://doi.org/10.1103/PhysRevLett.115.091601}{\emph{Phys. Rev. Lett.}
  {\bfseries 115} (2015) 091601}
  [\href{https://arxiv.org/abs/1504.08009}{{\ttfamily 1504.08009}}].

\bibitem{Guarino:2015vca}
A.~Guarino and O.~Varela, \emph{{Consistent $ \mathcal{N}=8 $ truncation of
  massive IIA on S$^{6}$}},
  \href{https://doi.org/10.1007/JHEP12(2015)020}{\emph{JHEP} {\bfseries 12}
  (2015) 020} [\href{https://arxiv.org/abs/1509.02526}{{\ttfamily
  1509.02526}}].

\bibitem{Cvetic:1999xp}
M.~Cvetic, M.~Duff, P.~Hoxha, J.~T. Liu, H.~Lu, J.~Lu et~al., \emph{{Embedding
  AdS black holes in ten-dimensions and eleven-dimensions}},
  \href{https://doi.org/10.1016/S0550-3213(99)00419-8}{\emph{Nucl. Phys. B}
  {\bfseries 558} (1999) 96}
  [\href{https://arxiv.org/abs/hep-th/9903214}{{\ttfamily hep-th/9903214}}].

\bibitem{Cvetic:2000dm}
M.~Cvetic, H.~Lu and C.~Pope, \emph{{Consistent Kaluza-Klein sphere
  reductions}}, \href{https://doi.org/10.1103/PhysRevD.62.064028}{\emph{Phys.
  Rev. D} {\bfseries 62} (2000) 064028}
  [\href{https://arxiv.org/abs/hep-th/0003286}{{\ttfamily hep-th/0003286}}].

\bibitem{Nastase:2000tu}
H.~Nastase and D.~Vaman, \emph{{On the nonlinear KK reductions on spheres of
  supergravity theories}},
  \href{https://doi.org/10.1016/S0550-3213(00)00214-5}{\emph{Nucl. Phys. B}
  {\bfseries 583} (2000) 211}
  [\href{https://arxiv.org/abs/hep-th/0002028}{{\ttfamily hep-th/0002028}}].

\bibitem{Samtleben:2019zrh}
H.~Samtleben and O.~Sarıoglu, \emph{{Consistent $S^3$ reductions of
  six-dimensional supergravity}},
  \href{https://doi.org/10.1103/PhysRevD.100.086002}{\emph{Phys. Rev.}
  {\bfseries D100} (2019) 086002}
  [\href{https://arxiv.org/abs/1907.08413}{{\ttfamily 1907.08413}}].

\bibitem{Bena:2017upb}
I.~Bena, D.~Turton, R.~Walker and N.~P. Warner, \emph{{Integrability and
  Black-Hole Microstate Geometries}},
  \href{https://doi.org/10.1007/JHEP11(2017)021}{\emph{JHEP} {\bfseries 11}
  (2017) 021} [\href{https://arxiv.org/abs/1709.01107}{{\ttfamily
  1709.01107}}].

\bibitem{Walker:2019ntz}
R.~Walker, \emph{{D1-D5-P superstrata in 5 and 6 dimensions: separable wave
  equations and prepotentials}},
  \href{https://doi.org/10.1007/JHEP09(2019)117}{\emph{JHEP} {\bfseries 09}
  (2019) 117} [\href{https://arxiv.org/abs/1906.04200}{{\ttfamily
  1906.04200}}].

\bibitem{Mayerson:2020tcl}
D.~R. Mayerson, R.~A. Walker and N.~P. Warner, \emph{{Microstate Geometries
  from Gauged Supergravity in Three Dimensions}},
  \href{https://doi.org/10.1007/JHEP10(2020)030}{\emph{JHEP} {\bfseries 10}
  (2020) 030} [\href{https://arxiv.org/abs/2004.13031}{{\ttfamily
  2004.13031}}].

\bibitem{Kalnins1980a}
E.~G. Kalnins and J.~Willard~Miller, \emph{Killing tensors and variable
  separation for hamilton-jacobi and helmholtz equations},
  \href{https://doi.org/10.1137/0511089}{\emph{{SIAM} Journal on Mathematical
  Analysis} {\bfseries 11} (1980) 1011}.

\bibitem{sumitomo1981}
T.~Sumitomo and K.~Tandai, \emph{Killing tensor fields on the standard sphere
  and spectra of so(n+1)/(so(n-1) x so(2)) and o(n+1)/(o(n-1) x o(2))},
  {\emph{Osaka Journal of Mathematics} {\bfseries 20} (1983) 51}.

\bibitem{Takeuchi1983}
M.~Takeuchi, \emph{Killing tensor fields on spaces of constant curvature},
  {\emph{Tsukuba Journal of Mathematics} {\bfseries 7} (1983) 233}.

\bibitem{Thompson1986}
G.~Thompson, \emph{Killing tensors in spaces of constant curvature},
  \href{https://doi.org/10.1063/1.527288}{\emph{Journal of Mathematical
  Physics} {\bfseries 27} (1986) 2693}.

\bibitem{Stackel1893}
P.~St{\"a}ckel, \emph{Ueber die bewegung eines punktes in einern-fachen
  mannigfaltigkeit}, {\emph{Mathematische Annalen} {\bfseries 42} (1893) 537}.

\bibitem{LeviCivita1904a}
T.~Levi-Civita, \emph{Sulla integrazione della equazione di hamilton-jacobi per
  separazione di variabili},
  \href{https://doi.org/10.1007/bf01445149}{\emph{Mathematische Annalen}
  {\bfseries 59} (1904) 383}.

\bibitem{10.2307/1968433}
L.~P. Eisenhart, \emph{Separable systems of {St\"ackel}}, {\emph{Annals of
  Mathematics} {\bfseries 35} (1934) 284}.

\bibitem{eisenhart1997riemannian}
L.~Eisenhart, \emph{Riemannian Geometry}, Princeton Landmarks in Mathematics
  and Physics. Princeton University Press, 1997.

\bibitem{Benenti2002}
S.~Benenti, C.~Chanu and G.~Rastelli, \emph{Remarks on the connection between
  the additive separation of the hamilton{\textendash}jacobi equation and the
  multiplicative separation of the schroedinger equation. i. the completeness
  and robertson conditions},
  \href{https://doi.org/10.1063/1.1506180}{\emph{Journal of Mathematical
  Physics} {\bfseries 43} (2002) 5183}.

\bibitem{KMWBook}
E.~G. Kalnins, J.~M. Kress and W.~Miller, \emph{Separation of Variables and
  Superintegrability}, 2053-2563. IOP Publishing, 2018,
  \href{https://doi.org/10.1088/978-0-7503-1314-8}{10.1088/978-0-7503-1314-8}.

\bibitem{Benenti2016}
S.~Benenti, \emph{Separability in riemannian manifolds},
  \href{https://doi.org/10.3842/sigma.2016.013}{\emph{Symmetry, Integrability
  and Geometry: Methods and Applications} (2016) }
  [\href{https://arxiv.org/abs/1512.07833}{{\ttfamily 1512.07833}}].

\bibitem{Kalnins1986}
E.~G. Kalnins and W.~Miller, \emph{Separation of variables on n-dimensional
  riemannian manifolds. i. the n-sphere sn and euclidean n-space rn},
  \href{https://doi.org/10.1063/1.527088}{\emph{Journal of Mathematical
  Physics} {\bfseries 27} (1986) 1721}.

\bibitem{Kalnins1981}
E.~G. Kalnins and J.~Willard~Miller, \emph{Killing tensors and nonorthogonal
  variable separation for hamilton{\textendash}jacobi equations},
  \href{https://doi.org/10.1137/0512054}{\emph{{SIAM} Journal on Mathematical
  Analysis} {\bfseries 12} (1981) 617}.

\bibitem{deWit:2013ija}
B.~de~Wit and H.~Nicolai, \emph{{Deformations of gauged SO(8) supergravity and
  supergravity in eleven dimensions}},
  \href{https://doi.org/10.1007/JHEP05(2013)077}{\emph{JHEP} {\bfseries 05}
  (2013) 077} [\href{https://arxiv.org/abs/1302.6219}{{\ttfamily 1302.6219}}].

\bibitem{Hohm:2014qga}
O.~Hohm and H.~Samtleben, \emph{{Consistent Kaluza-Klein Truncations via
  Exceptional Field Theory}},
  \href{https://doi.org/10.1007/JHEP01(2015)131}{\emph{JHEP} {\bfseries 01}
  (2015) 131} [\href{https://arxiv.org/abs/1410.8145}{{\ttfamily 1410.8145}}].

\bibitem{Godazgar:2015qia}
H.~Godazgar, M.~Godazgar, O.~Kr\"uger and H.~Nicolai, \emph{{Consistent 4-form
  fluxes for maximal supergravity}},
  \href{https://doi.org/10.1007/JHEP10(2015)169}{\emph{JHEP} {\bfseries 10}
  (2015) 169} [\href{https://arxiv.org/abs/1507.07684}{{\ttfamily
  1507.07684}}].

\bibitem{Schoebel2015a}
K.~Sch\"obel, \emph{An Algebraic Geometric Approach to Separation of
  Variables}. Springer Fachmedien Wiesbaden, 2015,
  \href{https://doi.org/10.1007/978-3-658-11408-4}{10.1007/978-3-658-11408-4}.

\bibitem{McLenaghan2004}
R.~G. McLenaghan, R.~Milson and R.~G. Smirnov, \emph{Killing tensors as
  irreducible representations of the general linear group},
  \href{https://doi.org/https://doi.org/10.1016/j.crma.2004.07.017}{\emph{Comptes
  Rendus Mathematique} {\bfseries 339} (2004) 621 }.

\bibitem{delong1982killing}
R.~Delong, \emph{Killing Tensors and the Hamilton-Jacobi Equation}. University
  of Minnesota, 1982.

\bibitem{Bobev:2020fon}
N.~Bobev, F.~F. Gautason, K.~Pilch, M.~Suh and J.~van Muiden,
  \emph{{Holographic interfaces in $ \mathcal{N} $ = 4 SYM: Janus and
  J-folds}}, \href{https://doi.org/10.1007/JHEP05(2020)134}{\emph{JHEP}
  {\bfseries 05} (2020) 134}
  [\href{https://arxiv.org/abs/2003.09154}{{\ttfamily 2003.09154}}].

\bibitem{DHoker:2007zhm}
E.~D'Hoker, J.~Estes and M.~Gutperle, \emph{{Exact half-BPS Type IIB interface
  solutions. I. Local solution and supersymmetric Janus}},
  \href{https://doi.org/10.1088/1126-6708/2007/06/021}{\emph{JHEP} {\bfseries
  06} (2007) 021} [\href{https://arxiv.org/abs/0705.0022}{{\ttfamily
  0705.0022}}].

\bibitem{Pope:2003jp}
C.~N. Pope and N.~P. Warner, \emph{{A Dielectric flow solution with maximal
  supersymmetry}},
  \href{https://doi.org/10.1088/1126-6708/2004/04/011}{\emph{JHEP} {\bfseries
  04} (2004) 011} [\href{https://arxiv.org/abs/hep-th/0304132}{{\ttfamily
  hep-th/0304132}}].

\bibitem{DHoker:2009lky}
E.~D'Hoker, J.~Estes, M.~Gutperle and D.~Krym, \emph{{Janus solutions in
  M-theory}}, \href{https://doi.org/10.1088/1126-6708/2009/06/018}{\emph{JHEP}
  {\bfseries 06} (2009) 018} [\href{https://arxiv.org/abs/0904.3313}{{\ttfamily
  0904.3313}}].

\bibitem{Bobev:2013yra}
N.~Bobev, K.~Pilch and N.~P. Warner, \emph{{Supersymmetric Janus Solutions in
  Four Dimensions}}, \href{https://doi.org/10.1007/JHEP06(2014)058}{\emph{JHEP}
  {\bfseries 06} (2014) 058} [\href{https://arxiv.org/abs/1311.4883}{{\ttfamily
  1311.4883}}].

\bibitem{Slansky:1981yr}
R.~Slansky, \emph{Group theory for unified model building},
  \href{https://doi.org/10.1016/0370-1573(81)90092-2}{\emph{Phys.Rept.}
  {\bfseries 79} (1981) 1}.

\bibitem{Feger2020a}
R.~Feger, T.~W. Kephart and R.~J. Saskowski, \emph{{LieART} 2.0 {\textendash} a
  mathematica application for lie algebras and representation theory},
  \href{https://doi.org/10.1016/j.cpc.2020.107490}{\emph{Computer Physics
  Communications} {\bfseries 257} (2020) 107490}.

\bibitem{vilenkin1978special}
N.~Vilenkin, \emph{Special Functions and the Theory of Group Representations},
  Translations of mathematical monographs. American Mathematical Soc., 1978.

\bibitem{Freedman:1999gp}
D.~Freedman, S.~Gubser, K.~Pilch and N.~Warner, \emph{{Renormalization group
  flows from holography supersymmetry and a c theorem}},
  \href{https://doi.org/10.4310/ATMP.1999.v3.n2.a7}{\emph{Adv. Theor. Math.
  Phys.} {\bfseries 3} (1999) 363}
  [\href{https://arxiv.org/abs/hep-th/9904017}{{\ttfamily hep-th/9904017}}].

\bibitem{Pilch:2000fu}
K.~Pilch and N.~P. Warner, \emph{{N=1 supersymmetric renormalization group
  flows from IIB supergravity}}, {\emph{Adv.Theor.Math.Phys.} {\bfseries 4}
  (2002) 627} [\href{https://arxiv.org/abs/hep-th/0006066}{{\ttfamily
  hep-th/0006066}}].

\bibitem{Brandhuber:2000ct}
A.~Brandhuber and K.~Sfetsos, \emph{{An N=2 gauge theory and its supergravity
  dual}}, \href{https://doi.org/10.1016/S0370-2693(00)00896-0}{\emph{Phys.
  Lett. B} {\bfseries 488} (2000) 373}
  [\href{https://arxiv.org/abs/hep-th/0004148}{{\ttfamily hep-th/0004148}}].

\bibitem{Wit1982}
B.~de~Wit and H.~Nicolai, \emph{N=8 supergravity},
  \href{https://doi.org/10.1016/0550-3213(82)90120-1}{\emph{Nucl. Phys.}
  {\bfseries B208} (1982) 323}.

\bibitem{Cremmer:1979up}
E.~Cremmer and B.~Julia, \emph{{The SO(8) Supergravity}},
  \href{https://doi.org/10.1016/0550-3213(79)90331-6}{\emph{Nucl. Phys. B}
  {\bfseries 159} (1979) 141}.

\bibitem{Varela:2015ywx}
O.~Varela, \emph{{Complete $D=11$ embedding of SO(8) supergravity}},
  \href{https://doi.org/10.1103/PhysRevD.97.045010}{\emph{Phys. Rev. D}
  {\bfseries 97} (2018) 045010}
  [\href{https://arxiv.org/abs/1512.04943}{{\ttfamily 1512.04943}}].

\bibitem{Kruger:2016agp}
O.~Kruger, \emph{{Non-linear uplift Ansatze for the internal metric and the
  four-form field-strength of maximal supergravity}},
  \href{https://doi.org/10.1007/JHEP05(2016)145}{\emph{JHEP} {\bfseries 05}
  (2016) 145} [\href{https://arxiv.org/abs/1602.03327}{{\ttfamily
  1602.03327}}].

\bibitem{Aharony:2008ug}
O.~Aharony, O.~Bergman, D.~L. Jafferis and J.~Maldacena, \emph{{N=6
  superconformal Chern-Simons-matter theories, M2-branes and their gravity
  duals}}, \href{https://doi.org/10.1088/1126-6708/2008/10/091}{\emph{JHEP}
  {\bfseries 10} (2008) 091} [\href{https://arxiv.org/abs/0806.1218}{{\ttfamily
  0806.1218}}].

\bibitem{Warner:1983vz}
N.~Warner, \emph{{Some New Extrema of the Scalar Potential of Gauged $N=8$
  Supergravity}},
  \href{https://doi.org/10.1016/0370-2693(83)90383-0}{\emph{Phys. Lett. B}
  {\bfseries 128} (1983) 169}.

\bibitem{Nicolai:1985hs}
H.~Nicolai and N.~Warner, \emph{{The SU(3) X U(1) Invariant Breaking of Gauged
  $N=8$ Supergravity}},
  \href{https://doi.org/10.1016/0550-3213(85)90643-1}{\emph{Nucl. Phys. B}
  {\bfseries 259} (1985) 412}.

\bibitem{Benna:2008zy}
M.~Benna, I.~Klebanov, T.~Klose and M.~Smedback, \emph{{Superconformal
  Chern-Simons Theories and AdS(4)/CFT(3) Correspondence}},
  \href{https://doi.org/10.1088/1126-6708/2008/09/072}{\emph{JHEP} {\bfseries
  09} (2008) 072} [\href{https://arxiv.org/abs/0806.1519}{{\ttfamily
  0806.1519}}].

\bibitem{Klebanov:2008vq}
I.~Klebanov, T.~Klose and A.~Murugan, \emph{{AdS(4)/CFT(3) Squashed, Stretched
  and Warped}},
  \href{https://doi.org/10.1088/1126-6708/2009/03/140}{\emph{JHEP} {\bfseries
  03} (2009) 140} [\href{https://arxiv.org/abs/0809.3773}{{\ttfamily
  0809.3773}}].

\bibitem{Ahn:2000aq}
C.-h. Ahn and J.~Paeng, \emph{{Three-dimensional SCFTs, supersymmetric domain
  wall and renormalization group flow}},
  \href{https://doi.org/10.1016/S0550-3213(00)00687-8}{\emph{Nucl. Phys. B}
  {\bfseries 595} (2001) 119}
  [\href{https://arxiv.org/abs/hep-th/0008065}{{\ttfamily hep-th/0008065}}].

\bibitem{Ahn:2000mf}
C.-h. Ahn and K.~Woo, \emph{{Supersymmetric domain wall and RG flow from
  4-dimensional gauged N=8 supergravity}},
  \href{https://doi.org/10.1016/S0550-3213(01)00008-6}{\emph{Nucl. Phys. B}
  {\bfseries 599} (2001) 83}
  [\href{https://arxiv.org/abs/hep-th/0011121}{{\ttfamily hep-th/0011121}}].

\bibitem{Corrado:2001nv}
R.~Corrado, K.~Pilch and N.~P. Warner, \emph{{An N=2 supersymmetric membrane
  flow}}, \href{https://doi.org/10.1016/S0550-3213(02)00134-7}{\emph{Nucl.
  Phys. B} {\bfseries 629} (2002) 74}
  [\href{https://arxiv.org/abs/hep-th/0107220}{{\ttfamily hep-th/0107220}}].

\bibitem{Pilch:2015dwa}
K.~Pilch, A.~Tyukov and N.~P. Warner, \emph{{$\mathcal{N}=2$ Supersymmetric
  Janus Solutions and Flows: From Gauged Supergravity to M Theory}},
  \href{https://doi.org/10.1007/JHEP05(2016)005}{\emph{JHEP} {\bfseries 05}
  (2016) 005} [\href{https://arxiv.org/abs/1510.08090}{{\ttfamily
  1510.08090}}].

\bibitem{Freedman:1999gk}
D.~Freedman, S.~Gubser, K.~Pilch and N.~Warner, \emph{{Continuous distributions
  of D3-branes and gauged supergravity}},
  \href{https://doi.org/10.1088/1126-6708/2000/07/038}{\emph{JHEP} {\bfseries
  07} (2000) 038} [\href{https://arxiv.org/abs/hep-th/9906194}{{\ttfamily
  hep-th/9906194}}].

\bibitem{Cvetic:1999xx}
M.~Cvetic, S.~Gubser, H.~Lu and C.~Pope, \emph{{Symmetric potentials of gauged
  supergravities in diverse dimensions and Coulomb branch of gauge theories}},
  \href{https://doi.org/10.1103/PhysRevD.62.086003}{\emph{Phys. Rev. D}
  {\bfseries 62} (2000) 086003}
  [\href{https://arxiv.org/abs/hep-th/9909121}{{\ttfamily hep-th/9909121}}].

\bibitem{Khavaev:2000gb}
A.~Khavaev and N.~P. Warner, \emph{{A Class of N=1 supersymmetric RG flows from
  five-dimensional N=8 supergravity}},
  \href{https://doi.org/10.1016/S0370-2693(00)01228-4}{\emph{Phys. Lett. B}
  {\bfseries 495} (2000) 215}
  [\href{https://arxiv.org/abs/hep-th/0009159}{{\ttfamily hep-th/0009159}}].

\bibitem{Gowdigere:2005wq}
C.~N. Gowdigere and N.~P. Warner, \emph{{Holographic Coulomb branch flows with
  N=1 supersymmetry}},
  \href{https://doi.org/10.1088/1126-6708/2006/03/049}{\emph{JHEP} {\bfseries
  03} (2006) 049} [\href{https://arxiv.org/abs/hep-th/0505019}{{\ttfamily
  hep-th/0505019}}].

\bibitem{Cvetic:2000tb}
M.~Cvetic, H.~Lu and C.~Pope, \emph{{Geometry of the embedding of supergravity
  scalar manifolds in D = 11 and D = 10}},
  \href{https://doi.org/10.1016/S0550-3213(00)00215-7}{\emph{Nucl. Phys. B}
  {\bfseries 584} (2000) 149}
  [\href{https://arxiv.org/abs/hep-th/0002099}{{\ttfamily hep-th/0002099}}].

\bibitem{Bena:2018bbd}
I.~Bena, P.~Heidmann and D.~Turton, \emph{{AdS$_{2}$ holography: mind the
  cap}}, \href{https://doi.org/10.1007/JHEP12(2018)028}{\emph{JHEP} {\bfseries
  12} (2018) 028} [\href{https://arxiv.org/abs/1806.02834}{{\ttfamily
  1806.02834}}].

\bibitem{Bena:2019azk}
I.~Bena, P.~Heidmann, R.~Monten and N.~P. Warner, \emph{{Thermal Decay without
  Information Loss in Horizonless Microstate Geometries}},
  \href{https://arxiv.org/abs/1905.05194}{{\ttfamily 1905.05194}}.

\bibitem{Bena:2020yii}
I.~Bena, F.~Eperon, P.~Heidmann and N.~P. Warner, \emph{{The Great Escape:
  Tunneling out of Microstate Geometries}},
  \href{https://arxiv.org/abs/2005.11323}{{\ttfamily 2005.11323}}.

\bibitem{Martinec:2020cml}
E.~J. Martinec and N.~P. Warner, \emph{{The Harder They Fall, the Bigger They
  Become: Tidal Trapping of Strings by Microstate Geometries}},
  \href{https://arxiv.org/abs/2009.07847}{{\ttfamily 2009.07847}}.

\bibitem{Bobev:2020lsk}
N.~Bobev, E.~Malek, B.~Robinson, H.~Samtleben and J.~van Muiden,
  \emph{{Kaluza-Klein Spectroscopy for the Leigh-Strassler SCFT}},
  \href{https://arxiv.org/abs/2012.07089}{{\ttfamily 2012.07089}}.

\bibitem{Malek:2019eaz}
E.~Malek and H.~Samtleben, \emph{{Kaluza-Klein Spectrometry for Supergravity}},
  \href{https://doi.org/10.1103/PhysRevLett.124.101601}{\emph{Phys. Rev. Lett.}
  {\bfseries 124} (2020) 101601}
  [\href{https://arxiv.org/abs/1911.12640}{{\ttfamily 1911.12640}}].

\bibitem{Malek:2020yue}
E.~Malek and H.~Samtleben, \emph{{Kaluza-Klein Spectrometry from Exceptional
  Field Theory}},
  \href{https://doi.org/10.1103/PhysRevD.102.106016}{\emph{Phys. Rev. D}
  {\bfseries 102} (2020) 10}
  [\href{https://arxiv.org/abs/2009.03347}{{\ttfamily 2009.03347}}].

\bibitem{Klebanov:2009kp}
I.~R. Klebanov, S.~S. Pufu and F.~D. Rocha, \emph{{The Squashed, Stretched, and
  Warped Gets Perturbed}},
  \href{https://doi.org/10.1088/1126-6708/2009/06/019}{\emph{JHEP} {\bfseries
  06} (2009) 019} [\href{https://arxiv.org/abs/0904.1009}{{\ttfamily
  0904.1009}}].

\bibitem{Guarino:2020flh}
A.~Guarino, E.~Malek and H.~Samtleben, \emph{{Stable non-supersymmetric Anti-de
  Sitter vacua of massive IIA supergravity}},
  \href{https://arxiv.org/abs/2011.06600}{{\ttfamily 2011.06600}}.

\bibitem{Pufu:2010ie}
S.~S. Pufu, I.~R. Klebanov, T.~Klose and J.~Lin, \emph{{Green's Functions and
  Non-Singlet Glueballs on Deformed Conifolds}},
  \href{https://doi.org/10.1088/1751-8113/44/5/055404}{\emph{J. Phys. A}
  {\bfseries 44} (2011) 055404}
  [\href{https://arxiv.org/abs/1009.2763}{{\ttfamily 1009.2763}}].

\bibitem{Houppe:2020oqp}
A.~Houppe and N.~P. Warner, \emph{{Supersymmetry and Superstrata in Three
  Dimensions}},  \href{https://arxiv.org/abs/2012.07850}{{\ttfamily
  2012.07850}}.

\end{thebibliography}\endgroup

\end{adjustwidth}


\end{document}